\documentclass[amsfonts, amssymb, amsmath,  showkeys, nofootinbib, superscriptaddress, longbibliography, 11pt, floatfix,tightenlines]{revtex4-2}
\usepackage{float}
\makeatletter
\let\newfloat\newfloat@ltx
\makeatother
\usepackage[english]{babel}
\usepackage[utf8]{inputenc}
\usepackage{graphics}
\usepackage{selinput}
\usepackage[normalem]{ulem}
\usepackage[shortlabels]{enumitem}
\usepackage{wrapfig}
\usepackage{array}
\usepackage{braket}
\usepackage{amsthm}
\usepackage{mathtools}
\usepackage{physics}
\usepackage{xcolor}
\usepackage{graphicx}
\usepackage[left=16mm,right=16mm,top=35mm,columnsep=15pt]{geometry} 
\usepackage{adjustbox}
\usepackage{placeins}
\usepackage[T1]{fontenc}
\usepackage{lipsum}
\usepackage{csquotes}
\usepackage{bm}
\usepackage{bbm}
\usepackage{booktabs}
\usepackage{mathrsfs}
\usepackage{soul}

\newcommand{\DC}{\mathcal{D}}
\newcommand{\EC}{\mathcal{E}}

\newcommand{\dya}[1]{\ket{#1}\!\bra{#1}}
\newcommand*{\id}{\openone}

\newcommand{\CMSE}{\operatorname{CMSE}}
\newcommand{\MSE}{\operatorname{MSE}}
\newcommand{\BMSE}{\operatorname{BMSE}}
\newcommand{\thetaest}{\hat{\theta}^*}
\newcommand{\Bias}{\operatorname{Bias}}
\newcommand{\Var}{\operatorname{Var}}

\newtheorem{theorem}{Theorem}
\newtheorem{corollary}{Corollary}
\newtheorem{lemma}{Lemma}

\newenvironment{specialproof}{\textit{Proof:}}{\hfill$\square$}
\newenvironment{specialproofsketch}{\textit{Proof sketch:}}{\hfill$\square$}

\usepackage[makeroom]{cancel}
\usepackage[toc,page]{appendix}
\usepackage[colorlinks=true,citecolor=blue,linkcolor=magenta]{hyperref}

\begin{document}

\title{More buck-per-shot: Why learning trumps mitigation in noisy quantum sensing}

\author{Aroosa Ijaz}
\email{a4ijaz@uwaterloo.ca}
\affiliation{Theoretical Division, Los Alamos National Laboratory, Los Alamos, NM 87545, USA}
\affiliation{Department of Physics and Astronomy, University of Waterloo, ON N2L 3G1, Canada}
\affiliation{Vector Institute, Toronto, ON M5G 0C6, Canada}

\author{C. Huerta Alderete}
\affiliation{Information Sciences, Los Alamos National Laboratory, Los Alamos, NM 87545, USA}
\affiliation{Quantum Science Center, Oak Ridge, TN 37931, USA}

\author{Fr\'{e}d\'{e}ric Sauvage}
\affiliation{Theoretical Division, Los Alamos National Laboratory, Los Alamos, NM 87545, USA}
\affiliation{Quantinuum, Partnership House, Carlisle Place, London SW1P 1BX, United Kingdom}

\author{Lukasz Cincio}
\affiliation{Theoretical Division, Los Alamos National Laboratory, Los Alamos, NM 87545, USA}

\author{M. Cerezo}
\email{cerezo@lanl.gov}
\affiliation{Information Sciences, Los Alamos National Laboratory, Los Alamos, NM 87545, USA}
\affiliation{Quantum Science Center, Oak Ridge, TN 37931, USA}

\author{Matthew L. Goh}
\email{matt.goh@merton.ox.ac.uk}
\affiliation{Theoretical Division, Los Alamos National Laboratory, Los Alamos, NM 87545, USA}
\affiliation{Department of Materials, University of Oxford, Oxford OX1 3PH, United Kingdom}

\begin{abstract}
Quantum sensing is one of the most promising applications for quantum technologies. However, reaching the ultimate sensitivities enabled by the laws of quantum mechanics can be a challenging task in realistic scenarios where noise is present. While several strategies have been proposed to deal with the detrimental effects of noise, these come at the cost of an extra shot budget. Given that shots are a precious resource for sensing --as infinite measurements could lead to infinite precision-- care must be taken to truly guarantee that any shot not being used for sensing is actually leading to some metrological improvement. In this work, we study whether investing shots in error-mitigation, inference techniques, or combinations thereof, can improve the sensitivity of a noisy quantum sensor on a (shot) budget. We present a detailed bias-variance error analysis for various sensing protocols. Our results show that the costs of zero-noise extrapolation techniques outweigh their benefits. We also find that pre-characterizing a quantum sensor via inference techniques leads to the best performance, under the assumption that the sensor is sufficiently stable. 
\end{abstract}

\maketitle

\section{Introduction}

The tremendous progress in state-of-the-art technologies for quantum control has enabled the exciting possibility of using quantum mechanical systems as sensors for high-precision measurements~\cite{giovannetti2011advances,paris2009quantum, Ye2024Essay, degen2017quantum, szigeti2021improving, huang2024entanglementenhanced}. In a typical single-parameter quantum sensing setting, one allows a quantum probe to interact with some environment that encodes in the state an unknown parameter of interest. Then, by performing --a finite number of-- measurements on the system, one aims to estimate the value of the encoded parameter. When the parameter is encoded as a phase, it is well known that in the absence of quantum entanglement, the measurement precision is capped by the so-called Standard Quantum Limit (SQL). However, by allowing for entanglement in the quantum system, one can beat the SQL and reach the so-called Heisenberg Limit (HL), thus estimating the parameter with a higher precision~\cite{degen2017quantum}.

In practice, approaching the HL is a difficult task as several fundamental requirements must be met. First, one needs a precise knowledge of the \textit{system-environment interaction}, i.e., what the parameter encoding mechanism is.  Second, one needs to determine, as well as be able to prepare, the \textit{optimal probe state}, which is a state that is as sensitive as possible to the interaction of interest. Third, one has to determine the \textit{optimal measurement procedure}. Finally, one must characterize the  \textit{system's response functional relation}, i.e., the relation between the measurement outcome and the encoded parameter, as this is fundamental to actually extracting the unknown parameter value.

In the absence of hardware noise, the previous requirements can oftentimes be met for simple toy-model problems via theoretical calculations, making it possible to find schemes where the HL can be saturated ~\cite{escher2011general, demkowicz2012elusive,haine2015heisenberg}. The situation becomes significantly more challenging if noise is accounted for, as the determination of the optimal state, measurement, and corresponding system response becomes significantly more challenging and likely beyond analytical approaches. Moreover, it is well known that noise hinders a quantum state's capability as a sensor ~\cite{nolan2017optimal,muller2018noise,fiderer2019maximal,wang2020noise,cerezo2021sub,zhou2023optimal}, and that even the smallest noise levels make it impossible to reach the HL~\cite{garcia2023effects,kwon2023efficacy}. Despite recent advances in quantum error correction \cite{krinner2022realizing,zhao2022realization,acharya2022suppressing,bluvstein2023logical,da2024demonstration,acharya2024quantum,sundaresan2023demonstrating,putterman2024hardware}, an enormous amount of work remains to make scalable sub-threshold error correction practical. Consequently, all quantum sensing schemes for the foreseeable future will be afflicted by noise. This creates a pressing need for quantum sensing protocols that account for, mitigate, or otherwise acknowledge the presence of noise.

Recently, researchers have imported tools from quantum computing and learning theory to improve noisy-sensing schemes.  For instance,  variational techniques have been used to prepare the best possible probe state and measurement protocol~\cite{cerezo2020variationalreview,cerezo2022challenges,koczor2020variational, beckey2020variational, kaubruegger2021quantum, ma2020adaptive,  thurtell2022optimizing, liu2022variational, Le2023variational, meyer2020variational, kaubruegger2023optimal, marciniak2022optimal, direkci2024heisenberglimited, castro2024variational,maclellan2024end}, data-driven inference method have been leveraged to learn the system's response~\cite{huerta2022inference}, and error mitigation tools have been used to denoise the quantum probe state~\cite{yamamoto2022error,kwon2023efficacy,hama2023quantum, chen2024qubitassisted,dyke2024mitigating}.
 Although some prior works have shown utility for certain error mitigation techniques in quantum sensing to varying degrees of success~\cite{yamamoto2022error,kwon2023efficacy,hama2023quantum, chen2024qubitassisted}, these error mitigation techniques are `active' techniques which attempt to coherently reduce the noise levels during the quantum circuit execution~\cite{koczor2020exponential,huggins2020virtual}. While these methods have shown promising success, they have the downside of requiring additional quantum resources (e.g., multiple copies of the quantum probe state), which limits their applicability in qubit-limited devices. On the other hand  `passive' error mitigation techniques attempt to deal with noise incoherenrently, leveraging  classical postprocessing tools and therefore requiring an additional shot budget for their implementation. While passive error mitigation techniques are easier to implement in the near-term,  comparatively little study of their application to quantum sensing exists, and recent work suggests it may be more limited~\cite{dyke2024mitigating}. For any such method, care must be taken when assessing its true performance, especially if it comes at the cost of some number of shots. 
This is due to the combination of two facts. First, even in the presence of noise, a quantum sensor's sensitivity is directly proportional to the number of measurements, as per the Quantum Cram\'er-Rao bound~\cite{hayashi2004quantum,liu2020quantum}. As such, more measurements directly implies better precision. Second, many of the aforementioned techniques require a non-zero shot investment to work~\cite{cai2022quantum,li2017efficient, temme2017error, giurgica2020digital, endo2018practical, czarnik2020error, seif2022shadow, otten2019recovering, montanaro2021error, acampora2021genetic, lolur2023reference, tsubouchi2024symmetric, saxena2024error, huerta2022inference}. Hence, if one assumes that the  sensing experiment has a capped number of shots --which is the case in all realistic scenarios-- one must split these shots between those used for parameter estimation, and those used for tools such as error mitigation or inference. Hence, every shot not being used for parameter estimation must be truly guaranteed to improve the sensor's capabilities.

\begin{figure}[t!]
\centering
\includegraphics[width=.6\columnwidth]{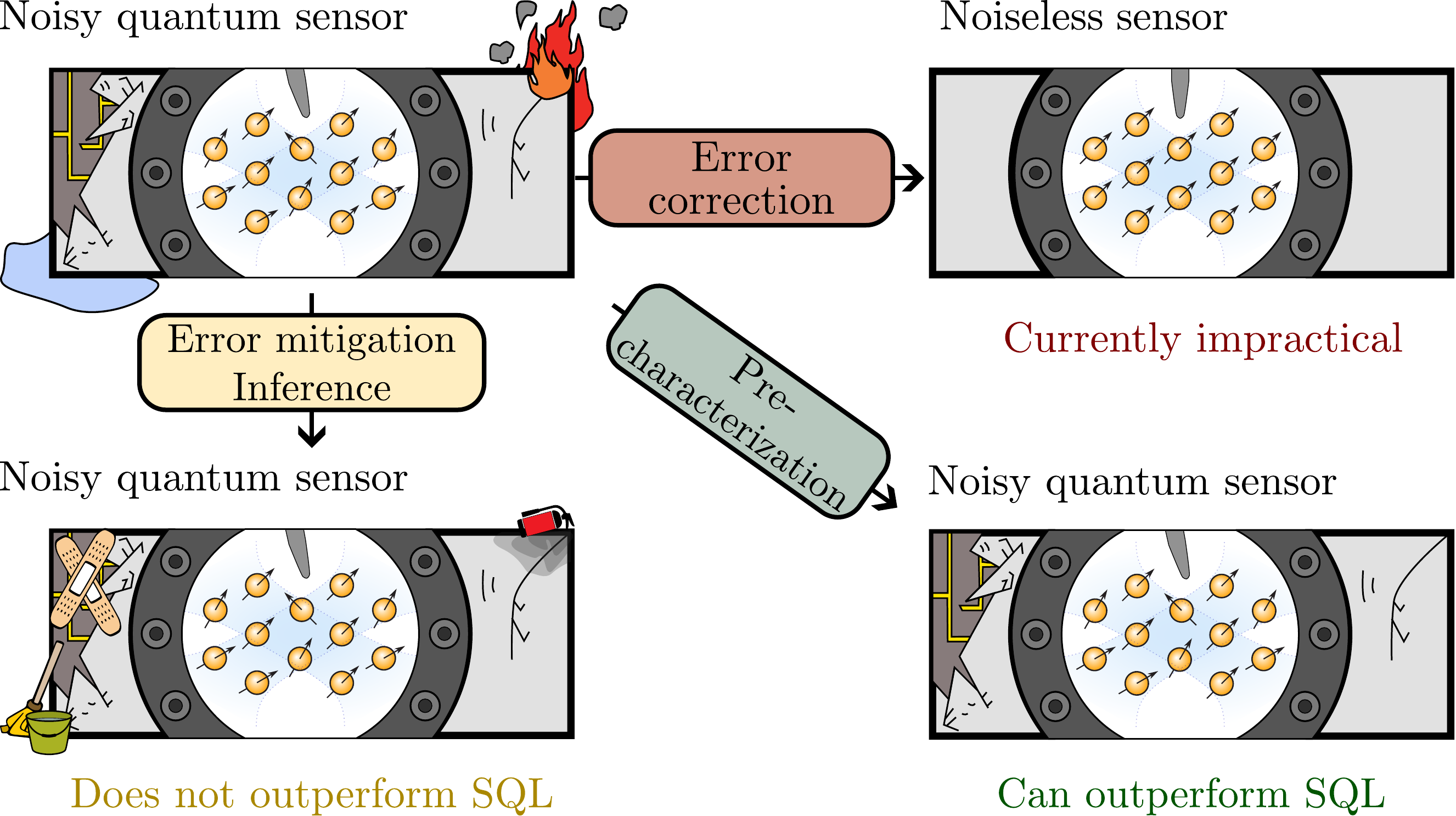}
\caption{\textbf{Summary of results.} We consider a noisy quantum sensing task, under the assumption that error correction is not available. We study whether the use of error-mitigation techniques (specifically ZNE) and inference tools can be used to improve the sensor's performance. Our results show that in order to beat the SQL, pre-characterization of the system's response is a necessary condition.}
\label{fig:Introduction_figure}
\end{figure}

As schematically shown in Fig.~\ref{fig:Introduction_figure}, in this work, we consider a noisy quantum sensing task of phase estimation and study whether it is worth investing shots in zero-noise extrapolation (ZNE) \cite{li2017efficient,temme2017error,giurgica2020digital} (the earliest, and most popular passive error-mitigation technique), inference-based methods~\cite{huerta2022inference} or their combinations. We therefore consider several sensing schemes that assume different levels of knowledge about the quantum sensor and analyze the expected sensitivity of each protocol by focusing on its bias-variance trade-off. We also consider the setup with \textit{pre-characterization} of the sensor where one pre-learns the system's response-functional relation -- assuming that the system-environment interaction is stable enough so that any information gained during the pre-characterization phase is still useful during the sensing phase.

Our theoretical and numerical results show that regardless of the existing knowledge about the system, ZNE is likely to lead to worse performance of the sensor. We find that \textit{noise-aware} sensing protocols (where the system's noisy response function is known) are most performant if the investment strategy is to use all the shots for noisy sensing. Hence, when the system response is not known, we argue that pre-characterization of the sensor is the right strategy. As such, we unveil that stability is a crucial (but often overlooked) criterion for a useful quantum sensor. This conclusion is summarized in Fig.~\ref{fig:Introduction_figure}.

\section{Framework}
\label{section:framework}

In this section, we present definitions and tools that will be used throughout this work. We begin by introducing a framework for noisy quantum sensing with unitary families.  We then show --via a toy model-- how and why noise can be detrimental to a quantum sensor's precision. Then, we recall the basics of ZNE for error mitigation. We finish by presenting inference tools that can be used to learn a noisy response function.

\subsection{Noisy-state quantum sensing}
\label{section:noisy_sensing_introduction}

We consider a noisy sensing setting where the goal is to use a quantum sensor to learn an unknown parameter $\alpha$ encoded into a quantum system by some environment. To begin, one prepares a probe state $\rho$ by sending some fiducial $n$-qubit state $\rho_{in}$ (such as the all-zero state) through a  (noisy) state preparation channel $\mathcal{E}_\lambda$. Here, $\lambda$ denotes a characteristic noise parameter\footnote{In practice, $\lambda$ can be a more complex tensor that characterizes the noise acting throughout the sensing protocol. However, for simplicity of notation, we will stick to the case when $\lambda$ is a unique real-valued parameter.}, such that $\lambda=0$ corresponds to a noiseless setting. In this work, we will not concern ourselves with how $\mathcal{E}_\lambda$ is determined. In particular, while in some noiseless cases, the optimal preparation channel $\mathcal{E}_0$ can be analytically derived, these scenarios are limited and rare when noise is accounted for. Thus, if $\lambda\neq0$, one generally needs to either heuristically train a probe state preparation channel~\cite{koczor2020variational, beckey2020variational, kaubruegger2021quantum, ma2020adaptive, marciniak2022optimal, thurtell2022optimizing, liu2022variational, Le2023variational, meyer2020variational, kaubruegger2023optimal, direkci2024heisenberglimited, castro2024variational} or simply implement $\mathcal{E}_0$ --assuming that one knows it-- and hope that the noise will not be too detrimental. 

The probe state is then made to interact with an external field that encodes an unknown parameter $\alpha$ in the probe state through the action of a Hamiltonian that characterizes the system-environment interaction
\begin{equation}
    H=\alpha H_{\text{field}}\,.
\end{equation}
As such, the parameter encoding channel is 
\begin{align}\label{eq:encoding-channel}
    \mathcal{S}_{\theta}(\rho) = e^{-i \theta H}\rho e^{i \theta H}\,,
\end{align}
where $\theta$ is a phase imprinted on the state by the interaction. Here, it is important to note that it is sometimes standard to conflate $\theta$ and $\alpha$. However, by measuring the phase $\theta$, one can obtain information about the parameter $\alpha$. For instance, in the simplest case, one has $\theta=\alpha T$ where $T$ is the interaction time (in Sec.~\ref{sec:parameter_measurement_and_phase_measurement} we discuss how additional effects can lead to more intricate functional relations between  $\theta$ and $\alpha$). 

Finally, one sends the state through a noisy pre-measurement channel\footnote{Note that $\mathcal{D}_\lambda$ includes any noise which acts immediately after $\mathcal{S}_{\theta}$.} $\mathcal{D}_\lambda$, after which one computes the expectation value of some suitable observable $O$. We again note that analytically obtaining $\mathcal{D}_\lambda$ and an optimal $O$ for any $\lambda\neq 0$ can be difficult, and their determination is usually performed heuristically. 

Putting it all together, the system's response $R_{\lambda}(\theta)$ at noise level $\lambda$ is given by
\begin{align}\label{eq:response}
    R_{\lambda}(\theta) = \Tr[\mathcal{D}_\lambda \circ \mathcal{S}_{\theta} \circ \mathcal{E}_\lambda (\rho_{in}) O]\,.
\end{align} 
In what follows, we will denote $R(\theta)\equiv R_{0}(\theta)$ as the noiseless response, and $\theta^*$ as the unknown phase of interest that one wants to estimate. Moreover, since in practice one estimates the system response via $N$ shots (i.e., $N$ independent repetitions of the experimental measurement), we will use $\overline{R}$ to indicate that the response was estimated via a finite number of measurement shots. Appendix~\ref{appendix:notation_summary_table} summarizes the notation used in this work.

At this point, we find it important to make several remarks. First, we note that in order to extract the value of a given $\theta^*$ from the $N$-shot estimated $\overline{R}_{\lambda}(\theta^*)$, one needs to invert the response function. That is, given $R_{\lambda}(\theta)$ and a known domain to which $\theta^*$ belongs (and in which $R_{\lambda}$ is invertible),  one can obtain an estimate of $\theta^*$ as $\thetaest=R_{\lambda}^{-1}(\overline{R}_{\lambda}(\theta^*))$. As previously mentioned, although in some limited cases one might know $R_{0}$, we highlight that in general one does not have access to the functional form of the noisy response function. To address this issue, inference techniques have been proposed to learn the response ~\cite{huerta2022inference} (see below for more details).

Next, we recall that the sensitivity of the noisy quantum sensing scheme, i.e., the precision to which the unknown phase is estimated, can be obtained via the error propagation formula~\cite{giovannetti2006quantum} 
\begin{equation}\label{eq:precision}
    \Delta^2\thetaest=\frac{\Delta^2 \overline{R}_{\lambda}(\theta^*)}{(\partial_\theta R_{\lambda}(\theta)|_{\theta=\theta^*})^2}=\frac{\Delta^2 R_{\lambda}(\theta^*)}{N(\partial_\theta R_{\lambda}(\theta)|_{\theta=\theta^*})^2}\,.
\end{equation}
Here, $\Delta^2\thetaest$ denotes the variance of the estimated phase, while $\Delta^2 \overline{R}_{\lambda}(\theta^*) =\Delta^2 R_{\lambda}(\theta^*)/N$ and $\Delta^2 R_{\lambda}(\theta^*)= \Tr[\mathcal{D}_\lambda \circ \mathcal{S}_{\theta^*} \circ \mathcal{E}_\lambda (\rho_{in}) O^2]-\Tr[\mathcal{D}_\lambda \circ \mathcal{S}_{\theta^*} \circ \mathcal{E}_\lambda (\rho_{in}) O]^2$ respectively correspond to the variances of the response mean  (obtained via $N$ samples) and variance of the response's distribution.  Finally, $\partial_\theta R_{\lambda}(\theta)$ represents the partial derivative of the response.  

As shown by Eq.~\eqref{eq:precision}, the precision to which one can estimate $\theta^*$ depends on three terms. First, the variance $\Delta^2 R_{\lambda}(\theta^*)$ which quantifies the degree to which the measured state is not an eigenstate of the measured observable. The larger this variance, the larger the parameter estimation error, and thus, the smaller the precision. Second, the partial derivative indicates how sensitive the response is to the encoded parameter so that larger derivatives imply a more sensitive sensor. Finally, increasing the number of measurement shots $N$ increases the precision of approximating $\theta^*$, making it clear that shots are a veritable ressource for quantum sensing. 

\subsection{Toy model: Noisy magnetic field sensing with GHZ states under global depolarizing noise}
\label{sec:global_depol_toy_model}

In this section, we present a simple toy model which illustrates the detrimental effects that quantum noise has on a quantum sensor. We focus on a magnetometry task where one wants  to estimate the value $\alpha$ of an unknown magnetic field that is encoded via a channel $\mathcal{S}_{\theta}(\rho)$ as in Eq.~\eqref{eq:encoding-channel} with
\begin{align}
    H=\frac{\alpha}{2}\sum_{j=1}^nZ_j\,,
\end{align}
where $Z_j$ denotes the Pauli $Z$ operator acting on the \mbox{$j$-th} qubit. For this example, we assume that the dominant source of noise is in state preparation, and thus assume that $\mathcal{D}_\lambda$ is the identity map. 

In a completely noiseless setting, and when the state preparation channel cannot generate any entanglement, the highest reachable precision is given by the SQL, $\Delta^2\theta=\frac{1}{N n}$. This limit is reached by the probe state $\mathcal{E}_0 (\rho_{\text{in}})=\dya{+}^{\otimes n}$ and observable $O=\sum_{j=1}^n X_j$; which together lead to the response function $R(\theta)=n\cos(\theta)$.  However, if $\mathcal{E}_0$ is allowed to create entanglement between the qubits, then it is known that the optimal probe state is the GHZ state $\mathcal{E}_0 (\rho_{\text{in}})=\dya{\text{GHZ}}$ with $\ket{\text{GHZ}}=\frac{1}{\sqrt{2}}(\ket{0}^{\otimes n}+\ket{1}^{\otimes n})$, the optimal measurement is the parity operator $O=X^{\otimes n}$, so that the response function is given by $R(\theta)=\cos(n\theta)$. In this case, one finds $\Delta^2 R(\theta)=1-\cos^2(n\theta)$, $|\partial_\theta R(\theta)|=n|\sin(n\theta)|$, and the precision is $\Delta^2\theta=\frac{1}{N n^2}$, 
which corresponds to the ultimate HL.  Hence,  allowing for entanglement in the probe states enables us to estimate the unknown parameter with a precision (error) that is a factor of $n$  larger (smaller) than without entanglement.  

When noise is accounted for, one can generically expect the sensitivity and precision to decrease. As an example, let us assume that global depolarizing noise acts throughout the $\EC_\lambda$ and $\DC_\lambda$ channels so that the parameter encoded state becomes
$(1-p_\lambda) \mathcal{S}_{\theta}(\rho_{\text{in}})+p_\lambda\id/2^n$. Here $p_\lambda$ is the total depolarization probability and strictly increases with the noise parameter $\lambda$, while $\id$ denotes the $(2^n\times 2^n)$-dimensional identity matrix. Straightforward calculations show that  $R_{\lambda}(\theta)=(1-p_\lambda) \cos(n\theta)$, which leads to  $\Delta^2R_\lambda(\theta)=1- (1-p_\lambda)^2 \cos^2(n\theta)$, $|\partial_\theta R_{\lambda}(\theta)|=(1-p_\lambda)n|\sin(n\theta)|$, and thus
\begin{equation}\label{eq:noise-GHZ-delta}
    \Delta^2\theta=\frac{2p_\lambda- p_\lambda^2}{Nn^2 (1-p_\lambda)^{2} \sin^2(n\theta^*) } + \frac{1}{Nn^2}\,,
\end{equation}
which is strictly larger than $\frac{1}{Nn^2}$ for any $0<p_\lambda< 1$. Here we will parameterize $p_\lambda=1-e^{-\lambda}$ such that $\lambda$ is a fault rate. Lastly, note that it is entirely possible to have a system that is so noisy that  $\Delta^2\theta$ in Eq.~\eqref{eq:noise-GHZ-delta} becomes larger than the SQL value of $\frac{1}{Nn}$, thus losing all advantages allowed by entanglement. Given that creating entangled states requires more operations than creating separable states, one can expect that noise will be particularly detrimental when trying to reach the HL via entangled states. 

This example perfectly illustrates that the detrimental effect of noise can be two-fold. First, it can make the state more spread across the eigenvalues of $O$, which increases the response's variance. Second, the signal usually gets flattened, and thus the derivatives get smaller. That is, one can expect that for $\lambda>\lambda '$, $\Delta^2 R_{\lambda}(\theta^*)\geq \Delta^2 R_{\lambda'}(\theta^*)$ and $\partial_\theta R_{\lambda}(\theta)|_{\theta=\theta^*}\leq \partial_\theta R_{\lambda'}(\theta)|_{\theta=\theta^*} $.

Furthermore, this example highlights a crucial principle of noisy quantum sensing. Although a larger system size $n$ improves the precision of the HL, if the noise level $\lambda$ grows cumulatively in $n$ (as it most typically will, due to e.g. additional entangling gates required), then there will be a point beyond which larger sensors are too noisy to provide additional metrological advantage. Without scalable quantum error correction, one cannot expect to indefinitely achieve the HL scaling $\Delta^2 \theta \propto 1/n^2$. Nonetheless, a metrological advantage from entanglement may still be found at limited scale if one can achieve a precision $\Delta^2\theta \leq \frac{1}{Nn}$ for some $n$, and making optimal use of noisy quantum sensors will require the use of techniques that can extend this window.

We note that while the previous results were derived for global depolarizing noise, they will generally hold for a wide range of Pauli noise models which have the maximally mixed state as their fixed point~\cite{wang2020noise,franca2020limitations,muller2016relative}. In fact, in~\cite{garcia2023effects} it was shown that for a wide range of realistic noise models, the eigenvalues of the quantum Fisher information --and thus the quantum probe sensitivity (through the Quantum Cram\'er-Rao bound~\cite{hayashi2004quantum,liu2020quantum})-- become exponentially suppressed with the noise strength. 

The previous results pose the question: ``\textit{What is the best strategy to deal with these negative effects?}'', and as we will see there are several options available. On the one hand, one can simply opt to use separable states and try to reach the SQL, as circuits for preparing separable states can be expected to be less noisy than circuits for preparing entangled states. Second, one can try to use the noisy sensor and hope that the noise levels still allow for sub-SQL errors. Finally, we ask whether one can employ error mitigation techniques to somehow tame the effects of noise. Error mitigation techniques have already been applied to quantum sensing, including virtual distillation~\cite{yamamoto2022error, kwon2023efficacy}, and even combining quantum metrology with quantum error correction tools~\cite{hama2023quantum, chen2024qubitassisted, koczor2020exponential}. Here, we instead study if ZNE~\cite{li2017efficient, temme2017error, giurgica2020digital} can be used to obtain an error-mitigated response function and a more accurate parameter estimate.

\begin{figure*}
	\centering
	\includegraphics[width=0.95\textwidth]{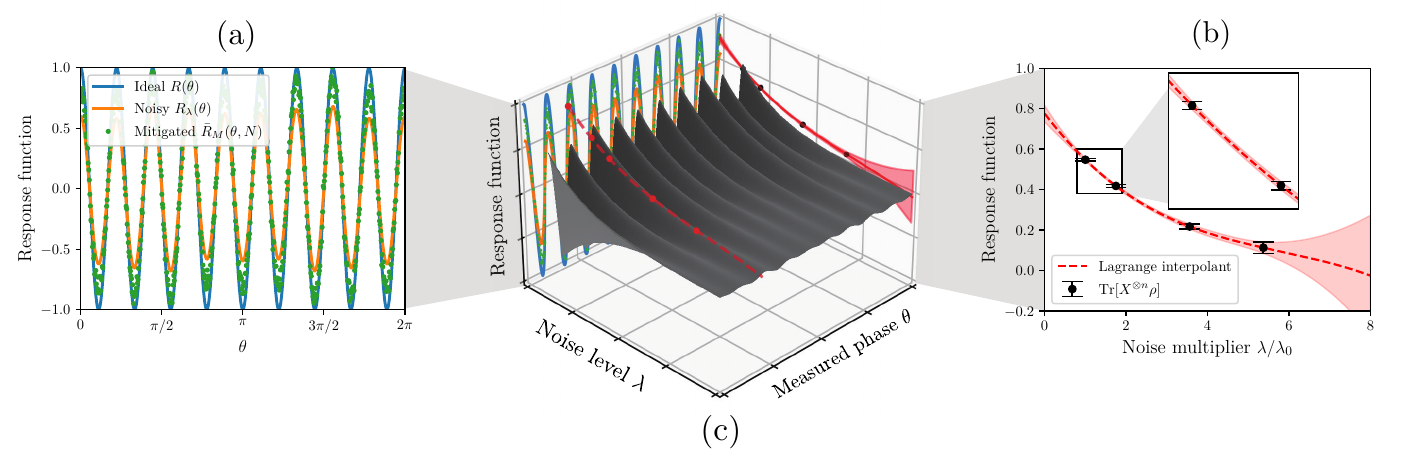}
	\caption{\textbf{Using zero-noise extrapolation for noisy quantum sensing.} Here we illustrate the use of ZNE to mitigate errors in a GHZ-state magnetometry task. (a) The ideal response function is $R(\theta)=\cos(n\theta)$ (blue). When realistic noise is included in state preparation, a noisy response function $R_{\lambda}(\theta)$ (orange) of smaller amplitude and differing periodicity is obtained. Applying ZNE to the noisy response function yields an imperfect approximation $\overline{R}_M(\theta, N)$ (green) of the ideal response function. (b) Fixing the phase at $\theta = \theta^*$, a zero-noise estimate $\overline{R}_M(\theta^*, N)$ is computed by evaluating the noisy response function at multiple boosted noise levels (black nodes) and computing a polynomial interpolant (mean result red dashed line, shaded bars represent 95\% confidence intervals). Noise nodes and shots are allocated as described in Appendix~\ref{appendix:ZNE_hyperparameters}. Inset: the colored area represents the variance in response $\overline{R}_{\lambda_j}(\theta^*, N_j)$ at each noise node. (c) The zero-noise-extrapolated response function $\overline{R}_M(\theta, N)$ can be determined for any value of $\theta$ with the shot budget $N$ by applying the procedure shown in (b), resulting in the curve depicted in (a).}
	\label{fig:ZNE_demo}
\end{figure*}

\subsection{Zero-noise extrapolation}
\label{section:ZNE_introduction}

\subsubsection{The method}

The core primitive of extrapolation-based error mitigation methods is the observation of a noisy quantum system at various noise levels to obtain the zero-noise value by extrapolating a fitting function. In this work, we will focus on Richardson extrapolation which has been shown to achieve a more accurate estimation of the zero-noise value when the noise and system size are small \cite{temme2017error}, and can be applied independently of the underlying noise model \cite{krebsbach2022optimization}. Given the phase of interest $\theta^*$ and a minimum possible noise $\lambda$, one observes the system response at $m+1$ distinct noise levels. These nodes are set as $\lambda_j = x_j \lambda$, where $x_j$ is the noise amplifying parameter satisfying  $x_j \in [1, \infty)\; \forall j$ and $x_0< x_1 < ,\dots, < x_m$, and with $x_0 =1$ defining the base noise level, $\lambda_0=\lambda$. The Lagrange polynomial of degree $m$ that interpolates the set of noisy responses $\{ \lambda_j, R_{\lambda_j}(\theta^*) \}_{j=0}^{m}$ is then evaluated at zero noise to get the mitigated response, denoted as $R_{M}$, via the equation
\begin{align} \label{eq:EMpoly}
	R_{M}(\theta^*) = \sum_{j=0}^{m} \gamma_j \; R_{\lambda_j}(\theta^*)\,, 
\end{align}
where $\gamma_j$ are the Lagrange basis polynomial coefficients at $x=0$ (see Appendix~\ref{appendix:richardson_extrapolation} for additional details). Generally, optimizing the hyper-parameters of this protocol (e.g., the number and spacing of noise nodes) for a given noise model can be computationally expensive (see Appendix~\ref{appendix:ZNE_hyperparameters}).

\subsubsection{The shot cost}
Here it is important to note that in practice one is limited to performing error mitigation with a finite shot budget $N$ that is usually dictated by hardware (or even economical) constraints. Importantly, the way in which the available shots are distributed across the m+1 noise nodes (i.e., how many measurements per noise level) can affect ZNE scheme's performance. However, one can optimally distribute the measurements following Lemma 1 in~\cite{Hoel1994optimal} (see Eq.~\eqref{eqn:richardson_shot_distribution}). Given that the response at different noise levels could be estimated with different number of shots, we use $\overline{R}_{\lambda_j}(\theta^*, N_j)$ to explicitly denote the expectation value obtained at noise level $\lambda_j$, obtained using $N_j$ measurement shots. Therefore, the corresponding zero-noise estimate at $\theta^*$ obtained from Lagrange interpolating the set $\{ \lambda_j, \overline{R}_{\lambda_j}(\theta^*, N_j) \}_{j=0}^{m}$ can be written as
\begin{align} \label{eq:EMpolyEst}
    \overline{R}_{M}(\theta^*, N) &= \sum_{j=0}^{m} \gamma_j \; \overline{R}_{\lambda_j}(\theta^*, N_j)\,,
\end{align}
where $\sum_{j=0}^m N_j = N$ is satisfied.

Hence, one can see that the hope of using ZNE for noisy quantum sensing relies on obtaining a good estimate of $\theta^*$ as $\thetaest=R^{-1}(\overline{R}_{M}(\theta^*, N))$. Note that for phase estimation to work here, we have to assume that the noiseless response function $R(\theta)$ is known. In Fig.~\ref{fig:ZNE_demo}, we illustrate how ZNE may be applied to a noisy quantum sensor. In later sections, we evaluate whether ZNE can be expected to improve its sensitivity when the overall budget is fixed, and compare it to alternative protocols.

\medskip

\subsection{Inference for learning the system's response}
\label{Subsection:InferenceIntroduction}

\subsubsection{The method}
As previously mentioned, one of the main difficulties of noisy quantum sensing is the fact that the functional form of the noisy response is unknown. To address this issue, the work of Ref.~\cite{huerta2022inference} proposed an inference-based method to learn the underlying noisy response function. This technique is aimed at sensing schemes where the system-environment Hamiltonian takes the form $H= \sum_j h_j$ with $h^{2}_j = \id$ and $[h_j, h_{j^{'}}]=0$ $\forall j, j^{'}$. Under these conditions, the noisy response function is a $n$-degree trigonometric polynomial function (see Appendix~\ref{appendix:Inference-based-sensing}), i.e., 
\begin{align}
\label{eq:trigPolynomial}
	R_{\lambda}(\theta)= \sum_{s=1}^{n} [a_s \cos(s\theta)+ b_s \sin(s\theta)] + c \,.
\end{align}
Now, one can measure the response at a set of $2n+1$ distinct phase values $\{\theta_k\}$ that are uniformly sampled from $[0,2 \pi)$, acquiring an inference training set $\{\theta_k,  R_{\lambda}(\theta_k) \}_{k=1}^{2n+1}$. The ensuing measurement outcomes can be used to solve a linear system of equations and therefore learn the unknown coefficients $\{a_s,b_s,c\}$ in Eq.~\eqref{eq:trigPolynomial}. The outcome of this procedure is an inferred noisy response function, denoted as  $\Tilde{R}_{\lambda}(\theta)$.

Note that the inference method described does not require exhaustive characterization of the noise, but only of its effect on the response function, which is a far less demanding requirement. As such, in realistic scenarios where the system dynamics are not easily accessible, it is much cheaper to use inference to obtain the functional form of $R_{\lambda}(\theta)$ than to perform full tomography of the noise channels. Moreover, for a general class of unitary families, this data-driven framework can be used to learn the response function and predict the sensitivity for arbitrary probe states, measurement schemes, and a wide class of quantum noise models.

\subsubsection{The shot cost}
\begin{figure}
	\centering
	\includegraphics[width=0.45\textwidth]{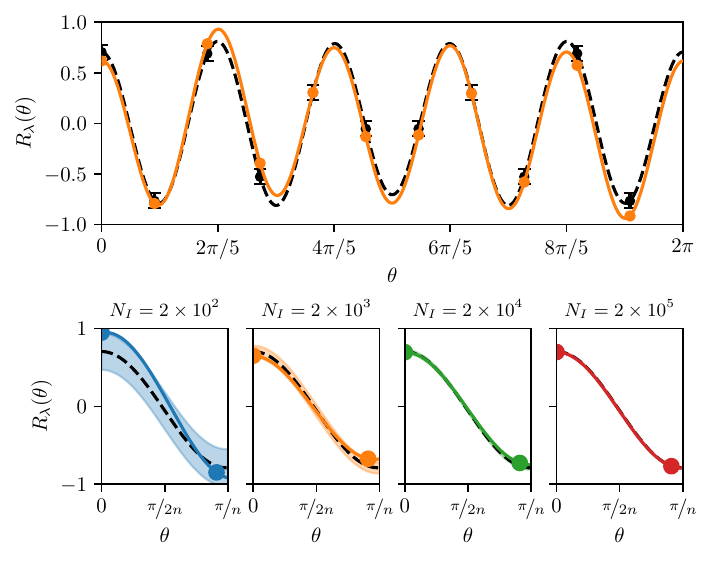}
	\caption{\textbf{The effect of shot budget on response function inference.} Here we illustrate the inference of a response function for a 5-qubit system following the noise model of Appendix~\ref{appendix:ibm_eagle_model}, at varying inference shot budgets $N_I$. (Top) Evaluating exactly $R_\lambda(\theta_k)$ at $2n+1$ nodes $\theta_k$ (black dots) and performing trigonometric interpolation exactly recovers the noisy response function $R_\lambda(\theta)$ (black dashed line). In practice, one has a limited inference shot budget $N_I=(2n+1)N_k$ and obtains $N_k$-shot estimated $\overline{R}_\lambda(\theta_k, N_k)$ (black error bars indicate uncertainty at $N_I=2\times 10^3$, orange dots indicate a single realization of sampling at this inference budget). Performing trigonometric interpolation yields an approximate inferred response function $\tilde{R}_\lambda(\theta, N_I)$ (orange line). (Bottom) We compare the inferred response function on a single fringe at four inference budgets, where colored dots again represent single realizations of inference nodes, the black dashed line is the exact response $R_\lambda(\theta)$, and shaded regions represent the uncertainty of the inferred response function. The error in inference clearly decreases with increasing $N_I$.}
	\label{fig:inference_illustration}
\end{figure}

When implementing the aforementioned inference scheme in practice, one must work with a finite shot budget $N_I$. To illustrate this fact, we denote the corresponding inferred function as $\tilde{R}_\lambda(\theta, N_I)$ which is obtained via the finite-sample inference training set $\{ \theta_k, \overline{R}_{\lambda}(\theta_k, N_k) \}_{k=1}^{2n+1}$. We will set the shots used for measurement at each $\theta_k$ as $ N_k \equiv \frac{N_I}{2n+1} $.  Hence, inference can only be implemented up to an error due to sampling errors as illustrated in Fig.~\ref{fig:inference_illustration}.

Next, for the parameter estimation step, a noisy observation is made at $\theta^*$ using a shot budget of $N_E$; namely $\overline{R}_\lambda(\theta^*,N_E)$ is measured. The inferred response function can now be used to invert this measurement obtaining a phase estimate $\thetaest =\tilde{R}_\lambda^{-1}(\overline{R}_\lambda(\theta^*, N_E))$, assuming $\tilde{R}_{\lambda}(\theta, N_I)$ is bijective in the interval of interest. Hence, inference-based quantum sensing relies on obtaining a good estimate of $\theta^*$ given the inferred function closely approximates the noisy function. 

In the present setting, there are shot costs associated with both inference (obtaining $\tilde{R}_\lambda(\theta,N_I)$) and estimation (measuring $\overline{R}_\lambda(\theta^*,N_E)$), and the total shot budget must be divided amongst the two steps; $N= N_I + N_E$. However, we note that while the estimation step requires the sensor to be exposed to the target field, the inference step does not. Therefore, one could treat inference as a pre-characterization step for the sensor, in which case the whole shot budget is allocated for estimation, i.e., $N=N_E$. A single characterization can then be used for many parameter estimation steps, so long as the noise model remains relatively stable. To draw fair comparisons to other sensing protocols, we consider both of these scenarios below.

At this point we note that while both error mitigation and inference techniques have an intrinsic show budget associated to them, there are some important difference that one needs to  highlight. As outlined throughout Sec.~\ref{section:framework}, noisy quantum sensing depends on both the noise present in the sensor's response $R_\lambda(\theta^*)$ to the field being measured, as well as on correct knowledge of the system's overall response-functional relationship $R_\lambda(\theta)$ on some relevant domain of $\theta$ such that this can be inverted. 
The latter is a property of the sensor itself that is completely independent of the parameter being estimated. As such, one can use inference to pre-characterize the system response and use this knowledge for posterior parameter estimations. In contrast, error mitigation techniques are used to denoise the sensor's response $R_\lambda(\theta^*)$ for a fixed field imprinting a phase $\theta^*$, and therefore the additional measurement resources must be repeated for every new field one wishes to measure.

\subsection{Parameter measurement via phase measurement}\label{sec:parameter_measurement_and_phase_measurement}

At this point, it is important to highlight a few subtleties regarding how to actually extract the field strength from the measurement outcomes. Given that the simple case of magnetometry via GHZ state plus parity measurement already showcases some of the key issues, we will start by evaluating this case. To begin, note that the phase imprinted on the probe quantum state $\theta^*$ is not the quantity of interest, the field strength $\alpha$ is. For this simple case, $\alpha$ is related to the imprinted phase by $\theta^*=\alpha T$. Next, since the response function is periodic $(R(\theta)=\cos(n\theta)= R(\theta+2\pi/n))$, it is not invertible over its whole domain. Along the same lines, to uniquely map noisy measurements of expectation values to phases, one needs to restrict to some domain $\theta\in[\theta_{\text{min}},\theta_{\text{max}}]$ over which $R_\lambda(\theta)$ is bijective. Therefore, estimating the phase as  $\thetaest=R_{\lambda}^{-1}(\overline{R}_{\lambda}(\theta^*))$ only allows us to obtain a value of $\theta^*$ which can be used, at best, to determine the remainder of $\alpha T$ modulo $2\pi/n$. 

This illustrates the fact that a quantum sensor, on its own, cannot unequivocally determine $\alpha$. This issue can be resolved by realizing that quantum sensors should be used to improve one's knowledge of a parameter via high-precision measurements. That is, quantum sensors can reduce the error in the determination of a given parameter for which one has a prior estimate $\alpha_{\text{prior}}$, usually obtained by classical means. Indeed, having access to $\alpha_{\text{prior}}$ enables us to determine suitable $\theta_{\text{min}}$ and $\theta_{\text{max}}$, and perform the response function inversion. 

Generally, when noise is accounted for, the period of the function might change and the sensitivity $\Delta^2\theta$ of the sensor can vary with $\theta$~\cite{huerta2022inference}. In this case, one needs to determine the domain over which the response will be inverted, as well as guarantee that the sensor is tuned to the point of optimal sensitivity. If knowledge of the functional form of the response function is available (e.g., because one has used inference techniques for pre-characterization), then one can identify its most sensitive region \eqref{eq:precision} and intentionally imprint an additional phase $\theta_{\text{bias}}$ to shift the sensor towards that highly sensitive region, if required. The target phase would then become $\theta^*=\alpha T+\theta_{\text{bias}}$. 
Physically, this could correspond to the application of continuous rotation gates, or the application of a bias field or frequency chirp as seen in quantum sensing experiments \cite{degen2017quantum,  huang2024entanglementenhanced, saywell2023enhancing, borras2024aquantumsensing}. 

Since the prior (classical) knowledge of the field strength $\alpha$ directly affects sensitivity, our modeling must make assumptions on how well the field is already known. Throughout this work, we will assume that whenever an $N$-shot quantum measurement protocol is executed, it is initialized with an optimal $N$-shot classical measurement $\alpha_{\text{prior}}$ that saturates the SQL. Due to the (classical) measurement uncertainty in $\alpha_{\text{prior}}$, the target phase for the quantum sensing protocol will have an innate randomness, and we acknowledge this fact by introducing the phase random variable $\Theta^*$. Additionally, in practice one will not be able to apply $\theta_{\text{bias}}$ to arbitrary precision and will at best be limited by, e.g., the resolution of the digital-to-analog converter used in the control system, leading to some rounding error $\epsilon_B$. Combining these effects, we assume that for a single run of any sensing protocol, the target phase is drawn from a normal distribution as
\begin{equation}
    \Theta^* \sim \mathcal{N}\left(\alpha_{\text{prior}}T+\theta_{\text{bias}}+\epsilon_B,\frac{1}{Nn}\right),
    \label{eqn:theta_star_distribution}
\end{equation}
where the rounding error is $\epsilon_B=2\pi/2^B$ for $B$ bits of precision, and we assume a precision $B=1 0$ throughout (which is a generous assumption for rotation gates on even late-NISQ hardware~\cite{koczor2024probabilistic}). We emphasize that the randomness of this distribution originates entirely from uncertainty in the classical prior measurement, and not e.g., instability in the field itself, which we do not model. Crucially, each run of a quantum sensing protocol will be conditioned on a particular value of $\Theta^*$.

\section{Analytic estimates of error bounds}
\label{section:theoretical}

\begin{figure}[t!]
\centering
\includegraphics[width=.6\linewidth]{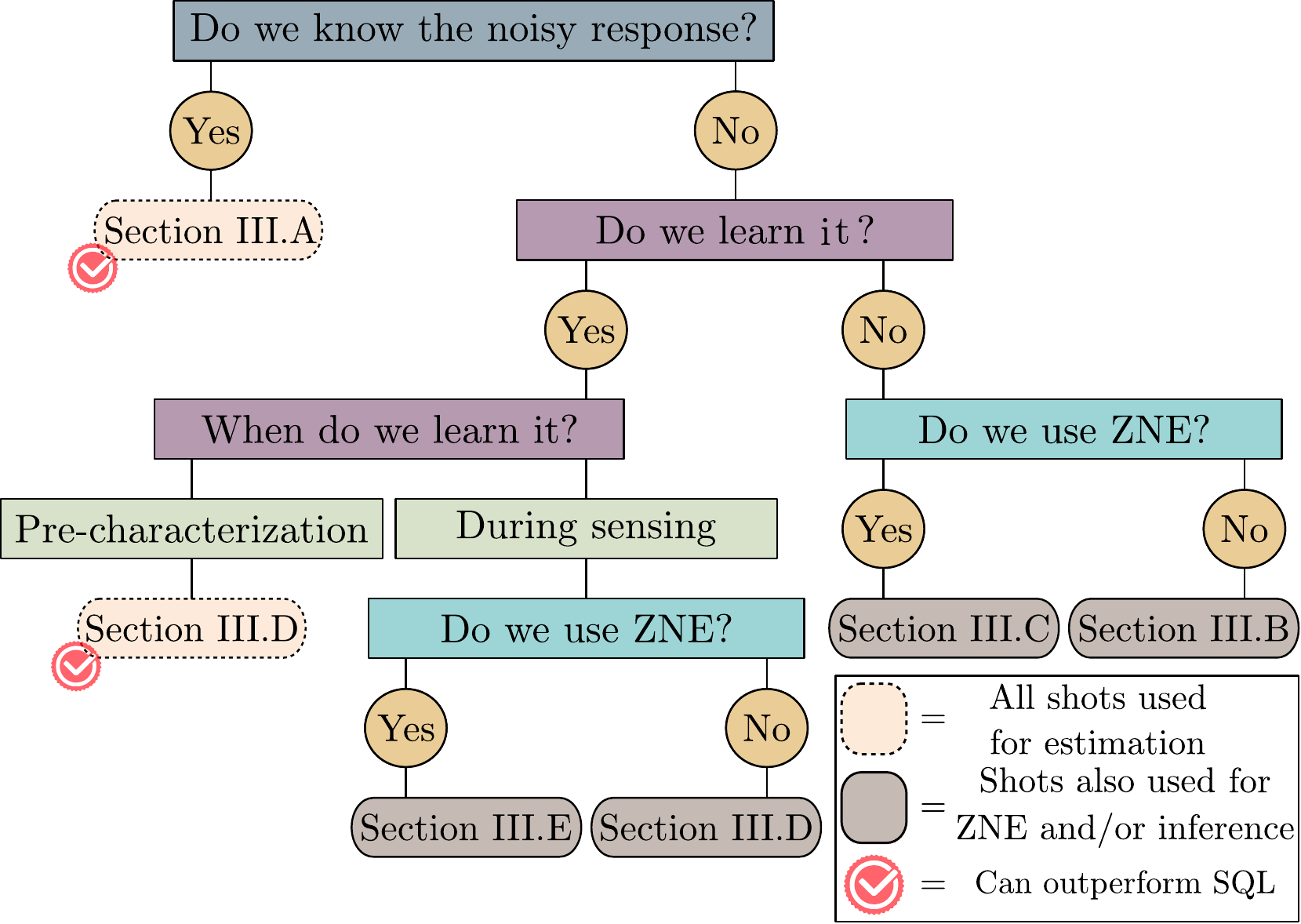}
\caption{\textbf{Flowchart of the protocols studied in this work.} We show a flowchart that explains the connection between the different sensing protocols presented in Sec.~\ref{section:theoretical}.}
\label{fig:flowchart}
\end{figure}

In this section, we present different noisy quantum sensing protocols based on error mitigation, inference, and combinations thereof. The connection between the different protocols is schematically shown in Fig.~\ref{fig:flowchart}.  Our goal will be to derive and compare error bounds for each setting, as these bounds will provide qualitative insight into their limiting factors. Such analysis also allows us to approximately compare shot-for-shot the performance of each protocol, and determine which (if any)  can outperform the SQL and achieve a genuine quantum advantage. As noted previously, the level of noise afflicting an entangled quantum sensor will typically grow cumulatively with its size $n$, and one should therefore not expect to asymptotically beat the SQL. Nonetheless, a genuine advantage can be obtained at some $n$ if a parameter can be estimated with a precision less than $1/Nn$; throughout this work, when we state that a protocol outperforms the SQL, we mean in that particular regime.

To begin, let us recall that in the previous section, we have defined the phase random variable $\Theta^*$.
Consider that in a single execution of the N-shot sensing protocol, a single prior classical estimate is used, and therefore $\Theta^*$ takes on a value of $\theta^*$ from its sample space. 
As such, we define the conditional mean squared error (CMSE) in the corresponding phase estimate $\thetaest$  as
\begin{equation}
    \label{eqn:mse_conditional}
    \CMSE[\thetaest |\theta^*] \equiv \mathbb{E}_N\left[(\thetaest-\theta^*)^2 \big| ~ \Theta^* = \theta^* \right]\,,
\end{equation}
where $\mathbb{E}_N$ indicates averaging over multiple runs of phase estimation (each run has a N-shot budget). Following the example of Refs.~\cite{marciniak2022optimal,kaubruegger2023optimal, direkci2024heisenberglimited}, we define the Bayesian mean-squared error (BMSE) as
\begin{align}
\BMSE[\thetaest]&\equiv\mathbb{E}_{\Theta^*}\left[\CMSE[\thetaest | \theta^*]\right]\nonumber \\
&= \int d\theta^* \mathcal{P}_{\Theta^*}( \theta^*) \CMSE[\thetaest | \theta^*]\,, 
\label{eqn:mse_general}
\end{align}
where $\mathbb{E}_{\Theta^*}$ indicates an average over possible random values of the target phase and $\mathcal{P}_{\Theta^*}$ is its probability density (for our purposes, $\Theta^*$ is normally distributed as in Eq.~\eqref{eqn:theta_star_distribution}). We emphasize that in this framework, CMSE is itself a random variable and a function of $\Theta^*$, while BMSE is a posterior mean-squared error accounting for all randomness in the protocol (both that of the prior classical initialization and the shot noise realizations in the quantum sensing protocol). 

Hence, BMSE can be used to quantify the overall performance of the different sensing protocols. Importantly, it is known that minimizing the BMSE guarantees that one can asymptotically approach the Cram\'er-Rao bound~\cite{rubio2019quantummetrology}. In what follows, we will fix the value of $\Theta^*$ at $\theta^*$, derive the CMSE for each considered protocol, and compare their performance. Crucially, since the CMSE decomposes into a variance and a bias term (see Appendix~\ref{appendix:conditional_MSE} for proof)
\small
\begin{equation}
    \CMSE[\thetaest |\theta^*] =\underbrace{\mathbb{E}_N\left[(\thetaest-\mathbb{E}_N[\thetaest | \theta^*])^2\right]}_{\Var_N[\thetaest | \theta^*]} + \bigg( \underbrace{\mathbb{E}_N[\thetaest | \theta^*]-\theta^*}_{\Bias_N[\thetaest | \theta^*]} \bigg)^2\,,\nonumber
\end{equation}
\normalsize
we will focus on keeping track of their individual contributions and show that the trade-off between conditional bias and conditional variance underpins a sensor's performance. Moreover, to be able to compare different sensing protocols, we assume that the terms in the parameter-encoding Hamiltonian are commuting, $\| O \|_{\infty} \leq 1$, and that the optimal point of sensitivity for the sensor is close to the point of maximum gradient.

Since we compare many similar, but subtly different protocols, this section is notationally dense; we provide a summary of important notation in Table \ref{table:notation} (an expanded table of definitions is also presented in Appendix~\ref{appendix:notation_summary_table}).

\begin{table}[t]
  \centering
  \begin{tabular}{@{}cc@{}}
    \toprule
    Symbol  & Meaning \\
    \midrule
    $R(\theta)$  &  Noiseless response at $\theta$\\
    $R_{\lambda}(\theta)$  &  Noisy response at $\theta$\\
    $\overline{R}_{\lambda}(\theta,N)$  &  $N$-shot noisy response estimate at $\theta$\\
    $R_{M}(\theta)$  &  Error-mitigated response at $\theta$\\
    $\Tilde{R}_{\lambda}(\theta)$  &  Inferred noisy response at $\theta$ \\
    $\Tilde{R}_{M}(\theta)$  &  Mitigated Inferred response at $\theta$ \\
    $\Theta^*$   &  Random variable for the unknown phase \\
    $\theta^*$ & Members of the sample space of $\Theta^*$ \\
    $\thetaest$  &  Estimated value of the unknown phase\\
    $\CMSE[\thetaest|\theta^*]$ &  Conditional MSE in $\thetaest$ given $\Theta^* = \theta^*$ \\
    $\BMSE[\thetaest]$  & Bayesian MSE in $\thetaest$  \\
    \bottomrule
  \end{tabular}
  \caption{\textbf{Summary of notation.} We also refer the reader to  Appendix~\ref{appendix:notation_summary_table} for additional details.}
  \label{table:notation}
\end{table}

\subsection{Noise-aware noisy quantum sensing}
\label{sec:ideal_noisy_sensing_bound}

In noise-aware noisy quantum sensing protocol, one assumes that the functional form of the noisy response $R_\lambda(\theta)$ is perfectly known. Such a case could arise when the noise model is sufficiently simple and one can derive the response analytically, or with an inference-based scheme where the system has been pre-characterized. Here, for $\Theta^* = \theta^*$ one estimates the response via $N$-shots, denoted as $\overline{R}_\lambda(\theta^*,N)$, and obtains the phase of interest via $\thetaest=R_\lambda^{-1}(\overline{R}_\lambda(\theta^*,N))$. Since the response function is known exactly by assumption, the bias in this phase estimate, which can be interpreted as a measure of erroneous assumptions, is zero. Its variance can be locally approximated by the error propagation formula of Eq.~\eqref{eq:precision}, leading to the following lower bound on the conditional mean squared error
\begin{align}
    \CMSE[\thetaest | \theta^*] &= \frac{\Delta^2 R_{\lambda}(\theta^*)}{N(\partial_\theta R_{\lambda}(\theta)|_{\theta=\theta^*})^2} \nonumber \\
    &\geq \underbrace{\frac{\Delta^2 R_{\lambda}(\theta^*)}{N L_{\lambda}^2}}_{\Var_N [\thetaest | \theta^*]} \,,
\label{eqn:noisy_meas_noisy_inv_error}
\end{align}
where
\begin{align*}
    L_{\lambda} \equiv \sup_{\theta\in[\theta_{\text{min}},\theta_{\text{max}}]} |\partial_{\theta}R_{\lambda}(\theta)|
\end{align*}
is the Lipschitz constant of $R_{\lambda}(\theta)$ on the relevant invertible domain. Since we assumed  $ \lVert O \rVert_{\infty} \leq 1$, then $L_{\lambda}\in \Theta(n)$ (see proof in Appendix~\ref{appendix:noiseless_lipschitz_bound}). 

\subsection{Naive noisy quantum sensing}
\label{sec:noisy_measurement_noiseless_inversion_error}

In this setting, one knows that noise acts throughout the sensing protocol but the noisy response $R_\lambda(\theta)$ is not known. Without access to $R_\lambda(\theta)$, one might naively attempt to invert an $N$-shot estimated $\overline{R}_\lambda(\theta^*,N)$ via the noiseless response function $R(\theta)$, as the latter can be analytically determined in some cases (e.g., $R(\theta)=\cos(n\theta)$ is known for GHZ-state magnetometry). Here, the phase estimate is given by $\thetaest=R^{-1}(\overline{R}_\lambda(\theta^*,N))$. This introduces a (conditional) bias in the phase estimate
\small
\begin{align} 
	\Bias_N[\thetaest | \theta^*] = \biggl| \mathbb{E}_{N}[R^{-1}(\overline{R}_{\lambda}(\theta^*,N))] ~ - ~ R_{\lambda}^{-1}(\overline{R}_{\lambda}(\theta^*,N)) \biggr| \,. \nonumber
\end{align}
\normalsize
Since inversion is performed via the noiseless response function $R(\theta)$, a lower bound on this bias under local linearization assumption is given by
\begin{align}
\label{eqn:bias_noiseless_func_sensing}
    \Bias_N [\thetaest | \theta^*] &= \biggl| \mathbb{E}_{N}\left[ \frac{\overline{R}_{\lambda}(\theta^*,N) - R(\theta^*)}{\partial_{\theta}R(\theta)|_{\theta=\theta^*}} \right] \biggr| \nonumber \\
    & \geq \frac{ | R_{\lambda}(\theta^*) - R(\theta^*) |}{L} \,,
\end{align}
where
\begin{align*}
    L \equiv \sup_{\theta\in[\theta_{\text{min}},\theta_{\text{max}}]} |\partial_{\theta}R(\theta)|
\end{align*}
is the Lipschitz constant of $R(\theta)$ on the relevant invertible domain, which again satisfies $L\in \Theta(n)$ (see Appendix~\ref{appendix:noiseless_lipschitz_bound}).

As before, the variance in the phase estimate is obtained via the error propagation formula with the noiseless response $R(\theta)$, yielding
\small
\begin{align}
	\label{eq:mse_noisy_lower_bound}
	\CMSE[\thetaest |\theta^*]  \geq \underbrace{\frac{\Delta^2 R_{\lambda}(\theta^*)}{N L^2}}_{\Var_N[\thetaest | \theta^*]} ~ + ~ \bigg( \underbrace{ \frac{ | R_{\lambda}(\theta^*) - R(\theta^*) |}{L} }_{\Bias_N[\thetaest | \theta^*]} \bigg)^2\,.
\end{align}
\normalsize
Crucially, we note that the bias term does not depend on the overall shot budget, imposing a hard limit on the sensitivity of this protocol.

\subsection{Noisy sensing mitigated by zero-noise extrapolation}
\label{sec:zne_error}

In the previous section, we found that when the noisy response function $R_\lambda(\theta)$ is unknown, the bias in the phase estimate imposes an ultimate limit on the precision obtained by naively inverting a noisy measurement with the noiseless response function. To address this issue, we consider a setting where ZNE is used (via $N$-shots) to obtain an estimate of the noiseless response at the phase of interest $\overline{R}_M(\theta^*,N)$ and use the noiseless response function (assumed to be known) for inversion. That is, $\thetaest=R^{-1}(\overline{R}_M(\theta^*,N))$. 

By exploiting properties of Lagrange basis polynomials, one can show that the zero-noise response estimate at the fixed phase value of $\theta^*$ satisfies\footnote{Here we are assuming that the $\lambda^{m+1}$-th contribution in the Richardson extrapolation is not zero.}
\begin{equation}
R_M(\theta^*)=R(\theta^*)+\Theta(\lambda^{m+1}) \,,	\label{eqn:richardson_remainder_bias}
\end{equation}
where $m+1$ noise nodes are utilized during ZNE (see Appendix~\ref{appendix:zne} for full derivation). Thus, once again lower-bounding the bias using the Lipschitz constant $L$, we obtain
\begin{align}
\label{eqn:bias_mitigated_sensing}
    \Bias_N[\thetaest | \theta^*] &= \biggl| \mathbb{E}_{N}\left[ \frac{\overline{R}_{M}(\theta^*,N) - R(\theta^*)}{\partial_{\theta}R(\theta)|_{\theta=\theta^*}} \right] \biggr| \\
    &\geq \frac{ | R_M(\theta^*) - R(\theta^*) |}{L} \nonumber \in \Omega\left(\frac{\lambda^{m+1}}{L}\right) \nonumber \,.
\end{align}
Comparing Eq.~\eqref{eqn:bias_noiseless_func_sensing} to Eq.~\eqref{eqn:bias_mitigated_sensing}, we can see that ZNE reduces the bias from $\Omega(\lambda)$ in the unmitigated case to $\Omega(\lambda^{m+1})$ in the mitigated case. 

Next, we compute the variance term as
\begin{equation}
    \Var_N[\thetaest | \theta^*]=\frac{\Delta^2\overline{R}_M(\theta^*,N)}{(\partial_\theta R(\theta)|_{\theta=\theta^*})^2} \,,
    \label{eqn:error_mitigated_variance_defn}
\end{equation}
where we can use the fact that 
\begin{align} 
\label{eq:varRM}
\Delta^2 \overline{R}_{M}(\theta^*,N) = \sum_{j=0}^{m} \gamma_j^2 \; \frac{\Delta^2 R_{\lambda_j}(\theta^*, N_j)}{N_j} \,.
\end{align}
The previous equation shows that the variance of $\overline{R}_{M}(\theta^*, N)$ depends on the choice of noise levels $\lambda_j$ and the relative allocation of shot budget to each noise node. If $\lambda$ is small, the variance in response function at each noise node will be approximately equal, i.e., $\Delta^2 R_{\lambda_j}(\theta^*) \approx \Delta^2 R_{\lambda}(\theta^*)$ (see Appendix~\ref{appendix:Variance_R_M_noise_nodes} for a proof). Hence, we can simplify Eq.~\eqref{eq:varRM} as
\begin{align} 
\label{eq:varRMSimplfied}
\Delta^2 \overline{R}_{M}(\theta^*,N) & \approx \sum_{j=0}^{m} \gamma_j^2 \; \frac{\Delta^2 R_{\lambda}(\theta^*)}{N_j} = \frac{\Lambda^2 \Delta^2 R_{\lambda}(\theta^*)}{N} \,,
\end{align}
where in the last step we have used the optimal shot allocation $N_j = N \frac{|\gamma_j|}{\sum_{j=0}^{m}|\gamma_j|}$ of~\cite{krebsbach2022optimization} (see Appendix~\ref{appendix:ZNE_hyperparameters}) and defined $\Lambda \equiv \sum_{j=0}^{m
} |\gamma_j|$. Since $\Lambda^2 \geq 1$, the variance of the error-mitigated protocol of Eq.~\eqref{eq:varRMSimplfied} is larger than that of the realizable noisy protocol in Eq.~\eqref{eqn:noisy_meas_noisy_inv_error} by a factor of $\Lambda^2$, which we refer to as the ``sampling overhead''.

Combining the bias and variance terms, the conditional error for this protocol is lower bounded by
\begin{align}
\label{eq:mse_mitigated_lower_bound}
\CMSE[\thetaest | \theta^*] \geq \underbrace{\frac{ \Lambda^2 \Delta^2 R_{\lambda}(\theta^*)}{N L^2}}_{\Var_N[\thetaest | \theta^*]} ~ + ~ 
 \bigg( \underbrace{  \Omega\left(\frac{\lambda^{m+1}}{L}\right) }_{\Bias_N[\thetaest | \theta^*]} \bigg)^2 \,.
\end{align}
Comparing this expression to the one in Eq.~\eqref{eq:mse_noisy_lower_bound}, it is clear that using ZNE reduces the bias of phase estimation at the cost of increasing the variance by a factor of $\Lambda^2$, which is a consequence of the well-known bias-variance trade-off arising in quantum error mitigation techniques~\cite{cai2022quantum}.

\subsection{Inference-based noisy sensing}
\label{sec:inference_error}

In the previous two sections, we have presented protocols where one needs to use the noiseless response, as knowledge about the noisy one is unavailable. We now study the error in a scheme where one approximately learns the noisy response via inference techniques at the cost of some quantum resources. Here, one obtains the phase of interest from $\thetaest=\Tilde{R}_{\lambda}^{-1}(\overline{R}_\lambda(\theta^*,N))$, where we recall that $\Tilde{R}_{\lambda}(\theta, N_I)$ denotes the inferred response function. Hence, the error in phase estimation will depend on the accuracy of inference, i.e., on $|R_\lambda(\theta) - \Tilde{R}_{\lambda}(\theta, N_I)|$.

To begin, we note that the inferred function $\Tilde{R}_{\lambda}(\theta, N_I)$ is a random variable since it is derived using a set of random variables, i.e., each $\overline{R}_{\lambda}(\theta_k, N_k)$ has an uncertainty of $\frac{\Delta^2R_{\lambda}(\theta_k)}{N_k}$ which propagates to the inferred function. If we define $\epsilon$ as the largest sampling error in the observed responses in a single inference protocol run, i.e., 
\begin{align}
\epsilon \equiv \underset{\theta_k\in \{ \theta_k \}}{\max} |R_{\lambda}(\theta_k) - \overline{R}_{\lambda}(\theta_k, N_k)| \,,
\end{align}
then it is known that $\Delta \Tilde{R}_{\lambda} \leq 5 \epsilon \log(n)$ (see Appendix~\ref{appendix:inference_error_bound_theorems}). However, $\epsilon$ will vary across different protocol runs. Consequently, if we let $\epsilon_{\text{max}}$ represent the largest possible sampling error in the noisy responses in any run, then we know that with high probability
\begin{align}
\epsilon_{\text{max}}  \equiv \max[ ~ \epsilon ~ ] \approx \underset{\theta_k \in \{ \theta_k \}}{\max} \frac{\Delta R_{\lambda}(\theta_k)}{\sqrt{N_k}} \,.
\end{align}

The maximum uncertainty in the inferred function cannot exceed $5 \epsilon_{\text{max}} \log(n)$, i.e., $\Delta \Tilde{R}_{\lambda} \leq 5 \epsilon \log(n) \leq 5 \epsilon_{\text{max}} \log(n)$. Since the problem setup of Sec.~\ref{Subsection:InferenceIntroduction} implies that the inference error originates only from sampling errors, $|R_\lambda(\theta) - \Tilde{R}_{\lambda}(\theta, N_I)| \leq 5 \epsilon_{\text{max}} \log(n)$. Hence, one can directly control this source of uncertainty by choosing a suitably large inference shot budget $N_I$. In the limit $N_I \rightarrow \infty$, one gets $\epsilon \rightarrow 0$, which implies that the error in inference satisfies $|R_\lambda(\theta) - \Tilde{R}_{\lambda}(\theta, N_I)| \rightarrow 0$.    

At this point, we note that one can estimate the resources required to design an inference-based sensing protocol that guarantees a desired accuracy, e.g., one might want to choose an $\epsilon$ such that near-Heisenberg scaling is achieved. To ensure that (with a high probability) a given $\epsilon$ is not exceeded in a single protocol run, the number of shots $N_k$ used at any $\theta_k$ during inference should satisfy $N_k \in \Omega \biggr( \frac{1}{\epsilon^2} \log (\frac{n}{\epsilon}) \biggl)$, as proved in Appendix~\ref{appendix:N_k_lower_bound_general_proof}. We highlight that this bound is tighter than that of Corollary 1 in Ref.~\cite{huerta2022inference}. 

Now, we derive the error bounds in the phase estimate for inference-based protocols. Let $F$ be a random variable that represents the inferred function and assume that  $\Theta^*$ takes the value  $\theta^*$. Next, we sample $F = f$ from its distribution, i.e., we measure the shot-limited responses at $2n+1$ distinct phase values and obtain the corresponding $\Tilde{R}_{\lambda}(\theta, N_I)$ function. The BMSE in $\thetaest$ can be rewritten using the generalized law of total expectation as~\cite{champ2022generalized}
\small
\begin{align}
\label{eqn:mse_inference_averaging}
    \BMSE[\thetaest] & = \mathbb{E}_{\Theta^*}\left[\mathbb{E}_{N_I}\left[\mathbb{E}_{N_E} \left[(\thetaest-\theta^*)^2 \big| \theta^*, f \right] \big| ~ \theta^* \right]\right] \,,
\end{align}
\normalsize
where $\mathbb{E}_{N_I}$ represents an average over multiple realizations of the inference step and $\mathbb{E}_{N_E}$ represents an average over multiple realizations of the parameter estimation step. The total error can again be decomposed into its conditional bias and variance components. In what follows, we derive bounds on these bias and variance errors in terms of $\epsilon_{\text{max}}$. We crucially note that by doing so we are considering the worst-case performance for this protocol compared to all other protocols analyzed here. However, it is easier to work with upper bounds in interpolation errors than to derive the exact errors for each protocol run.

The worst-case conditional bias in the estimate of $\theta^*$ is given by
\small
\begin{align}
\label{eqn:inference_bias_derivation_steps}
\Bias[\thetaest | \theta^*] &= \mathbb{E}_{N_I}\left[\mathbb{E}_{N_E} \left[\thetaest \big| \theta^*, f \right] \big|~ \theta^* \right] ~ - ~ \theta^* \\
&= \mathbb{E}_{N_I}\left[ \biggl| \mathbb{E}_{N_E} \left[ \Tilde{R}_{\lambda}^{-1}(\overline{R}_{\lambda}(\theta^*,N_E)) \right] ~ - ~{R}_{\lambda}^{-1}({R}_{\lambda}(\theta^*)) \biggr| \right] \nonumber \\
&= \mathbb{E}_{N_I}\left[ \biggl| \mathbb{E}_{N_E}\left[ \frac{\overline{R}_{\lambda}(\theta^*, N_E) - \Tilde{R}_{\lambda}(\theta^*,N_I)}{\partial_{\theta}\Tilde{R}_{\lambda}(\theta)|_{\theta=\theta^*}} \right] \biggr| \right] \nonumber \\
&= \mathbb{E}_{N_I}\left[ \biggl| \frac{R_{\lambda}(\theta^*, N_E) - \Tilde{R}_{\lambda}(\theta^*, N_I)}{\partial_{\theta}\Tilde{R}_{\lambda}(\theta)|_{\theta=\theta^*}} \biggr| \right] \nonumber \\
& \gtrsim \frac{5\epsilon_{\text{max}} \log(n)}{L_{\lambda}} \nonumber \,,
\end{align}
\normalsize
where in the last line, the largest possible inference error is assumed in the numerator and  $\partial_\theta \Tilde{R}_{\lambda}(\theta) \sim L_{\lambda}$ holds given inference error is relatively small. As expected, since the bias originates from how well the inferred function approximates the noisy function, it will decrease if we increase the resources allocated to inference.

Then, applying the law of total variance to the conditional variance in $\CMSE$ leads to 
\begin{equation}
\begin{split}\label{eqn:variance_breakdown_inference}
    \Var[\thetaest | \theta^*] &=  \mathbb{E}_{N_I}\left[ \Var_{N_E}\left[\thetaest \big| \theta^*, f \right]  \big| ~ \theta^* \right]  \\
    &+ \Var_{N_I}\left[ \mathbb{E}_{N_E}\left[\thetaest \big| \theta^*, f \right]  \big| ~ \theta^* \right] \,.
\end{split}
\end{equation}
The first term in Eq.~\eqref{eqn:variance_breakdown_inference} can be lower bounded as
\begin{align}
\label{eqn:estimation_variance_inference}
    \mathbb{E}_{N_I}\left[ \Var_{N_E} \left[ \thetaest \big| \theta^*, f \right] \big| ~ \theta^* \right] &= \mathbb{E}_{N_I}\left[ \frac{\Delta^2 R_{\lambda}(\theta^*)}{N_E (\partial_\theta \Tilde{R}_{\lambda}(\theta)|_{\theta=\theta^*})^2} \right] \nonumber \\
    &= \frac{\Delta^2 R_{\lambda}(\theta^*)}{N_E (\mathbb{E}_{N_I} [\partial_\theta \Tilde{R}_{\lambda}(\theta)|_{\theta=\theta^*}])^2}  \nonumber \\ 
    &\gtrsim \frac{\Delta^2 R_{\lambda}(\theta^*)}{N_E L_{\lambda}^2} \,,
\end{align}
while the  second term is lower bounded by 
\small
\begin{align}
\label{eqn:err_fluc_variance_inference}
    \Var_{N_I}\left[ \mathbb{E}_{N_E}\left[\thetaest \big| \theta^*, f \right] \big| ~ \theta^* \right] &= \frac{\Delta^2 \Tilde{R}_{\lambda}}{(\partial_\theta \Tilde{R}_{\lambda}(\theta)|_{\theta=\theta^*})^2} \\
    &\approx \bigg( \frac{5 \epsilon_{\text{max}} \log(n)}{\partial_\theta \Tilde{R}_{\lambda}(\theta)|_{\theta=\theta^*}} \bigg)^2 \nonumber \\
    & \gtrsim \frac{25 
    \log^2(n)\underset{\theta_k \in \{ \theta_k \}}{\max} \Delta^2 R_{\lambda}(\theta_k)}{N_k L_{\lambda}^2} \nonumber\,.
\end{align}
\normalsize

The last variance term captures the worst-case error in sensitivity due to fluctuations in the inferred function (also see a proof based on geometric arguments in Ref.~\cite{huerta2022inference}). Collecting all these sources of errors, we can obtain the worst-case lower bound on the conditional mean squared error in the phase estimate $\thetaest$ as   
\small
\begin{align}
\label{eq:mse_inferred_lower_bound}
\CMSE[\thetaest | \theta^*] &\gtrsim \underbrace{\frac{\Delta^2 R_{\lambda}(\theta^*)}{N_E L_{\lambda}^2} + \biggl( \frac{5 \epsilon_{\text{max}} \log(n)}{L_{\lambda}} \biggr)^2 }_{\Var[\thetaest | \theta^*]} \\
& + \bigg( \underbrace{\frac{\mathbb{E}_{N_I} \left[ 5 \epsilon \log(n) \right] }{L_{\lambda}} }_{\Bias[\thetaest | \theta^*]} \bigg)^2  \nonumber  \,.
\end{align}
\normalsize

\subsection{Inference-based noisy sensing mitigated by zero-noise extrapolation}
\label{sec:error_mitigated_inference_bound}

Previously we saw that applying ZNE to quantum sensing can reduce -- but not eliminate -- bias at cost of increased variance. As this bias places an ultimate limit on the efficacy of mitigation-based sensing protocols, one might question whether the protocol can be improved by combining error mitigation with inference-based techniques. Such a procedure allows one to infer a mitigated response function that is already shifted by the mitigation bias accrued when using ZNE. As such, one estimates the phase of interest here as $\thetaest=\Tilde{R}_M^{-1}(\overline{R}_M(\theta^*,N_E))$, where $\Tilde{R}_M(\theta, N_I)$ denotes the inferred $N_I$-shot approximate of the error-mitigated response function $R_M(\theta)$. In what follows, we assume that $R_M(\theta)$ is also a trigonometric polynomial of degree $n$. We will see later that this protocol is comparatively inefficient even with this simplification. 

Now, Richardson extrapolation is applied to obtain the zero-noise response estimates at each of the $2n+1$ inference nodes. Hence, at each $\theta_k$, a mitigated estimate $\overline{R}_{M}(\theta_k, N_k)$ is obtained following Eq.~\eqref{eq:EMpolyEst} using a set of responses at $m+1$ different noise levels $\{ \lambda_j, \overline{R}_{\lambda_j}(\theta_k, N_j) \}_{j=0}^{m}$, where ZNE shot allocation satisfies $\sum_{j=0}^{m} N_j = N_k $. The inferred function $\Tilde{R}_M(\theta, N_I)$ is then derived from the set of mitigated responses $\{\theta_k, \overline{R}_{M}(\theta_k, N_k) \}_{k=1}^{2n+1}$ following the protocol explained in Sec.~\ref{Subsection:InferenceIntroduction}, such that $\sum_k N_k = N_I$.

Now, the uncertainty due to sampling errors in $\overline{R}_{M}$ propagates to the inferred function. Under the same assumptions as those used for Eq.~\eqref{eq:varRMSimplfied}, we know that for each point in this set $\Delta^2 \overline{R}_M(\theta_k, N_k) \approx \frac{\Lambda^2 \Delta^2 R_{\lambda}(\theta_k)}{N_k}$. If we define $\chi$ as the largest sampling error in a set of mitigated responses for a single inference protocol run, i.e., 
\begin{equation} 
 \chi \equiv \underset{\theta_k\in \{ \theta_k \}}{\max} |R_M(\theta_k) - \overline{R}_{M}(\theta_k, N_k)| \,,
 \end{equation}
then from the accuracy of trigonometric interpolation, we know that $\Delta \Tilde{R}_M \leq 5 \chi \log(n)$ will hold for that run. As before, if $\chi_{\text{max}}$ is the largest possible sampling error in any run, then  $\Delta \Tilde{R}_M \leq 5 \chi \log(n) \leq 5 \chi_{\text{max}} \log(n)$, and with high probability
\begin{align}
\chi_{\text{max}}  = \max [ ~ \chi ~] \approx \underset{\theta_k \in \{ \theta_k \}}{\max} \frac{\Lambda \Delta R_{\lambda}(\theta_k)}{\sqrt{N_k}} \,. 
\end{align}
Hence fluctuations in inferred function across different protocol runs come from the fluctuations in the mitigated responses. Given the assumption that $R_M(\theta)$ is a trigonometric sum, the upper-bound on error in inference is given by $|R_M(\theta) - \Tilde{R}_M(\theta, N_I)| \leq 5 \chi_{\text{max}} \log(n)$. 

To avoid the large bias error encountered in Sec.~\ref{sec:noisy_measurement_noiseless_inversion_error}, once the inferred function $\Tilde{R}_M(\theta, N_I)$ is obtained, one also used ZNE during the estimation step. Here we take $\Theta^* = \theta^*$,  and assume that a shot budget $N_E$ is allocated to obtain a mitigated response $\overline{R}_{M}(\theta^*, N_E)$. Following Sec.~\ref{sec:inference_error}, we derive the error bounds in the phase estimate $\thetaest$. We keep $F$ to refer to the randomness in the inferred function. The worst-case $\CMSE$ can again be decomposed into its conditional bias and variance components. The worst-case conditional bias in the estimate is given by
\small
\begin{align}
\Bias[\thetaest | \theta^*] &= \mathbb{E}_{N_I}\left[\mathbb{E}_{N_E} \left[\thetaest \big| \theta^*, f \right] \big|~ \theta^* \right] ~ - ~ \theta^* \\
&= \mathbb{E}_{N_I}\left[ \biggl| \mathbb{E}_{N_E} \left[\Tilde{R}_M^{-1}(\overline{R}_{M}(\theta^*,N_E))\right] - {R}^{-1}({R}(\theta^*))  \biggr| \right] \nonumber \\
&= \mathbb{E}_{N_I}\left[ \biggl| \mathbb{E}_{N_E}\left[ \frac{\overline{R}_{M}(\theta^*, N_E) - \Tilde{R}_M(\theta^*, N_I)}{\partial_{\theta}\Tilde{R}_M(\theta)|_{\theta=\theta^*}} \right] \biggr| \right] \nonumber \\
&= \mathbb{E}_{N_I}\left[ \biggl| \frac{R_{M}(\theta^*, N_E) - \Tilde{R}_M(\theta^*, N_I)}{\partial_{\theta}\Tilde{R}_M(\theta)|_{\theta=\theta^*}} \biggr| \right] \nonumber \\
& \gtrsim \frac{5\chi_{\text{max}} \log(n)}{L} \nonumber \,,
\end{align}
\normalsize
where in the last line, $\partial_\theta \Tilde{R}_M(\theta) \sim L$ holds given mitigation bias is relatively small. Note that, as expected, $\Bias[\thetaest | \theta^*]$ will decrease as we increase the quantum resources.

The conditional variance will have a similar breakdown into two terms as in Sec.~\ref{sec:inference_error}. The first term in the conditional variance in $\CMSE$ can be lower bounded as
\small
\begin{align}
\label{eqn:estimation_variance_error_mitigated_inference}
    \mathbb{E}_{N_I}\left[ \Var_{N_E}\left[\thetaest \big| \theta^*, f \right]  \big| ~ \theta^* \right] &= \mathbb{E}_{N_I}\left[ \frac{\Lambda^2 \Delta^2 R_{\lambda}(\theta^*)}{N_E (\partial_\theta \Tilde{R}_M(\theta)|_{\theta=\theta^*})^2} \right] \nonumber\\
    &\gtrsim \frac{\Lambda^2 \Delta^2 R_{\lambda}(\theta^*)}{N_E L^2}  \,,
\end{align}
\normalsize
while we find the following lower bound for the second term in the conditional variance
\small
\begin{align}
\label{eqn:err_fluc_variance_error_mitigated_inference}
    \Var_{N_I}\left[ \mathbb{E}_{N_E}\left[\thetaest \big| \theta^*, f \right] \big| ~ \theta^* \right] &= \frac{\Delta^2 \Tilde{R}_M}{(\partial_\theta \Tilde{R}_M(\theta)|_{\theta=\theta^*})^2} \\
    &\approx \bigg( \frac{5 \chi_{\text{max}} \log(n)}{\partial_\theta \Tilde{R}_M(\theta)|_{\theta=\theta^*}} \bigg)^2 \nonumber \\
    & \gtrsim \frac{25 
    \log^2(n)\underset{\theta_k \in \{ \theta_k \}}{\max} \Lambda^2 \Delta^2 R_{\lambda}(\theta_k)}{N_k L^2} \nonumber\,.
\end{align}
\normalsize
Collecting all these sources of errors, we can now obtain that the worst-case lower bound on the conditional mean squared error in the phase estimate $\thetaest$ is 
\begin{align}
\label{eq:mse_inferred_mitigated_lower_bound}
\CMSE[\thetaest | \theta^*] &\gtrsim \underbrace{\frac{\Lambda^2 \Delta^2 R_{\lambda}(\theta^*)}{N_E L^2} +  \bigg(\frac{5 \chi_{\text{max}} \log(n)}{ L^2}\biggr)^2 }_{\Var[\thetaest | \theta^*]} \\
& + \bigg( \underbrace{\frac{\mathbb{E}_{N_I} \left[ 5 \chi \log(n) \right]}{L} }_{\Bias[\thetaest | \theta^*]} \bigg)^2 \nonumber  \,.
\end{align}

As before, we can estimate the shot budget required to design an error-mitigated, inference-based sensing protocol that guarantees a certain desired accuracy. To ensure that a given $\chi$ is not exceeded in a single protocol run with a high probability, the number of shots $N_k$ used at any $\theta_k$ during inference should satisfy $N_k \in \Omega \biggr( \frac{\Lambda^3 \log(n)}{\chi^2 \max_j |\gamma_j|}) \biggl)$ (see proof in Appendix~\ref{appendix:N_k_hoeffding_error_mitigation_inference_proof}).

\subsection{Comparison of error bounds}
\label{sec:analytic_error_comparison}

Having derived error bounds for the different protocols considered, we are now in a position to make an approximate comparison between them. Importantly, we note that our goal here is to gain intuition for which methods lead to the best performance (in terms of error versus system size $n$) as well as about their relative sensitivities. In what follows, we will assume that the global-depolarizing noise presented in Sec.~\ref{sec:global_depol_toy_model} acts throughout the sensing scheme. 

For convenience, let us recall the five protocols studied above (see also Fig.~\ref{fig:flowchart}): noise-aware noisy sensing (Sec.~\ref{sec:ideal_noisy_sensing_bound}), naive noisy sensing (Sec.~\ref{sec:noisy_measurement_noiseless_inversion_error}), error-mitigated sensing (Sec.~\ref{sec:zne_error}), inference-based sensing (Sec.~\ref{sec:inference_error}), and error-mitigated, inference-based sensing (Sec.~\ref{sec:error_mitigated_inference_bound}). Then, in the case of inference-based sensing, we further consider two cases: a case where the sensor is only operated at the time of parameter estimation (thus total shot budget $N=N_E+N_I$), and the `pre-characterized' case described in Sec.~\ref{Subsection:InferenceIntroduction} with $N_E = N$ (i.e., all budgeted shots are expended for parameter estimation) and some prior shot budget $N_I$ has already been expended to characterize the response function. Here, we set $N_I=C_{\text{pre}}\times n \times N$, where the constant $C_{\text{pre}}$ is the ``pre-characterization overhead''. We will use $C_{\text{pre}}=100$, but we note that this is a significant overestimate, and a substantially lower overhead was found to be sufficient in the numerical study of Sec.~\ref{sec:numerics}. 

\begin{figure}
	\centering
	\includegraphics[width=0.48\textwidth]{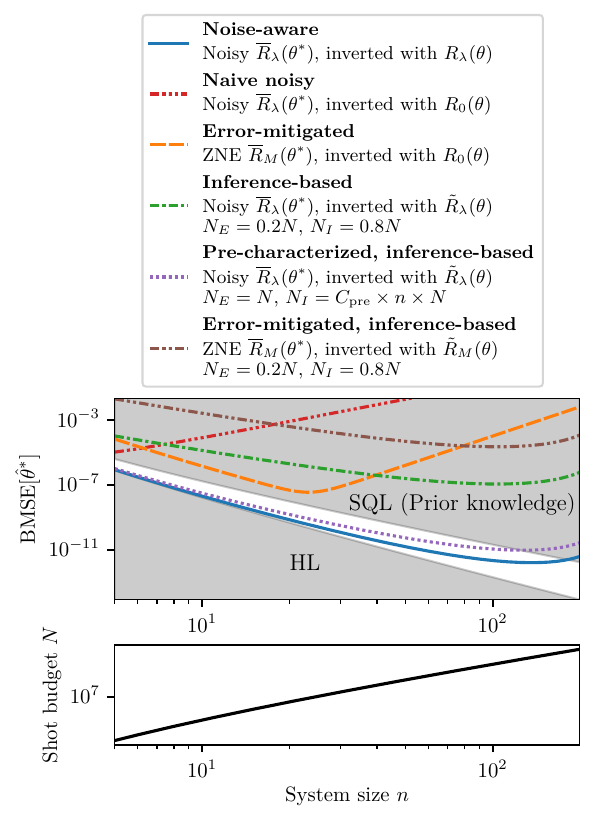}
\caption{\textbf{Comparison of analytically estimated error bounds.} We analytically estimate the errors (BMSE) of all protocols considered in Sec.~\ref{section:theoretical} (top panel), comparing errors to the SQL and the HL. We assume that all protocols are initialized with prior knowledge at the SQL, as outlined in Sec.~\ref{sec:parameter_measurement_and_phase_measurement}. We further assume a shot-budget scaling of $N\propto n^2 \log(n)^3$ (bottom panel), and a noise scaling of $\lambda=10^{-2}\times n$.} 
	\label{fig:analytic_error_bounds}
\end{figure}

In Fig.~\ref{fig:analytic_error_bounds}, we compare the analytical error bounds of all protocols versus the system size $n$. A summary of the exact expressions used to evaluate these is given in Appendix~\ref{appendix:error_bound_exact_expressions} under the assumption of depolarizing noise.   A shot budget scaling of $N\propto n^2 \log(n)^3$ is used, in line with the requirements to achieve Heisenberg-like scaling with an inferred response function (see Appendix~\ref{appendix:N_k_lower_bound_hoeffding_proof}). We additionally apply a noise scaling of $\lambda=10^{-2}\times n$. Crucially, we note that all protocols that lack knowledge of the system's noisy response function fail to outperform the SQL, highlighting the importance of pre-characterization in achieving tangible advantage from entangled probe states.

As expected, the noise-aware sensing protocol (blue) is the best-performing protocol, as it achieves the lowest overall error and is capable of reaching sub-SQL errors within the considered parameter regime. In contrast, naive noisy sensing (red) has the overall worst performance due to its large bias, highlighting the importance of either learning a good approximation of the noisy response function (e.g., via inference) or ensuring the measurement approximately obeys a known noiseless response function (e.g., error mitigation or full error correction). Crucially, the error-mitigated protocol (orange) and noisy inference-based sensing (solid green) protocols also fail to outperform the SQL. We note also that error-mitigated inference (brown) is not performant. Although it eventually outperforms sensing with error mitigation alone (orange) at larger system sizes where the bias of ZNE is a bottleneck, its performance falls well short of the SQL and is always worse than noisy inference-based sensing (green). Indeed, one can see that this will always be the case by comparing Eqs.~\eqref{eq:mse_inferred_lower_bound} and~\eqref{eq:mse_inferred_mitigated_lower_bound}, as each error term is boosted by a sampling-overhead-dependent factor in the error-mitigated case. Overall, in the realistic scenario where one does not know $R_\lambda(\theta)$, then the only protocol to outperform the SQL anywhere is pre-characterized inference (purple). In line with our previous remarks, we see that even the most performant protocols do not outperform the SQL indefinitely in system size $n$, as the cumulative effect of noise is eventually too great to obtain a metrological advantage from entanglement.

\begin{figure*}
	\centering
	\includegraphics[width=0.9\textwidth]{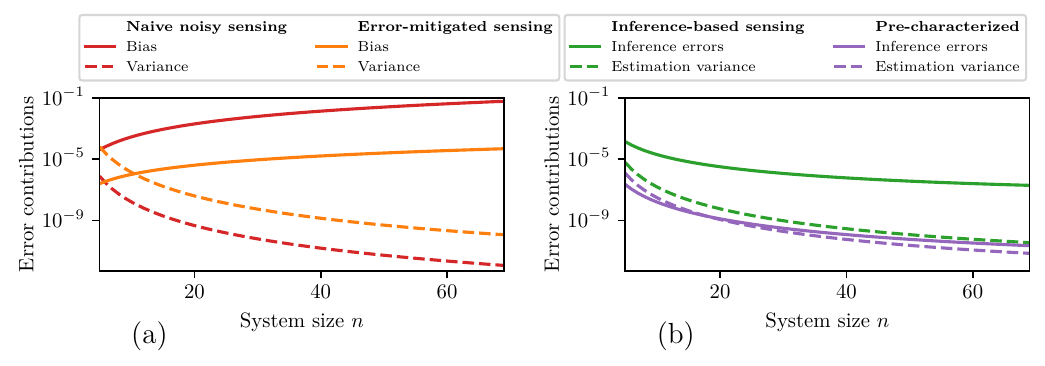}
	\caption{\textbf{Individual error contributions.} We show the different terms that affect the error estimations of Fig.~\ref{fig:analytic_error_bounds}. (a) We compare naive noisy sensing (red) to error-mitigated sensing (orange). Solid lines indicate bias terms, and dashed lines indicate variance terms. (b) We compare inference-based sensing (green) to pre-characterized, inference-based sensing (purple). Solid lines indicate error originating from the inference protocol (this includes both bias and inference variance), while dashed lines indicate estimation variance. }
	\label{fig:analytic_bias_variance}
\end{figure*}

Dividing the error into bias and variance terms enables additional insight into the limiting sources of error in each protocol, and the trade-offs involved. As shown in Fig.~\ref{fig:analytic_bias_variance}(a), we observe the bias-variance trade-off inherent to error mitigation methods, which ultimately limits the effectiveness of ZNE for sensing applications. In Fig.~\ref{fig:analytic_bias_variance}(b), we plot inference error (Eq.~\eqref{eqn:inference_bias_derivation_steps}) and estimation variance error (Eq.~\eqref{eqn:estimation_variance_inference}). Here, one can see that pre-characterization greatly reduces the error in the dominant terms of inference-based sensing. Furthermore, we note that in contrast to error mitigation, there is no trade-off associated with this improvement: All error terms are diminished by the application of pre-characterization, including bias. This provides an intuitive explanation for the comparatively good performance of pre-characterized inference methods.

We emphasize that the error comparisons of this subsection are intended only as a back-of-the-envelope, approximate, intuition for the factors affecting the performance of these protocols, and to qualitatively demonstrate that pre-characterization can in principle improve performance beyond the SQL. In particular, as stated previously (and outlined in Appendix~\ref{appendix:error_bound_exact_expressions}), the error of inference-based sensing is particularly overestimated. However, as we will see in the next section, in numerical simulations under realistic noise, the errors achieved via inference are much smaller, and therefore a quantum advantage can be obtained with substantially lower overhead $C_{\text{pre}}$.

\section{Numerical experiments}
\label{sec:numerics}

In the previous section, we derived error bounds for the considered quantum sensing protocols, enabling a back-of-the-envelope comparison of the performance for every scheme. While such analysis provided useful insight, one can only draw limited conclusions from such worst-case studies (e.g., the errors in inference-based protocols are substantially overestimated). Furthermore, the results plotted in Sec.~\ref{sec:analytic_error_comparison} are based on a simple global-depolarizing noise model. Such an oversimplified noise model preserves the cosine shape of the response function (merely reducing its amplitude), whereas more general and realistic noise models can significantly change the functional form of the noisy response function $R_\lambda(\theta)$, introducing shifts, breaking the symmetry of invertible regions and creating vast differences in sensitivity between them~\cite{huerta2022inference}. Inference-based sensing protocols are particularly well-equipped to deal with these more complex situations. We must therefore compare the considered protocols on more realistic noise models that capture essential features of response functions encountered on real hardware, which cannot be evaluated analytically. Therefore, in this section we support our analytic results with numerical experiments, explicitly simulating parameter estimation protocols for a noisy GHZ magnetometry task.
All results in this section are obtained by Monte Carlo sampling of Eq.~\eqref{eqn:mse_general} --- we explicitly simulate the sensing protocols described in Sec.~\ref{section:framework} for randomly sampled realizations of shot noise (at all protocol stages, including the classical prior measurement, inference and estimation steps) and estimate the BMSE by averaging over these samples. All nuances of the protocols excluded in Sec.~\ref{section:theoretical} are therefore accounted for here: for example, we explicitly perform Richardson extrapolation for each sample of ZNE, and fully construct the inferred response function $\tilde{R}_\lambda(\theta)$ for each realization of the inference step. 
ZNE is performed using hyper-parameters chosen for overall reliable performance (see Appendix~\ref{appendix:ZNE_hyperparameters}).

We focus here on noise-aware sensing (Sec.~\ref{sec:ideal_noisy_sensing_bound}), error-mitigated sensing (Sec.~\ref{sec:zne_error}), and inference-based sensing (Sec.~\ref{sec:inference_error}) protocols, excluding other protocols that were analytically demonstrated to perform poorly in Sec.~\ref{section:theoretical}. Our numerical experiments are performed with two noise models. In systems that are sufficiently small for full density-matrix simulation, we utilize a realistic NISQ noise model based on IBM's Eagle processor~\cite{goh2024direct}. This noise model, outlined in Appendix~\ref{appendix:ibm_eagle_model}, includes noise channels generated from the sparse Pauli-Lindblad model used for error mitigation in Ref.~\cite{kim2023evidence}. In systems that are too large for full-state simulation, we utilize a local depolarizing noise model, outlined in Appendix~\ref{appendix:local_depolarizing_noise}. All full-state simulations in this work were implemented using the Quantum Exact Simulation Toolkit (QuEST) \cite{jones2019quest} via the QuESTlink\,\cite{jones2020questlink} frontend.

\subsection{Numerical comparison of errors across varying regimes}
\label{sec:combined_numerics}
\begin{figure*}
	\centering
	\includegraphics[width=0.95\textwidth]{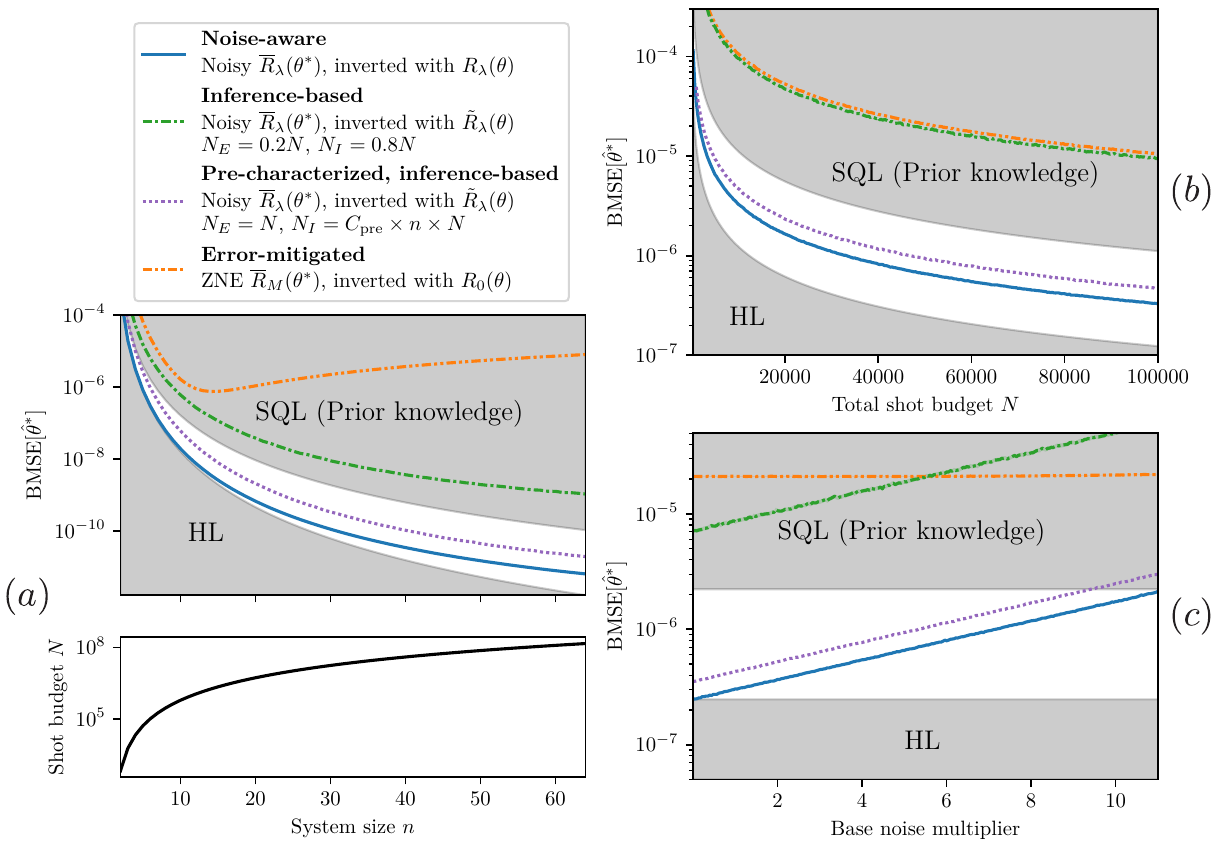}
	\caption{\textbf{Numerical experiments for magnetometry protocols.} (a) We consider the scaling of protocol errors in system size (top panel), with shot budget scaling $N\propto n^2\log(n)^3$ (bottom panel). This experiment is conducted with the local-depolarizing error model of Appendix~\ref{appendix:local_depolarizing_noise} at per-gate fault probability $p=9\times10^{-3}$. (b) Scaling of protocol errors with the shot budget, at fixed system size $n=9$ and base noise rate.  This experiment is conducted with the IBM Eagle error model outlined in Appendix~\ref{appendix:ibm_eagle_model}. (c) Scaling of protocol errors with base noise rate, at fixed system size $n=9$ and shot budget $N=5\times10^4$. This experiment is conducted with the IBM Eagle noise model outlined in Appendix~\ref{appendix:ibm_eagle_model}. Confidence intervals in all figures are too small to visualize.}
	\label{fig:numerics_combined}
\end{figure*}

The relative performance of the quantum sensing protocols under consideration could in principle depend on properties of the system such as system size, noise level, and shot budget. Here, we use numerical experiments to compare protocol errors across varied regimes, ultimately strengthening the previous conclusions that ZNE typically will not outperform the SQL. As such, in the absence of a priori knowledge of the noise model or the response function, only pre-characterized inference can reliably outperform the SQL (and thus obtain actual benefits from quantum entanglement). These numerical experiments are depicted in Fig.~\ref{fig:numerics_combined}.

In Fig.~\ref{fig:numerics_combined}(a), we compare the numerically computed protocol errors at varying system sizes. Since the comparison reaches system sizes well beyond what could be simulated with full-density-matrix methods, we utilize the local-depolarizing error model of Appendix~\ref{appendix:local_depolarizing_noise}, wherein every CNOT in state preparation induces local 2-qubit depolarizing noise at per-gate fault probability $p=9\times10^{-3}$. Once again, to ensure Heisenberg-like scaling in the inference-based protocol, we use a shot budget scaling of $N\propto n^2 \log(n)^3$. Qualitatively, these numerical results can be compared to those obtained from the analytical error bounds in Sec.~\ref{sec:analytic_error_comparison} (Fig.~\ref{fig:analytic_error_bounds}). A similar ordering of curves is obtained: ZNE and inference-based sensing fail to outperform the SQL at any system size, but pre-characterized inference does outperform the SQL. However, as previously suspected, we find that inference-based sensing is overall more performant than the analytic estimates would suggest since they pessimistically assume upper bounds are saturated in inference error. In contrast to Fig.~\ref{fig:analytic_error_bounds}, inference (without pre-characterization) strictly outperforms ZNE here, although it still does not outperform the SQL. Encouragingly, pre-characterized inference outperforms the SQL with a much lower overhead - here, similar performance is obtained with only $C_{\text{pre}}=1$ (versus $C_{\text{pre}}=100$ in the analytic estimates of Sec.~\ref{sec:analytic_error_comparison}). Overall, we find that the qualitative intuition of our analytic error estimates was sound, but inference performs better than one may have believed from the worst-case analytical bounds alone.

While the scaling in $n$ is most important here (i.e., SQL vs HL), it is also crucial to verify that our results are not an artifact of a particular regime -- e.g., one may naively expect that ZNE would perform relatively better at a larger shot budget $N$ (where the effect of increased variance is lower) or at a lower base noise level. Furthermore, real experimental noise models will typically differ significantly from the idealized ones already considered \cite{huerta2022inference}. We therefore perform further numerical experiments, using a realistic noise model based on real current-generation hardware. Noise is simulated using full Kraus maps applied in a density-matrix simulator, with a noise model based on data learned from IBM's Eagle processor with the base noise rate boosted by a factor of 5 (in order to operate in a regime where fringe contrast is sufficiently reduced to warrant the use of error mitigation) (full details in Appendix~\ref{appendix:ibm_eagle_model}).

In Fig.~\ref{fig:numerics_combined}(b), we compare numerically simulated protocol errors at a fixed system size $n$ and varying shot budget $N$. Here, we find again that inference-based sensing and ZNE fail to outperform the SQL, inference (slightly) outperforms ZNE, and overall, only pre-characterized inference outperforms the SQL, indicating that our conclusions are not a product of considering a shot budget regime unfavorable to ZNE. 

Finally, in Fig.~\ref{fig:numerics_combined}(c), we compare these errors at a fixed system size $n$ and shot budget $N$, and varying base noise level. We once again see that inference-based and error-mitigated sensing do not outperform the SQL, although the error in inference-based sensing does worsen beyond that of error-mitigated sensing at very high noise levels (exceeding those of current generation IBM hardware, so of lesser relevance). Pre-characterized inference performs within a constant factor of noise-aware sensing and is again the only protocol that obtains a clear quantum advantage without assuming \emph{a priori} knowledge of the system.

Interestingly,  in the limit of a noiseless system, inference-based sensing (without pre-characterization) does not outperform the SQL. Hence,  even in the absence of noise, there is a cost associated with learning the system's response function. Although in this case (GHZ probe state) the noiseless response function $R_0(\theta)$ is known, there is much ongoing investigation into the use of non-trivial optimized probe states \cite{koczor2020variational, beckey2020variational, kaubruegger2021quantum, ma2020adaptive, marciniak2022optimal, thurtell2022optimizing, liu2022variational, Le2023variational, meyer2020variational, kaubruegger2023optimal, direkci2024heisenberglimited, castro2024variational} for which even the noiseless response function may not be known \emph{a priori}. Such results expands the relevance of our conclusions regarding the utility of pre-characterization and learning.

\begin{figure}
	\centering
	\includegraphics[width=0.5\textwidth]{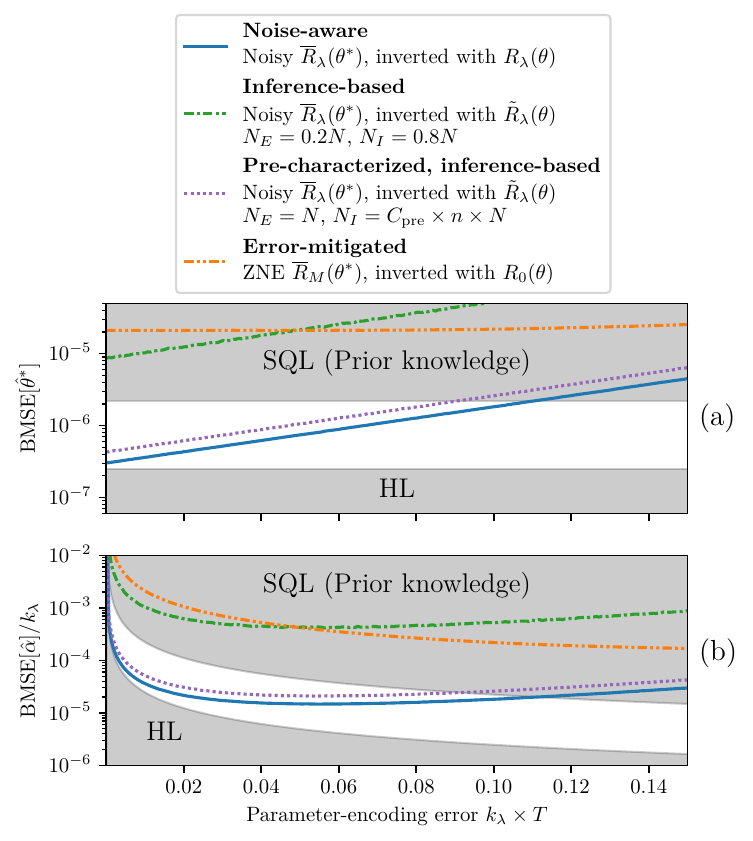}
	\caption{\textbf{Effects of noise in the parameter-encoding stage.} We consider the scaling of protocol errors with the noise level $k_\lambda\times T$ in the parameter-encoding stage, at fixed system size $n=9$, shot budget $N=5\times10^4$, and state-preparation base noise level. This experiment is conducted using the IBM Eagle noise model of Appendix~\ref{appendix:ibm_eagle_model} for the noisy probe state preparation channel $\mathcal{E}_\lambda$, and depolarizing noise for the pre-measurement noise channel $\mathcal{D}_\lambda$. Confidence intervals are too small to visualize.
    (a) Scaling of phase estimate errors with parameter-encoding error. (b) Scaling of parameter estimate errors $\operatorname{BMSE}[\hat{\alpha}]$ with parameter-encoding error, relative to the overall interaction noise level $k_\lambda$. A tradeoff is observed --- larger interaction time improves achievable precision, but also increases exposure to error.}
	\label{fig:interrogation_scaling}
\end{figure}

Noisy quantum sensing is not only afflicted with noise in the preparation of entangled probe states, but also noise during the parameter encoding process~\cite{giovannetti2011advances} (i.e. noise that acts during the system-environment interaction, modelled by the pre-measurement noise channel $\mathcal{D}_\lambda$ in our framework). Therefore, we conduct additional numerical experiments including this additional noise source in the parameter encoding stage. In Fig.~\ref{fig:interrogation_scaling}, we conduct simulations at fixed system size $n$, fixed shot budget $N$, and fixed (base level) state-preparation noise in the state preparation channel $\mathcal{E}_\lambda$ (per the model of Appendix~\ref{appendix:ibm_eagle_model}). We additionally burden the system with depolarizing noise applied to each qubit in the parameter-encoding stage. These noise processes, included in the pre-measurement noise channel $\mathcal{D}_\lambda$, act with a probability $p_\lambda = 1-e^{-\lambda}$, where $\lambda=k_\lambda T$ is a cumulative noise rate that scales with the interaction time $T$, with a strength parametrized by $k_\lambda$.

Much like the scaling of probe-state-preparation noise, in Fig.~\ref{fig:interrogation_scaling}(a) we see that inference-based and error-mitigated sensing do not outperform the SQL, but noise-aware and pre-characterized inference are capable of doing so if the noise levels are not excessively high. In principle, the achievable precision of parameter estimates $\hat{\alpha}$ is inversely proportional to the interaction time $T$, however in practice a longer $T$ also increases exposure to noise that acts during parameter encoding. We depict this effect in Fig.~\ref{fig:interrogation_scaling}(b), showing the scaling of the average error in parameter estimates $\hat{\alpha}$ with the interaction time $T$. A clear tradeoff is observed, setting an optimal interaction time (which may in general be protocol- or regime-dependent). Overall, we see that pre-characterized inference is equally capable of ameliorating the presence of noise induced by the parameter-encoding step as it is for probe-state-preparation noise --- since the response function $R_\lambda$ depends on all noise processes, expending resources to learn it is broadly useful.

\subsection{Requirements for pre-characterized inference-based sensing}
Throughout the error comparisons of Sec.~\ref{sec:analytic_error_comparison} and Sec.~\ref{sec:combined_numerics}, we consistently observed that in the absence of a priori knowledge of the noisy response function $R_\lambda(\theta)$ (which is generally not a practical assumption since device noise is challenging to characterize, and full-process tomography is costly), the only protocol that could reliably outperform the SQL was pre-characterized inference. This can prove crucial in achieving asymptotic advantage from entanglement in quantum sensing.

However, it is natural to question whether the costs of pre-characterized sensing are indeed practical. In previous sections, we considered a simple overhead $N_I = C_{\text{pre}}\times n \times N$, finding that overall $N_I\in \mathcal{O}(nN)$ shots were more than sufficient to pre-characterize a sensor to the point of outperforming the SQL, across a range of regimes spanning different system sizes, shot budgets, noise rates, and error models. Hence, here we study the effect of varying pre-characterization shot budgets to better understand the associated overhead.

\begin{figure}
	\centering
	\includegraphics[width=0.5\textwidth]{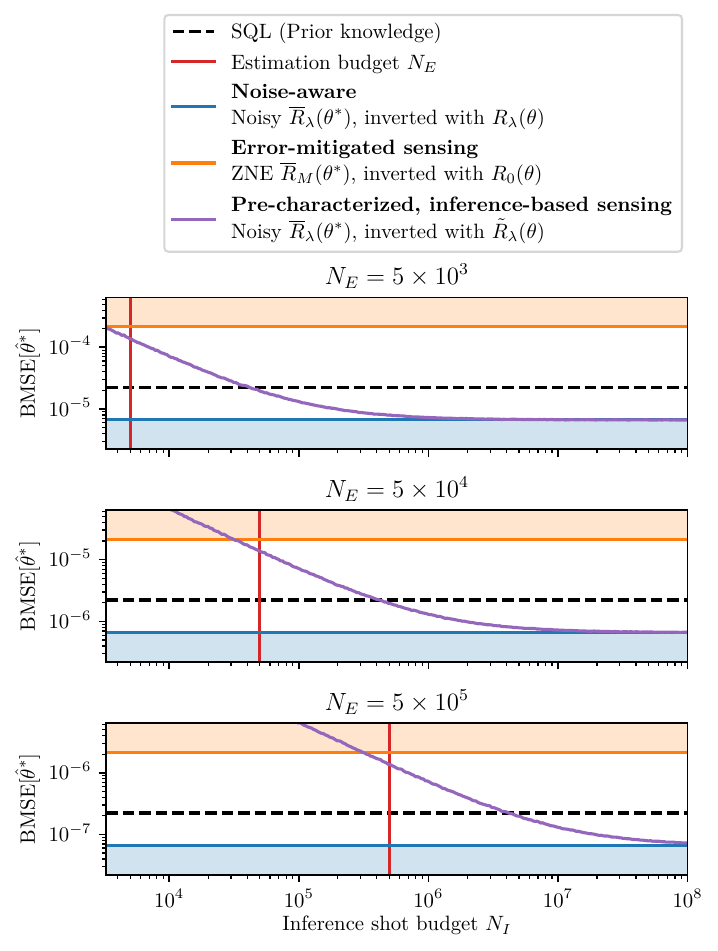}
	\caption{\textbf{Error convergence of pre-characterized, inference-based sensing.} We numerically study a 9-qubit sensor afflicted by realistic noise in state preparation. At three fixed estimation budgets $N_E\in 5\times\{10^3,10^4,10^5\}$ (red vertical lines), we numerically compute the error for pre-characterize inference-based sensing at a range of inference budgets $N_I$. We further compare this error to that of error-mitigated sensing (orange), noise-aware sensing (blue), and the SQL (black dashed).}
	\label{fig:inference_scaling}
\end{figure}

In Fig.~\ref{fig:inference_scaling}, we consider the use of pre-characterized, inference-based sensing on a $9$-qubit device, with state preparation afflicted by the IBM Eagle noise model of Appendix~\ref{appendix:ibm_eagle_model} at a base noise rate boosted by a factor of 5. We consider, at three fixed estimation shot budgets $N_E\in 5\times\{10^3,10^4,10^5\}$, the dependence of inference-based sensing error on the pre-characterization budget $N_I$. We further compare these to the error achieved with error-mitigated sensing at $N=N_E$ and noise-aware sensing at $N=N_E$. As $N_I\to\infty$, we observe that the inference-based protocol converges to the noise-aware sensing protocol, as one would expect (since $\tilde{R}_\lambda$ becomes an infinitely good approximation to $R_\lambda$). 
In this regime, we find that the threshold $N_I$ for the protocol to outperform the SQL is less than an order of magnitude greater than $N_E$, and $N_I$ two orders of magnitude larger than $N_E$ is already sufficient to be extremely close to the noise-aware limit. Furthermore, we find for all three estimation budgets $N_E$ that when $N_I$ is pessimistically chosen to be slightly less than $N_E$, one can still outperform error-mitigated sensing. Overall, we see that with modest pre-characterization overheads, a single pre-characterization can be reused for as many parameter estimations as desired (within the window of noise model stability). This supports the practicality of pre-characterization to achieve quantum advantage with inference-based sensing.

\section{Conclusions}

In this work, we have studied whether using ZNE and inference techniques can enhance a quantum sensor when one is working with a finite measurement shot budget. Indeed, by studying the shot-by-shot performance of various protocols where different levels of prior knowledge are assumed, we find that the remaining bias and increased variance arising from the use of ZNE makes it not amenable for sensing. Then, we also found that when learning the system response via inference, the required shot budget can still be prohibitively large. However, if the system is stable enough so that one can pre-characterize and pre-learn the response via inference once, and then use this model for posterior sensing experiments, one can beat the SQL and achieve a quantum advantage. The extent to which this approach can be used in realistic scenarios will be experiment-dependent but it hints at the fact that the more stable the system and noise are, the better it will perform under constrained budget scenarios.

In the larger picture of quantum sensing, we hope that our work will raise awareness of the important fact that information about the system and shots are both crucial resources for quantum sensing. From the framework we have developed, it is clear that noise is prohibitively detrimental to quantum sensing in the absence of device characterization, as we have to spend precious quantum resources to characterize and learn from the noisy sensing scheme (or suffer the large biases that come as a consequence of not doing so). Our work demonstrates that learning-based techniques are extremely valuable in the pursuit of quantum advantage for sensing problems. With this realization, we anticipate that further advances can be made in quantum sensing by incorporating the wealth of learning tools that have been developed for Bayesian inference, classical machine learning, and quantum machine learning.  

\section*{Acknowledgements}
We thank Kristan Temme, Youngseok Kim and Andrew Eddins for supplying the parameters of the noise model outlined in Appendix F1. 
A.I. acknowledges support by the U.S. Department of Energy (DOE) through a quantum computing program sponsored by the LANL Information Science \& Technology Institute.
C. H. A.  acknowledges support from the Laboratory Directed Research and Development (LDRD) program of Los Alamos National Laboratory (LANL) under project number 20220803PRD3.
L.C. and M.C. were supported by the U.S. DOE, Office of Science, Office of Advanced Scientific Computing Research, under the Computational Partnerships program. 
M.L.G. acknowledges the Rhodes Trust for the support of a Rhodes Scholarship. M.L.G. was also (partially) supported by the LDRD program of LANL under project number 20230049DR; and the Engineering and Physical Sciences Research Council under EPSRC project EP/Y004655/1. 
This work was also supported by the Quantum Science Center (QSC), a National Quantum Information Science Research Center of the U.S. Department of Energy (DOE).
The authors acknowledge the use of the University of Oxford Advanced Research Computing (ARC) facility \cite{oxfordARC} in carrying out this work.


\begin{thebibliography}{86}%
\makeatletter
\providecommand \@ifxundefined [1]{%
 \@ifx{#1\undefined}
}%
\providecommand \@ifnum [1]{%
 \ifnum #1\expandafter \@firstoftwo
 \else \expandafter \@secondoftwo
 \fi
}%
\providecommand \@ifx [1]{%
 \ifx #1\expandafter \@firstoftwo
 \else \expandafter \@secondoftwo
 \fi
}%
\providecommand \natexlab [1]{#1}%
\providecommand \enquote  [1]{``#1''}%
\providecommand \bibnamefont  [1]{#1}%
\providecommand \bibfnamefont [1]{#1}%
\providecommand \citenamefont [1]{#1}%
\providecommand \href@noop [0]{\@secondoftwo}%
\providecommand \href [0]{\begingroup \@sanitize@url \@href}%
\providecommand \@href[1]{\@@startlink{#1}\@@href}%
\providecommand \@@href[1]{\endgroup#1\@@endlink}%
\providecommand \@sanitize@url [0]{\catcode `\\12\catcode `\$12\catcode
  `\&12\catcode `\#12\catcode `\^12\catcode `\_12\catcode `\%12\relax}%
\providecommand \@@startlink[1]{}%
\providecommand \@@endlink[0]{}%
\providecommand \url  [0]{\begingroup\@sanitize@url \@url }%
\providecommand \@url [1]{\endgroup\@href {#1}{\urlprefix }}%
\providecommand \urlprefix  [0]{URL }%
\providecommand \Eprint [0]{\href }%
\providecommand \doibase [0]{https://doi.org/}%
\providecommand \selectlanguage [0]{\@gobble}%
\providecommand \bibinfo  [0]{\@secondoftwo}%
\providecommand \bibfield  [0]{\@secondoftwo}%
\providecommand \translation [1]{[#1]}%
\providecommand \BibitemOpen [0]{}%
\providecommand \bibitemStop [0]{}%
\providecommand \bibitemNoStop [0]{.\EOS\space}%
\providecommand \EOS [0]{\spacefactor3000\relax}%
\providecommand \BibitemShut  [1]{\csname bibitem#1\endcsname}%
\let\auto@bib@innerbib\@empty
\bibitem [{\citenamefont {Giovannetti}\ \emph {et~al.}(2011)\citenamefont
  {Giovannetti}, \citenamefont {Lloyd},\ and\ \citenamefont
  {Maccone}}]{giovannetti2011advances}%
  \BibitemOpen
  \bibfield  {author} {\bibinfo {author} {\bibfnamefont {V.}~\bibnamefont
  {Giovannetti}}, \bibinfo {author} {\bibfnamefont {S.}~\bibnamefont {Lloyd}},\
  and\ \bibinfo {author} {\bibfnamefont {L.}~\bibnamefont {Maccone}},\
  }\bibfield  {title} {\bibinfo {title} {Advances in quantum metrology},\
  }\href {https://www.nature.com/articles/nphoton.2011.35} {\bibfield
  {journal} {\bibinfo  {journal} {Nat. Photonics}\ }\textbf {\bibinfo {volume}
  {5}},\ \bibinfo {pages} {222} (\bibinfo {year} {2011})}\BibitemShut {NoStop}%
\bibitem [{\citenamefont {Paris}(2009)}]{paris2009quantum}%
  \BibitemOpen
  \bibfield  {author} {\bibinfo {author} {\bibfnamefont {M.~G.}\ \bibnamefont
  {Paris}},\ }\bibfield  {title} {\bibinfo {title} {Quantum estimation for
  quantum technology},\ }\href {https://doi.org/10.1142/S0219749909004839}
  {\bibfield  {journal} {\bibinfo  {journal} {International Journal of Quantum
  Information}\ }\textbf {\bibinfo {volume} {7}},\ \bibinfo {pages} {125}
  (\bibinfo {year} {2009})}\BibitemShut {NoStop}%
\bibitem [{\citenamefont {Ye}\ and\ \citenamefont
  {Zoller}(2024)}]{Ye2024Essay}%
  \BibitemOpen
  \bibfield  {author} {\bibinfo {author} {\bibfnamefont {J.}~\bibnamefont
  {Ye}}\ and\ \bibinfo {author} {\bibfnamefont {P.}~\bibnamefont {Zoller}},\
  }\bibfield  {title} {\bibinfo {title} {Essay: Quantum sensing with atomic,
  molecular, and optical platforms for fundamental physics},\ }\href
  {https://doi.org/10.1103/PhysRevLett.132.190001} {\bibfield  {journal}
  {\bibinfo  {journal} {Phys. Rev. Lett.}\ }\textbf {\bibinfo {volume} {132}},\
  \bibinfo {pages} {190001} (\bibinfo {year} {2024})}\BibitemShut {NoStop}%
\bibitem [{\citenamefont {Degen}\ \emph {et~al.}(2017)\citenamefont {Degen},
  \citenamefont {Reinhard},\ and\ \citenamefont
  {Cappellaro}}]{degen2017quantum}%
  \BibitemOpen
  \bibfield  {author} {\bibinfo {author} {\bibfnamefont {C.~L.}\ \bibnamefont
  {Degen}}, \bibinfo {author} {\bibfnamefont {F.}~\bibnamefont {Reinhard}},\
  and\ \bibinfo {author} {\bibfnamefont {P.}~\bibnamefont {Cappellaro}},\
  }\bibfield  {title} {\bibinfo {title} {Quantum sensing},\ }\href
  {https://doi.org/10.1103/RevModPhys.89.035002} {\bibfield  {journal}
  {\bibinfo  {journal} {Rev. Mod. Phys.}\ }\textbf {\bibinfo {volume} {89}},\
  \bibinfo {pages} {035002} (\bibinfo {year} {2017})}\BibitemShut {NoStop}%
\bibitem [{\citenamefont {Szigeti}\ \emph {et~al.}(2021)\citenamefont
  {Szigeti}, \citenamefont {Hosten},\ and\ \citenamefont
  {Haine}}]{szigeti2021improving}%
  \BibitemOpen
  \bibfield  {author} {\bibinfo {author} {\bibfnamefont {S.~S.}\ \bibnamefont
  {Szigeti}}, \bibinfo {author} {\bibfnamefont {O.}~\bibnamefont {Hosten}},\
  and\ \bibinfo {author} {\bibfnamefont {S.~A.}\ \bibnamefont {Haine}},\
  }\bibfield  {title} {\bibinfo {title} {Improving cold-atom sensors with
  quantum entanglement: Prospects and challenges},\ }\bibfield  {journal}
  {\bibinfo  {journal} {Applied Physics Letters}\ }\textbf {\bibinfo {volume}
  {118}},\ \href {https://doi.org/10.1063/5.0050235} {10.1063/5.0050235}
  (\bibinfo {year} {2021})\BibitemShut {NoStop}%
\bibitem [{\citenamefont {Huang}\ \emph {et~al.}(2024)\citenamefont {Huang},
  \citenamefont {Zhuang},\ and\ \citenamefont
  {Lee}}]{huang2024entanglementenhanced}%
  \BibitemOpen
  \bibfield  {author} {\bibinfo {author} {\bibfnamefont {J.}~\bibnamefont
  {Huang}}, \bibinfo {author} {\bibfnamefont {M.}~\bibnamefont {Zhuang}},\ and\
  \bibinfo {author} {\bibfnamefont {C.}~\bibnamefont {Lee}},\ }\bibfield
  {title} {\bibinfo {title} {Entanglement-enhanced quantum metrology: From
  standard quantum limit to heisenberg limit},\ }\bibfield  {journal} {\bibinfo
   {journal} {Applied Physics Reviews}\ }\textbf {\bibinfo {volume} {11}},\
  \href {https://doi.org/10.1063/5.0204102} {10.1063/5.0204102} (\bibinfo
  {year} {2024})\BibitemShut {NoStop}%
\bibitem [{\citenamefont {Escher}\ \emph {et~al.}(2011)\citenamefont {Escher},
  \citenamefont {de~Matos~Filho},\ and\ \citenamefont
  {Davidovich}}]{escher2011general}%
  \BibitemOpen
  \bibfield  {author} {\bibinfo {author} {\bibfnamefont {B.}~\bibnamefont
  {Escher}}, \bibinfo {author} {\bibfnamefont {R.~L.}\ \bibnamefont
  {de~Matos~Filho}},\ and\ \bibinfo {author} {\bibfnamefont {L.}~\bibnamefont
  {Davidovich}},\ }\bibfield  {title} {\bibinfo {title} {General framework for
  estimating the ultimate precision limit in noisy quantum-enhanced
  metrology},\ }\href {https://doi.org/10.1038/nphys1958} {\bibfield  {journal}
  {\bibinfo  {journal} {Nature Physics}\ }\textbf {\bibinfo {volume} {7}},\
  \bibinfo {pages} {406} (\bibinfo {year} {2011})}\BibitemShut {NoStop}%
\bibitem [{\citenamefont {Demkowicz-Dobrza{\'n}ski}\ \emph
  {et~al.}(2012)\citenamefont {Demkowicz-Dobrza{\'n}ski}, \citenamefont
  {Ko{\l}ody{\'n}ski},\ and\ \citenamefont
  {Gu{\c{t}}{\u{a}}}}]{demkowicz2012elusive}%
  \BibitemOpen
  \bibfield  {author} {\bibinfo {author} {\bibfnamefont {R.}~\bibnamefont
  {Demkowicz-Dobrza{\'n}ski}}, \bibinfo {author} {\bibfnamefont
  {J.}~\bibnamefont {Ko{\l}ody{\'n}ski}},\ and\ \bibinfo {author}
  {\bibfnamefont {M.}~\bibnamefont {Gu{\c{t}}{\u{a}}}},\ }\bibfield  {title}
  {\bibinfo {title} {The elusive heisenberg limit in quantum-enhanced
  metrology},\ }\href {https://doi.org/10.1038/ncomms2067} {\bibfield
  {journal} {\bibinfo  {journal} {Nature communications}\ }\textbf {\bibinfo
  {volume} {3}},\ \bibinfo {pages} {1063} (\bibinfo {year} {2012})}\BibitemShut
  {NoStop}%
\bibitem [{\citenamefont {Haine}\ \emph {et~al.}(2015)\citenamefont {Haine},
  \citenamefont {Szigeti}, \citenamefont {Lang},\ and\ \citenamefont
  {Caves}}]{haine2015heisenberg}%
  \BibitemOpen
  \bibfield  {author} {\bibinfo {author} {\bibfnamefont {S.~A.}\ \bibnamefont
  {Haine}}, \bibinfo {author} {\bibfnamefont {S.~S.}\ \bibnamefont {Szigeti}},
  \bibinfo {author} {\bibfnamefont {M.~D.}\ \bibnamefont {Lang}},\ and\
  \bibinfo {author} {\bibfnamefont {C.~M.}\ \bibnamefont {Caves}},\ }\bibfield
  {title} {\bibinfo {title} {Heisenberg-limited metrology with information
  recycling},\ }\href {https://doi.org/10.1103/PhysRevA.91.041802} {\bibfield
  {journal} {\bibinfo  {journal} {Physical Review A}\ }\textbf {\bibinfo
  {volume} {91}},\ \bibinfo {pages} {041802} (\bibinfo {year}
  {2015})}\BibitemShut {NoStop}%
\bibitem [{\citenamefont {Nolan}\ \emph {et~al.}(2017)\citenamefont {Nolan},
  \citenamefont {Szigeti},\ and\ \citenamefont {Haine}}]{nolan2017optimal}%
  \BibitemOpen
  \bibfield  {author} {\bibinfo {author} {\bibfnamefont {S.~P.}\ \bibnamefont
  {Nolan}}, \bibinfo {author} {\bibfnamefont {S.~S.}\ \bibnamefont {Szigeti}},\
  and\ \bibinfo {author} {\bibfnamefont {S.~A.}\ \bibnamefont {Haine}},\
  }\bibfield  {title} {\bibinfo {title} {Optimal and robust quantum metrology
  using interaction-based readouts},\ }\href
  {https://doi.org/10.1103/PhysRevLett.119.193601} {\bibfield  {journal}
  {\bibinfo  {journal} {Physical Review Letters}\ }\textbf {\bibinfo {volume}
  {119}},\ \bibinfo {pages} {193601} (\bibinfo {year} {2017})}\BibitemShut
  {NoStop}%
\bibitem [{\citenamefont {M{\"u}ller}\ \emph {et~al.}(2018)\citenamefont
  {M{\"u}ller}, \citenamefont {Gherardini},\ and\ \citenamefont
  {Caruso}}]{muller2018noise}%
  \BibitemOpen
  \bibfield  {author} {\bibinfo {author} {\bibfnamefont {M.~M.}\ \bibnamefont
  {M{\"u}ller}}, \bibinfo {author} {\bibfnamefont {S.}~\bibnamefont
  {Gherardini}},\ and\ \bibinfo {author} {\bibfnamefont {F.}~\bibnamefont
  {Caruso}},\ }\bibfield  {title} {\bibinfo {title} {Noise-robust quantum
  sensing via optimal multi-probe spectroscopy},\ }\href
  {https://doi.org/10.1038/s41598-018-32434-x} {\bibfield  {journal} {\bibinfo
  {journal} {Scientific reports}\ }\textbf {\bibinfo {volume} {8}},\ \bibinfo
  {pages} {14278} (\bibinfo {year} {2018})}\BibitemShut {NoStop}%
\bibitem [{\citenamefont {Fiderer}\ \emph {et~al.}(2019)\citenamefont
  {Fiderer}, \citenamefont {Fra{\"i}sse},\ and\ \citenamefont
  {Braun}}]{fiderer2019maximal}%
  \BibitemOpen
  \bibfield  {author} {\bibinfo {author} {\bibfnamefont {L.~J.}\ \bibnamefont
  {Fiderer}}, \bibinfo {author} {\bibfnamefont {J.~M.}\ \bibnamefont
  {Fra{\"i}sse}},\ and\ \bibinfo {author} {\bibfnamefont {D.}~\bibnamefont
  {Braun}},\ }\bibfield  {title} {\bibinfo {title} {Maximal quantum fisher
  information for mixed states},\ }\href
  {https://doi.org/10.1103/PhysRevLett.123.250502} {\bibfield  {journal}
  {\bibinfo  {journal} {Physical Review Letters}\ }\textbf {\bibinfo {volume}
  {123}},\ \bibinfo {pages} {250502} (\bibinfo {year} {2019})}\BibitemShut
  {NoStop}%
\bibitem [{\citenamefont {Wang}\ \emph {et~al.}(2021)\citenamefont {Wang},
  \citenamefont {Fontana}, \citenamefont {Cerezo}, \citenamefont {Sharma},
  \citenamefont {Sone}, \citenamefont {Cincio},\ and\ \citenamefont
  {Coles}}]{wang2020noise}%
  \BibitemOpen
  \bibfield  {author} {\bibinfo {author} {\bibfnamefont {S.}~\bibnamefont
  {Wang}}, \bibinfo {author} {\bibfnamefont {E.}~\bibnamefont {Fontana}},
  \bibinfo {author} {\bibfnamefont {M.}~\bibnamefont {Cerezo}}, \bibinfo
  {author} {\bibfnamefont {K.}~\bibnamefont {Sharma}}, \bibinfo {author}
  {\bibfnamefont {A.}~\bibnamefont {Sone}}, \bibinfo {author} {\bibfnamefont
  {L.}~\bibnamefont {Cincio}},\ and\ \bibinfo {author} {\bibfnamefont {P.~J.}\
  \bibnamefont {Coles}},\ }\bibfield  {title} {\bibinfo {title} {Noise-induced
  barren plateaus in variational quantum algorithms},\ }\href
  {https://doi.org/10.1038/s41467-021-27045-6} {\bibfield  {journal} {\bibinfo
  {journal} {Nature Communications}\ }\textbf {\bibinfo {volume} {12}},\
  \bibinfo {pages} {1} (\bibinfo {year} {2021})}\BibitemShut {NoStop}%
\bibitem [{\citenamefont {Cerezo}\ \emph
  {et~al.}(2021{\natexlab{a}})\citenamefont {Cerezo}, \citenamefont {Sone},
  \citenamefont {Beckey},\ and\ \citenamefont {Coles}}]{cerezo2021sub}%
  \BibitemOpen
  \bibfield  {author} {\bibinfo {author} {\bibfnamefont {M.}~\bibnamefont
  {Cerezo}}, \bibinfo {author} {\bibfnamefont {A.}~\bibnamefont {Sone}},
  \bibinfo {author} {\bibfnamefont {J.~L.}\ \bibnamefont {Beckey}},\ and\
  \bibinfo {author} {\bibfnamefont {P.~J.}\ \bibnamefont {Coles}},\ }\bibfield
  {title} {\bibinfo {title} {Sub-quantum fisher information},\ }\bibfield
  {journal} {\bibinfo  {journal} {Quantum Science and Technology}\ }\href
  {https://doi.org/10.1088/2058-9565/abfbef} {10.1088/2058-9565/abfbef}
  (\bibinfo {year} {2021}{\natexlab{a}})\BibitemShut {NoStop}%
\bibitem [{\citenamefont {Zhou}\ \emph {et~al.}(2023)\citenamefont {Zhou},
  \citenamefont {Michalakis},\ and\ \citenamefont {Gefen}}]{zhou2023optimal}%
  \BibitemOpen
  \bibfield  {author} {\bibinfo {author} {\bibfnamefont {S.}~\bibnamefont
  {Zhou}}, \bibinfo {author} {\bibfnamefont {S.}~\bibnamefont {Michalakis}},\
  and\ \bibinfo {author} {\bibfnamefont {T.}~\bibnamefont {Gefen}},\ }\bibfield
   {title} {\bibinfo {title} {Optimal protocols for quantum metrology with
  noisy measurements},\ }\href {https://doi.org/10.1103/PRXQuantum.4.040305}
  {\bibfield  {journal} {\bibinfo  {journal} {PRX Quantum}\ }\textbf {\bibinfo
  {volume} {4}},\ \bibinfo {pages} {040305} (\bibinfo {year}
  {2023})}\BibitemShut {NoStop}%
\bibitem [{\citenamefont {Garc{\'i}a-Mart{\'i}n}\ \emph
  {et~al.}(2024)\citenamefont {Garc{\'i}a-Mart{\'i}n}, \citenamefont
  {Larocca},\ and\ \citenamefont {Cerezo}}]{garcia2023effects}%
  \BibitemOpen
  \bibfield  {author} {\bibinfo {author} {\bibfnamefont {D.}~\bibnamefont
  {Garc{\'i}a-Mart{\'i}n}}, \bibinfo {author} {\bibfnamefont {M.}~\bibnamefont
  {Larocca}},\ and\ \bibinfo {author} {\bibfnamefont {M.}~\bibnamefont
  {Cerezo}},\ }\bibfield  {title} {\bibinfo {title} {Effects of noise on the
  overparametrization of quantum neural networks},\ }\href
  {https://doi.org/10.1103/PhysRevResearch.6.013295} {\bibfield  {journal}
  {\bibinfo  {journal} {Phys. Rev. Res.}\ }\textbf {\bibinfo {volume} {6}},\
  \bibinfo {pages} {013295} (\bibinfo {year} {2024})}\BibitemShut {NoStop}%
\bibitem [{\citenamefont {Kwon}\ \emph {et~al.}(2023)\citenamefont {Kwon},
  \citenamefont {Oh}, \citenamefont {Lim}, \citenamefont {Jeong},\ and\
  \citenamefont {Jiang}}]{kwon2023efficacy}%
  \BibitemOpen
  \bibfield  {author} {\bibinfo {author} {\bibfnamefont {H.}~\bibnamefont
  {Kwon}}, \bibinfo {author} {\bibfnamefont {C.}~\bibnamefont {Oh}}, \bibinfo
  {author} {\bibfnamefont {Y.}~\bibnamefont {Lim}}, \bibinfo {author}
  {\bibfnamefont {H.}~\bibnamefont {Jeong}},\ and\ \bibinfo {author}
  {\bibfnamefont {L.}~\bibnamefont {Jiang}},\ }\bibfield  {title} {\bibinfo
  {title} {Efficacy of virtual purification-based error mitigation on quantum
  metrology},\ }\href {https://arxiv.org/abs/2303.15838} {\bibfield  {journal}
  {\bibinfo  {journal} {arXiv preprint arXiv:2303.15838}\ } (\bibinfo {year}
  {2023})}\BibitemShut {NoStop}%
\bibitem [{\citenamefont {Krinner}\ \emph {et~al.}(2022)\citenamefont
  {Krinner}, \citenamefont {Lacroix}, \citenamefont {Remm}, \citenamefont
  {Di~Paolo}, \citenamefont {Genois}, \citenamefont {Leroux}, \citenamefont
  {Hellings}, \citenamefont {Lazar}, \citenamefont {Swiadek}, \citenamefont
  {Herrmann} \emph {et~al.}}]{krinner2022realizing}%
  \BibitemOpen
  \bibfield  {author} {\bibinfo {author} {\bibfnamefont {S.}~\bibnamefont
  {Krinner}}, \bibinfo {author} {\bibfnamefont {N.}~\bibnamefont {Lacroix}},
  \bibinfo {author} {\bibfnamefont {A.}~\bibnamefont {Remm}}, \bibinfo {author}
  {\bibfnamefont {A.}~\bibnamefont {Di~Paolo}}, \bibinfo {author}
  {\bibfnamefont {E.}~\bibnamefont {Genois}}, \bibinfo {author} {\bibfnamefont
  {C.}~\bibnamefont {Leroux}}, \bibinfo {author} {\bibfnamefont
  {C.}~\bibnamefont {Hellings}}, \bibinfo {author} {\bibfnamefont
  {S.}~\bibnamefont {Lazar}}, \bibinfo {author} {\bibfnamefont
  {F.}~\bibnamefont {Swiadek}}, \bibinfo {author} {\bibfnamefont
  {J.}~\bibnamefont {Herrmann}}, \emph {et~al.},\ }\bibfield  {title} {\bibinfo
  {title} {Realizing repeated quantum error correction in a distance-three
  surface code},\ }\href
  {https://doi.org/https://doi.org/10.1038/s41586-022-04566-8} {\bibfield
  {journal} {\bibinfo  {journal} {Nature}\ }\textbf {\bibinfo {volume} {605}},\
  \bibinfo {pages} {669} (\bibinfo {year} {2022})}\BibitemShut {NoStop}%
\bibitem [{\citenamefont {Zhao}\ \emph {et~al.}(2022)\citenamefont {Zhao},
  \citenamefont {Ye}, \citenamefont {Huang}, \citenamefont {Zhang},
  \citenamefont {Wu}, \citenamefont {Guan}, \citenamefont {Zhu}, \citenamefont
  {Wei}, \citenamefont {He}, \citenamefont {Cao} \emph
  {et~al.}}]{zhao2022realization}%
  \BibitemOpen
  \bibfield  {author} {\bibinfo {author} {\bibfnamefont {Y.}~\bibnamefont
  {Zhao}}, \bibinfo {author} {\bibfnamefont {Y.}~\bibnamefont {Ye}}, \bibinfo
  {author} {\bibfnamefont {H.-L.}\ \bibnamefont {Huang}}, \bibinfo {author}
  {\bibfnamefont {Y.}~\bibnamefont {Zhang}}, \bibinfo {author} {\bibfnamefont
  {D.}~\bibnamefont {Wu}}, \bibinfo {author} {\bibfnamefont {H.}~\bibnamefont
  {Guan}}, \bibinfo {author} {\bibfnamefont {Q.}~\bibnamefont {Zhu}}, \bibinfo
  {author} {\bibfnamefont {Z.}~\bibnamefont {Wei}}, \bibinfo {author}
  {\bibfnamefont {T.}~\bibnamefont {He}}, \bibinfo {author} {\bibfnamefont
  {S.}~\bibnamefont {Cao}}, \emph {et~al.},\ }\bibfield  {title} {\bibinfo
  {title} {Realization of an error-correcting surface code with superconducting
  qubits},\ }\href {https://doi.org/10.1103/PhysRevLett.129.030501} {\bibfield
  {journal} {\bibinfo  {journal} {Physical Review Letters}\ }\textbf {\bibinfo
  {volume} {129}},\ \bibinfo {pages} {030501} (\bibinfo {year}
  {2022})}\BibitemShut {NoStop}%
\bibitem [{\citenamefont {Acharya}\ \emph {et~al.}(2023)\citenamefont
  {Acharya}, \citenamefont {Aleiner}, \citenamefont {Allen}, \citenamefont
  {Andersen}, \citenamefont {Ansmann}, \citenamefont {Arute}, \citenamefont
  {Arya}, \citenamefont {Asfaw}, \citenamefont {Atalaya}, \citenamefont
  {Babbush} \emph {et~al.}}]{acharya2022suppressing}%
  \BibitemOpen
  \bibfield  {author} {\bibinfo {author} {\bibfnamefont {R.}~\bibnamefont
  {Acharya}}, \bibinfo {author} {\bibfnamefont {I.}~\bibnamefont {Aleiner}},
  \bibinfo {author} {\bibfnamefont {R.}~\bibnamefont {Allen}}, \bibinfo
  {author} {\bibfnamefont {T.~I.}\ \bibnamefont {Andersen}}, \bibinfo {author}
  {\bibfnamefont {M.}~\bibnamefont {Ansmann}}, \bibinfo {author} {\bibfnamefont
  {F.}~\bibnamefont {Arute}}, \bibinfo {author} {\bibfnamefont
  {K.}~\bibnamefont {Arya}}, \bibinfo {author} {\bibfnamefont {A.}~\bibnamefont
  {Asfaw}}, \bibinfo {author} {\bibfnamefont {J.}~\bibnamefont {Atalaya}},
  \bibinfo {author} {\bibfnamefont {R.}~\bibnamefont {Babbush}}, \emph
  {et~al.},\ }\bibfield  {title} {\bibinfo {title} {Suppressing quantum errors
  by scaling a surface code logical qubit},\ }\href
  {https://doi.org/10.1038/s41586-022-05434-1} {\bibfield  {journal} {\bibinfo
  {journal} {Nature}\ }\textbf {\bibinfo {volume} {614}},\ \bibinfo {pages}
  {676} (\bibinfo {year} {2023})}\BibitemShut {NoStop}%
\bibitem [{\citenamefont {Bluvstein}\ \emph {et~al.}(2023)\citenamefont
  {Bluvstein}, \citenamefont {Evered}, \citenamefont {Geim}, \citenamefont
  {Li}, \citenamefont {Zhou}, \citenamefont {Manovitz}, \citenamefont {Ebadi},
  \citenamefont {Cain}, \citenamefont {Kalinowski}, \citenamefont {Hangleiter}
  \emph {et~al.}}]{bluvstein2023logical}%
  \BibitemOpen
  \bibfield  {author} {\bibinfo {author} {\bibfnamefont {D.}~\bibnamefont
  {Bluvstein}}, \bibinfo {author} {\bibfnamefont {S.~J.}\ \bibnamefont
  {Evered}}, \bibinfo {author} {\bibfnamefont {A.~A.}\ \bibnamefont {Geim}},
  \bibinfo {author} {\bibfnamefont {S.~H.}\ \bibnamefont {Li}}, \bibinfo
  {author} {\bibfnamefont {H.}~\bibnamefont {Zhou}}, \bibinfo {author}
  {\bibfnamefont {T.}~\bibnamefont {Manovitz}}, \bibinfo {author}
  {\bibfnamefont {S.}~\bibnamefont {Ebadi}}, \bibinfo {author} {\bibfnamefont
  {M.}~\bibnamefont {Cain}}, \bibinfo {author} {\bibfnamefont {M.}~\bibnamefont
  {Kalinowski}}, \bibinfo {author} {\bibfnamefont {D.}~\bibnamefont
  {Hangleiter}}, \emph {et~al.},\ }\bibfield  {title} {\bibinfo {title}
  {Logical quantum processor based on reconfigurable atom arrays},\ }\href
  {https://www.nature.com/articles/s41586-023-06927-3#article-info} {\bibfield
  {journal} {\bibinfo  {journal} {Nature}\ ,\ \bibinfo {pages} {1}} (\bibinfo
  {year} {2023})}\BibitemShut {NoStop}%
\bibitem [{\citenamefont {Da~Silva}\ \emph {et~al.}(2024)\citenamefont
  {Da~Silva}, \citenamefont {Ryan-Anderson}, \citenamefont {Bello-Rivas},
  \citenamefont {Chernoguzov}, \citenamefont {Dreiling}, \citenamefont {Foltz},
  \citenamefont {Gaebler}, \citenamefont {Gatterman}, \citenamefont {Hayes},
  \citenamefont {Hewitt} \emph {et~al.}}]{da2024demonstration}%
  \BibitemOpen
  \bibfield  {author} {\bibinfo {author} {\bibfnamefont {M.}~\bibnamefont
  {Da~Silva}}, \bibinfo {author} {\bibfnamefont {C.}~\bibnamefont
  {Ryan-Anderson}}, \bibinfo {author} {\bibfnamefont {J.}~\bibnamefont
  {Bello-Rivas}}, \bibinfo {author} {\bibfnamefont {A.}~\bibnamefont
  {Chernoguzov}}, \bibinfo {author} {\bibfnamefont {J.}~\bibnamefont
  {Dreiling}}, \bibinfo {author} {\bibfnamefont {C.}~\bibnamefont {Foltz}},
  \bibinfo {author} {\bibfnamefont {J.}~\bibnamefont {Gaebler}}, \bibinfo
  {author} {\bibfnamefont {T.}~\bibnamefont {Gatterman}}, \bibinfo {author}
  {\bibfnamefont {D.}~\bibnamefont {Hayes}}, \bibinfo {author} {\bibfnamefont
  {N.}~\bibnamefont {Hewitt}}, \emph {et~al.},\ }\bibfield  {title} {\bibinfo
  {title} {Demonstration of logical qubits and repeated error correction with
  better-than-physical error rates},\ }\href {https://arxiv.org/abs/2404.02280}
  {\bibfield  {journal} {\bibinfo  {journal} {arXiv preprint arXiv:2404.02280}\
  } (\bibinfo {year} {2024})}\BibitemShut {NoStop}%
\bibitem [{\citenamefont {Acharya}\ \emph {et~al.}(2024)\citenamefont
  {Acharya}, \citenamefont {Aghababaie-Beni}, \citenamefont {Aleiner},
  \citenamefont {Andersen}, \citenamefont {Ansmann}, \citenamefont {Arute},
  \citenamefont {Arya}, \citenamefont {Asfaw}, \citenamefont {Astrakhantsev},
  \citenamefont {Atalaya} \emph {et~al.}}]{acharya2024quantum}%
  \BibitemOpen
  \bibfield  {author} {\bibinfo {author} {\bibfnamefont {R.}~\bibnamefont
  {Acharya}}, \bibinfo {author} {\bibfnamefont {L.}~\bibnamefont
  {Aghababaie-Beni}}, \bibinfo {author} {\bibfnamefont {I.}~\bibnamefont
  {Aleiner}}, \bibinfo {author} {\bibfnamefont {T.~I.}\ \bibnamefont
  {Andersen}}, \bibinfo {author} {\bibfnamefont {M.}~\bibnamefont {Ansmann}},
  \bibinfo {author} {\bibfnamefont {F.}~\bibnamefont {Arute}}, \bibinfo
  {author} {\bibfnamefont {K.}~\bibnamefont {Arya}}, \bibinfo {author}
  {\bibfnamefont {A.}~\bibnamefont {Asfaw}}, \bibinfo {author} {\bibfnamefont
  {N.}~\bibnamefont {Astrakhantsev}}, \bibinfo {author} {\bibfnamefont
  {J.}~\bibnamefont {Atalaya}}, \emph {et~al.},\ }\bibfield  {title} {\bibinfo
  {title} {Quantum error correction below the surface code threshold},\ }\href
  {https://arxiv.org/abs/2408.13687} {\bibfield  {journal} {\bibinfo  {journal}
  {arXiv preprint arXiv:2408.13687}\ } (\bibinfo {year} {2024})}\BibitemShut
  {NoStop}%
\bibitem [{\citenamefont {Sundaresan}\ \emph {et~al.}(2023)\citenamefont
  {Sundaresan}, \citenamefont {Yoder}, \citenamefont {Kim}, \citenamefont {Li},
  \citenamefont {Chen}, \citenamefont {Harper}, \citenamefont {Thorbeck},
  \citenamefont {Cross}, \citenamefont {C{\'o}rcoles},\ and\ \citenamefont
  {Takita}}]{sundaresan2023demonstrating}%
  \BibitemOpen
  \bibfield  {author} {\bibinfo {author} {\bibfnamefont {N.}~\bibnamefont
  {Sundaresan}}, \bibinfo {author} {\bibfnamefont {T.~J.}\ \bibnamefont
  {Yoder}}, \bibinfo {author} {\bibfnamefont {Y.}~\bibnamefont {Kim}}, \bibinfo
  {author} {\bibfnamefont {M.}~\bibnamefont {Li}}, \bibinfo {author}
  {\bibfnamefont {E.~H.}\ \bibnamefont {Chen}}, \bibinfo {author}
  {\bibfnamefont {G.}~\bibnamefont {Harper}}, \bibinfo {author} {\bibfnamefont
  {T.}~\bibnamefont {Thorbeck}}, \bibinfo {author} {\bibfnamefont {A.~W.}\
  \bibnamefont {Cross}}, \bibinfo {author} {\bibfnamefont {A.~D.}\ \bibnamefont
  {C{\'o}rcoles}},\ and\ \bibinfo {author} {\bibfnamefont {M.}~\bibnamefont
  {Takita}},\ }\bibfield  {title} {\bibinfo {title} {Demonstrating multi-round
  subsystem quantum error correction using matching and maximum likelihood
  decoders},\ }\href {https://doi.org/10.1038/s41467-023-38247-5} {\bibfield
  {journal} {\bibinfo  {journal} {Nature Communications}\ }\textbf {\bibinfo
  {volume} {14}},\ \bibinfo {pages} {2852} (\bibinfo {year}
  {2023})}\BibitemShut {NoStop}%
\bibitem [{\citenamefont {Putterman}\ \emph {et~al.}(2024)\citenamefont
  {Putterman}, \citenamefont {Noh}, \citenamefont {Hann}, \citenamefont
  {MacCabe}, \citenamefont {Aghaeimeibodi}, \citenamefont {Patel},
  \citenamefont {Lee}, \citenamefont {Jones}, \citenamefont {Moradinejad},
  \citenamefont {Rodriguez}, \citenamefont {Mahuli}, \citenamefont {Rose},
  \citenamefont {Owens}, \citenamefont {Levine}, \citenamefont {Rosenfeld},
  \citenamefont {Reinhold}, \citenamefont {Moncelsi}, \citenamefont {Alcid},
  \citenamefont {Alidoust}, \citenamefont {Arrangoiz-Arriola}, \citenamefont
  {Barnett}, \citenamefont {Bienias}, \citenamefont {Carson}, \citenamefont
  {Chen}, \citenamefont {Chen}, \citenamefont {Chinkezian}, \citenamefont
  {Chisholm}, \citenamefont {Chou}, \citenamefont {Clerk}, \citenamefont
  {Clifford}, \citenamefont {Cosmic}, \citenamefont {Curiel}, \citenamefont
  {Davis}, \citenamefont {DeLorenzo}, \citenamefont {D'Ewart}, \citenamefont
  {Diky}, \citenamefont {D'Souza}, \citenamefont {Dumitrescu}, \citenamefont
  {Eisenmann}, \citenamefont {Elkhouly}, \citenamefont {Evenbly}, \citenamefont
  {Fang}, \citenamefont {Fang}, \citenamefont {Fling}, \citenamefont {Fon},
  \citenamefont {Garcia}, \citenamefont {Gorshkov}, \citenamefont {Grant},
  \citenamefont {Gray}, \citenamefont {Grimberg}, \citenamefont {Grimsmo},
  \citenamefont {Haim}, \citenamefont {Hand}, \citenamefont {He}, \citenamefont
  {Hernandez}, \citenamefont {Hover}, \citenamefont {Hung}, \citenamefont
  {Hunt}, \citenamefont {Iverson}, \citenamefont {Jarrige}, \citenamefont
  {Jaskula}, \citenamefont {Jiang}, \citenamefont {Kalaee}, \citenamefont
  {Karabalin}, \citenamefont {Karalekas}, \citenamefont {Keller}, \citenamefont
  {Khalajhedayati}, \citenamefont {Kubica}, \citenamefont {Lee}, \citenamefont
  {Leroux}, \citenamefont {Lieu}, \citenamefont {Ly}, \citenamefont {Madrigal},
  \citenamefont {Marcaud}, \citenamefont {McCabe}, \citenamefont {Miles},
  \citenamefont {Milsted}, \citenamefont {Minguzzi}, \citenamefont {Mishra},
  \citenamefont {Mukherjee}, \citenamefont {Naghiloo}, \citenamefont
  {Oblepias}, \citenamefont {Ortuno}, \citenamefont {Pagdilao}, \citenamefont
  {Pancotti}, \citenamefont {Panduro}, \citenamefont {Paquette}, \citenamefont
  {Park}, \citenamefont {Peairs}, \citenamefont {Perello}, \citenamefont
  {Peterson}, \citenamefont {Ponte}, \citenamefont {Preskill}, \citenamefont
  {Qiao}, \citenamefont {Refael}, \citenamefont {Resnick}, \citenamefont
  {Retzker}, \citenamefont {Reyna}, \citenamefont {Runyan}, \citenamefont
  {Ryan}, \citenamefont {Sahmoud}, \citenamefont {Sanchez}, \citenamefont
  {Sanil}, \citenamefont {Sankar}, \citenamefont {Sato}, \citenamefont
  {Scaffidi}, \citenamefont {Siavoshi}, \citenamefont {Sivarajah},
  \citenamefont {Skogland}, \citenamefont {Su}, \citenamefont {Swenson},
  \citenamefont {Teo}, \citenamefont {Tomada}, \citenamefont {Torlai},
  \citenamefont {Wollack}, \citenamefont {Ye}, \citenamefont {Zerrudo},
  \citenamefont {Zhang}, \citenamefont {Brandão}, \citenamefont {Matheny},\
  and\ \citenamefont {Painter}}]{putterman2024hardware}%
  \BibitemOpen
  \bibfield  {author} {\bibinfo {author} {\bibfnamefont {H.}~\bibnamefont
  {Putterman}}, \bibinfo {author} {\bibfnamefont {K.}~\bibnamefont {Noh}},
  \bibinfo {author} {\bibfnamefont {C.~T.}\ \bibnamefont {Hann}}, \bibinfo
  {author} {\bibfnamefont {G.~S.}\ \bibnamefont {MacCabe}}, \bibinfo {author}
  {\bibfnamefont {S.}~\bibnamefont {Aghaeimeibodi}}, \bibinfo {author}
  {\bibfnamefont {R.~N.}\ \bibnamefont {Patel}}, \bibinfo {author}
  {\bibfnamefont {M.}~\bibnamefont {Lee}}, \bibinfo {author} {\bibfnamefont
  {W.~M.}\ \bibnamefont {Jones}}, \bibinfo {author} {\bibfnamefont
  {H.}~\bibnamefont {Moradinejad}}, \bibinfo {author} {\bibfnamefont
  {R.}~\bibnamefont {Rodriguez}}, \bibinfo {author} {\bibfnamefont
  {N.}~\bibnamefont {Mahuli}}, \bibinfo {author} {\bibfnamefont
  {J.}~\bibnamefont {Rose}}, \bibinfo {author} {\bibfnamefont {J.~C.}\
  \bibnamefont {Owens}}, \bibinfo {author} {\bibfnamefont {H.}~\bibnamefont
  {Levine}}, \bibinfo {author} {\bibfnamefont {E.}~\bibnamefont {Rosenfeld}},
  \bibinfo {author} {\bibfnamefont {P.}~\bibnamefont {Reinhold}}, \bibinfo
  {author} {\bibfnamefont {L.}~\bibnamefont {Moncelsi}}, \bibinfo {author}
  {\bibfnamefont {J.~A.}\ \bibnamefont {Alcid}}, \bibinfo {author}
  {\bibfnamefont {N.}~\bibnamefont {Alidoust}}, \bibinfo {author}
  {\bibfnamefont {P.}~\bibnamefont {Arrangoiz-Arriola}}, \bibinfo {author}
  {\bibfnamefont {J.}~\bibnamefont {Barnett}}, \bibinfo {author} {\bibfnamefont
  {P.}~\bibnamefont {Bienias}}, \bibinfo {author} {\bibfnamefont {H.~A.}\
  \bibnamefont {Carson}}, \bibinfo {author} {\bibfnamefont {C.}~\bibnamefont
  {Chen}}, \bibinfo {author} {\bibfnamefont {L.}~\bibnamefont {Chen}}, \bibinfo
  {author} {\bibfnamefont {H.}~\bibnamefont {Chinkezian}}, \bibinfo {author}
  {\bibfnamefont {E.~M.}\ \bibnamefont {Chisholm}}, \bibinfo {author}
  {\bibfnamefont {M.-H.}\ \bibnamefont {Chou}}, \bibinfo {author}
  {\bibfnamefont {A.}~\bibnamefont {Clerk}}, \bibinfo {author} {\bibfnamefont
  {A.}~\bibnamefont {Clifford}}, \bibinfo {author} {\bibfnamefont
  {R.}~\bibnamefont {Cosmic}}, \bibinfo {author} {\bibfnamefont {A.~V.}\
  \bibnamefont {Curiel}}, \bibinfo {author} {\bibfnamefont {E.}~\bibnamefont
  {Davis}}, \bibinfo {author} {\bibfnamefont {L.}~\bibnamefont {DeLorenzo}},
  \bibinfo {author} {\bibfnamefont {J.~M.}\ \bibnamefont {D'Ewart}}, \bibinfo
  {author} {\bibfnamefont {A.}~\bibnamefont {Diky}}, \bibinfo {author}
  {\bibfnamefont {N.}~\bibnamefont {D'Souza}}, \bibinfo {author} {\bibfnamefont
  {P.~T.}\ \bibnamefont {Dumitrescu}}, \bibinfo {author} {\bibfnamefont
  {S.}~\bibnamefont {Eisenmann}}, \bibinfo {author} {\bibfnamefont
  {E.}~\bibnamefont {Elkhouly}}, \bibinfo {author} {\bibfnamefont
  {G.}~\bibnamefont {Evenbly}}, \bibinfo {author} {\bibfnamefont {M.~T.}\
  \bibnamefont {Fang}}, \bibinfo {author} {\bibfnamefont {Y.}~\bibnamefont
  {Fang}}, \bibinfo {author} {\bibfnamefont {M.~J.}\ \bibnamefont {Fling}},
  \bibinfo {author} {\bibfnamefont {W.}~\bibnamefont {Fon}}, \bibinfo {author}
  {\bibfnamefont {G.}~\bibnamefont {Garcia}}, \bibinfo {author} {\bibfnamefont
  {A.~V.}\ \bibnamefont {Gorshkov}}, \bibinfo {author} {\bibfnamefont {J.~A.}\
  \bibnamefont {Grant}}, \bibinfo {author} {\bibfnamefont {M.~J.}\ \bibnamefont
  {Gray}}, \bibinfo {author} {\bibfnamefont {S.}~\bibnamefont {Grimberg}},
  \bibinfo {author} {\bibfnamefont {A.~L.}\ \bibnamefont {Grimsmo}}, \bibinfo
  {author} {\bibfnamefont {A.}~\bibnamefont {Haim}}, \bibinfo {author}
  {\bibfnamefont {J.}~\bibnamefont {Hand}}, \bibinfo {author} {\bibfnamefont
  {Y.}~\bibnamefont {He}}, \bibinfo {author} {\bibfnamefont {M.}~\bibnamefont
  {Hernandez}}, \bibinfo {author} {\bibfnamefont {D.}~\bibnamefont {Hover}},
  \bibinfo {author} {\bibfnamefont {J.~S.~C.}\ \bibnamefont {Hung}}, \bibinfo
  {author} {\bibfnamefont {M.}~\bibnamefont {Hunt}}, \bibinfo {author}
  {\bibfnamefont {J.}~\bibnamefont {Iverson}}, \bibinfo {author} {\bibfnamefont
  {I.}~\bibnamefont {Jarrige}}, \bibinfo {author} {\bibfnamefont {J.-C.}\
  \bibnamefont {Jaskula}}, \bibinfo {author} {\bibfnamefont {L.}~\bibnamefont
  {Jiang}}, \bibinfo {author} {\bibfnamefont {M.}~\bibnamefont {Kalaee}},
  \bibinfo {author} {\bibfnamefont {R.}~\bibnamefont {Karabalin}}, \bibinfo
  {author} {\bibfnamefont {P.~J.}\ \bibnamefont {Karalekas}}, \bibinfo {author}
  {\bibfnamefont {A.~J.}\ \bibnamefont {Keller}}, \bibinfo {author}
  {\bibfnamefont {A.}~\bibnamefont {Khalajhedayati}}, \bibinfo {author}
  {\bibfnamefont {A.}~\bibnamefont {Kubica}}, \bibinfo {author} {\bibfnamefont
  {H.}~\bibnamefont {Lee}}, \bibinfo {author} {\bibfnamefont {C.}~\bibnamefont
  {Leroux}}, \bibinfo {author} {\bibfnamefont {S.}~\bibnamefont {Lieu}},
  \bibinfo {author} {\bibfnamefont {V.}~\bibnamefont {Ly}}, \bibinfo {author}
  {\bibfnamefont {K.~V.}\ \bibnamefont {Madrigal}}, \bibinfo {author}
  {\bibfnamefont {G.}~\bibnamefont {Marcaud}}, \bibinfo {author} {\bibfnamefont
  {G.}~\bibnamefont {McCabe}}, \bibinfo {author} {\bibfnamefont
  {C.}~\bibnamefont {Miles}}, \bibinfo {author} {\bibfnamefont
  {A.}~\bibnamefont {Milsted}}, \bibinfo {author} {\bibfnamefont
  {J.}~\bibnamefont {Minguzzi}}, \bibinfo {author} {\bibfnamefont
  {A.}~\bibnamefont {Mishra}}, \bibinfo {author} {\bibfnamefont
  {B.}~\bibnamefont {Mukherjee}}, \bibinfo {author} {\bibfnamefont
  {M.}~\bibnamefont {Naghiloo}}, \bibinfo {author} {\bibfnamefont
  {E.}~\bibnamefont {Oblepias}}, \bibinfo {author} {\bibfnamefont
  {G.}~\bibnamefont {Ortuno}}, \bibinfo {author} {\bibfnamefont
  {J.}~\bibnamefont {Pagdilao}}, \bibinfo {author} {\bibfnamefont
  {N.}~\bibnamefont {Pancotti}}, \bibinfo {author} {\bibfnamefont
  {A.}~\bibnamefont {Panduro}}, \bibinfo {author} {\bibfnamefont
  {J.}~\bibnamefont {Paquette}}, \bibinfo {author} {\bibfnamefont
  {M.}~\bibnamefont {Park}}, \bibinfo {author} {\bibfnamefont {G.~A.}\
  \bibnamefont {Peairs}}, \bibinfo {author} {\bibfnamefont {D.}~\bibnamefont
  {Perello}}, \bibinfo {author} {\bibfnamefont {E.~C.}\ \bibnamefont
  {Peterson}}, \bibinfo {author} {\bibfnamefont {S.}~\bibnamefont {Ponte}},
  \bibinfo {author} {\bibfnamefont {J.}~\bibnamefont {Preskill}}, \bibinfo
  {author} {\bibfnamefont {J.}~\bibnamefont {Qiao}}, \bibinfo {author}
  {\bibfnamefont {G.}~\bibnamefont {Refael}}, \bibinfo {author} {\bibfnamefont
  {R.}~\bibnamefont {Resnick}}, \bibinfo {author} {\bibfnamefont
  {A.}~\bibnamefont {Retzker}}, \bibinfo {author} {\bibfnamefont {O.~A.}\
  \bibnamefont {Reyna}}, \bibinfo {author} {\bibfnamefont {M.}~\bibnamefont
  {Runyan}}, \bibinfo {author} {\bibfnamefont {C.~A.}\ \bibnamefont {Ryan}},
  \bibinfo {author} {\bibfnamefont {A.}~\bibnamefont {Sahmoud}}, \bibinfo
  {author} {\bibfnamefont {E.}~\bibnamefont {Sanchez}}, \bibinfo {author}
  {\bibfnamefont {R.}~\bibnamefont {Sanil}}, \bibinfo {author} {\bibfnamefont
  {K.}~\bibnamefont {Sankar}}, \bibinfo {author} {\bibfnamefont
  {Y.}~\bibnamefont {Sato}}, \bibinfo {author} {\bibfnamefont {T.}~\bibnamefont
  {Scaffidi}}, \bibinfo {author} {\bibfnamefont {S.}~\bibnamefont {Siavoshi}},
  \bibinfo {author} {\bibfnamefont {P.}~\bibnamefont {Sivarajah}}, \bibinfo
  {author} {\bibfnamefont {T.}~\bibnamefont {Skogland}}, \bibinfo {author}
  {\bibfnamefont {C.-J.}\ \bibnamefont {Su}}, \bibinfo {author} {\bibfnamefont
  {L.~J.}\ \bibnamefont {Swenson}}, \bibinfo {author} {\bibfnamefont {S.~M.}\
  \bibnamefont {Teo}}, \bibinfo {author} {\bibfnamefont {A.}~\bibnamefont
  {Tomada}}, \bibinfo {author} {\bibfnamefont {G.}~\bibnamefont {Torlai}},
  \bibinfo {author} {\bibfnamefont {E.~A.}\ \bibnamefont {Wollack}}, \bibinfo
  {author} {\bibfnamefont {Y.}~\bibnamefont {Ye}}, \bibinfo {author}
  {\bibfnamefont {J.~A.}\ \bibnamefont {Zerrudo}}, \bibinfo {author}
  {\bibfnamefont {K.}~\bibnamefont {Zhang}}, \bibinfo {author} {\bibfnamefont
  {F.~G. S.~L.}\ \bibnamefont {Brandão}}, \bibinfo {author} {\bibfnamefont
  {M.~H.}\ \bibnamefont {Matheny}},\ and\ \bibinfo {author} {\bibfnamefont
  {O.}~\bibnamefont {Painter}},\ }\href {https://arxiv.org/abs/2409.13025}
  {\bibinfo {title} {Hardware-efficient quantum error correction using
  concatenated bosonic qubits}} (\bibinfo {year} {2024})\BibitemShut {NoStop}%
\bibitem [{\citenamefont {Cerezo}\ \emph
  {et~al.}(2021{\natexlab{b}})\citenamefont {Cerezo}, \citenamefont
  {Arrasmith}, \citenamefont {Babbush}, \citenamefont {Benjamin}, \citenamefont
  {Endo}, \citenamefont {Fujii}, \citenamefont {McClean}, \citenamefont
  {Mitarai}, \citenamefont {Yuan}, \citenamefont {Cincio},\ and\ \citenamefont
  {Coles}}]{cerezo2020variationalreview}%
  \BibitemOpen
  \bibfield  {author} {\bibinfo {author} {\bibfnamefont {M.}~\bibnamefont
  {Cerezo}}, \bibinfo {author} {\bibfnamefont {A.}~\bibnamefont {Arrasmith}},
  \bibinfo {author} {\bibfnamefont {R.}~\bibnamefont {Babbush}}, \bibinfo
  {author} {\bibfnamefont {S.~C.}\ \bibnamefont {Benjamin}}, \bibinfo {author}
  {\bibfnamefont {S.}~\bibnamefont {Endo}}, \bibinfo {author} {\bibfnamefont
  {K.}~\bibnamefont {Fujii}}, \bibinfo {author} {\bibfnamefont {J.~R.}\
  \bibnamefont {McClean}}, \bibinfo {author} {\bibfnamefont {K.}~\bibnamefont
  {Mitarai}}, \bibinfo {author} {\bibfnamefont {X.}~\bibnamefont {Yuan}},
  \bibinfo {author} {\bibfnamefont {L.}~\bibnamefont {Cincio}},\ and\ \bibinfo
  {author} {\bibfnamefont {P.~J.}\ \bibnamefont {Coles}},\ }\bibfield  {title}
  {\bibinfo {title} {Variational quantum algorithms},\ }\href
  {https://doi.org/10.1038/s42254-021-00348-9} {\bibfield  {journal} {\bibinfo
  {journal} {Nature Reviews Physics}\ }\textbf {\bibinfo {volume} {3}},\
  \bibinfo {pages} {625–644} (\bibinfo {year}
  {2021}{\natexlab{b}})}\BibitemShut {NoStop}%
\bibitem [{\citenamefont {Cerezo}\ \emph {et~al.}(2022)\citenamefont {Cerezo},
  \citenamefont {Verdon}, \citenamefont {Huang}, \citenamefont {Cincio},\ and\
  \citenamefont {Coles}}]{cerezo2022challenges}%
  \BibitemOpen
  \bibfield  {author} {\bibinfo {author} {\bibfnamefont {M.}~\bibnamefont
  {Cerezo}}, \bibinfo {author} {\bibfnamefont {G.}~\bibnamefont {Verdon}},
  \bibinfo {author} {\bibfnamefont {H.-Y.}\ \bibnamefont {Huang}}, \bibinfo
  {author} {\bibfnamefont {L.}~\bibnamefont {Cincio}},\ and\ \bibinfo {author}
  {\bibfnamefont {P.~J.}\ \bibnamefont {Coles}},\ }\bibfield  {title} {\bibinfo
  {title} {Challenges and opportunities in quantum machine learning},\
  }\bibfield  {journal} {\bibinfo  {journal} {Nature Computational Science}\
  }\href {https://doi.org/10.1038/s43588-022-00311-3}
  {10.1038/s43588-022-00311-3} (\bibinfo {year} {2022})\BibitemShut {NoStop}%
\bibitem [{\citenamefont {Koczor}\ \emph {et~al.}(2020)\citenamefont {Koczor},
  \citenamefont {Endo}, \citenamefont {Jones}, \citenamefont {Matsuzaki},\ and\
  \citenamefont {Benjamin}}]{koczor2020variational}%
  \BibitemOpen
  \bibfield  {author} {\bibinfo {author} {\bibfnamefont {B.}~\bibnamefont
  {Koczor}}, \bibinfo {author} {\bibfnamefont {S.}~\bibnamefont {Endo}},
  \bibinfo {author} {\bibfnamefont {T.}~\bibnamefont {Jones}}, \bibinfo
  {author} {\bibfnamefont {Y.}~\bibnamefont {Matsuzaki}},\ and\ \bibinfo
  {author} {\bibfnamefont {S.~C.}\ \bibnamefont {Benjamin}},\ }\bibfield
  {title} {\bibinfo {title} {Variational-state quantum metrology},\ }\bibfield
  {journal} {\bibinfo  {journal} {New Journal of Physics}\ }\href
  {https://doi.org/10.1088/1367-2630/ab965e} {10.1088/1367-2630/ab965e}
  (\bibinfo {year} {2020})\BibitemShut {NoStop}%
\bibitem [{\citenamefont {Beckey}\ \emph {et~al.}(2022)\citenamefont {Beckey},
  \citenamefont {Cerezo}, \citenamefont {Sone},\ and\ \citenamefont
  {Coles}}]{beckey2020variational}%
  \BibitemOpen
  \bibfield  {author} {\bibinfo {author} {\bibfnamefont {J.~L.}\ \bibnamefont
  {Beckey}}, \bibinfo {author} {\bibfnamefont {M.}~\bibnamefont {Cerezo}},
  \bibinfo {author} {\bibfnamefont {A.}~\bibnamefont {Sone}},\ and\ \bibinfo
  {author} {\bibfnamefont {P.~J.}\ \bibnamefont {Coles}},\ }\bibfield  {title}
  {\bibinfo {title} {Variational quantum algorithm for estimating the quantum
  {F}isher information},\ }\href
  {https://doi.org/10.1103/PhysRevResearch.4.013083} {\bibfield  {journal}
  {\bibinfo  {journal} {Physical Review Research}\ }\textbf {\bibinfo {volume}
  {4}},\ \bibinfo {pages} {013083} (\bibinfo {year} {2022})}\BibitemShut
  {NoStop}%
\bibitem [{\citenamefont {Kaubruegger}\ \emph {et~al.}(2021)\citenamefont
  {Kaubruegger}, \citenamefont {Vasilyev}, \citenamefont {Schulte},
  \citenamefont {Hammerer},\ and\ \citenamefont
  {Zoller}}]{kaubruegger2021quantum}%
  \BibitemOpen
  \bibfield  {author} {\bibinfo {author} {\bibfnamefont {R.}~\bibnamefont
  {Kaubruegger}}, \bibinfo {author} {\bibfnamefont {D.~V.}\ \bibnamefont
  {Vasilyev}}, \bibinfo {author} {\bibfnamefont {M.}~\bibnamefont {Schulte}},
  \bibinfo {author} {\bibfnamefont {K.}~\bibnamefont {Hammerer}},\ and\
  \bibinfo {author} {\bibfnamefont {P.}~\bibnamefont {Zoller}},\ }\bibfield
  {title} {\bibinfo {title} {Quantum variational optimization of {R}amsey
  interferometry and atomic clocks},\ }\href
  {https://doi.org/10.1103/PhysRevX.11.041045} {\bibfield  {journal} {\bibinfo
  {journal} {Physical Review X}\ }\textbf {\bibinfo {volume} {11}},\ \bibinfo
  {pages} {041045} (\bibinfo {year} {2021})}\BibitemShut {NoStop}%
\bibitem [{\citenamefont {Ma}\ \emph {et~al.}(2021)\citenamefont {Ma},
  \citenamefont {Gokhale}, \citenamefont {Zheng}, \citenamefont {Zhou},
  \citenamefont {Yu}, \citenamefont {Jiang}, \citenamefont {Maurer},\ and\
  \citenamefont {Chong}}]{ma2020adaptive}%
  \BibitemOpen
  \bibfield  {author} {\bibinfo {author} {\bibfnamefont {Z.}~\bibnamefont
  {Ma}}, \bibinfo {author} {\bibfnamefont {P.}~\bibnamefont {Gokhale}},
  \bibinfo {author} {\bibfnamefont {T.-X.}\ \bibnamefont {Zheng}}, \bibinfo
  {author} {\bibfnamefont {S.}~\bibnamefont {Zhou}}, \bibinfo {author}
  {\bibfnamefont {X.}~\bibnamefont {Yu}}, \bibinfo {author} {\bibfnamefont
  {L.}~\bibnamefont {Jiang}}, \bibinfo {author} {\bibfnamefont
  {P.}~\bibnamefont {Maurer}},\ and\ \bibinfo {author} {\bibfnamefont {F.~T.}\
  \bibnamefont {Chong}},\ }\bibfield  {title} {\bibinfo {title} {Adaptive
  circuit learning for quantum metrology},\ }in\ \href
  {https://doi.org/10.1109/qce52317.2021.00063} {\emph {\bibinfo {booktitle}
  {2021 IEEE International Conference on Quantum Computing and Engineering
  (QCE)}}}\ (\bibinfo {organization} {IEEE},\ \bibinfo {year} {2021})\ pp.\
  \bibinfo {pages} {419--430}\BibitemShut {NoStop}%
\bibitem [{\citenamefont {Thurtell}\ and\ \citenamefont
  {Miyake}(2024)}]{thurtell2022optimizing}%
  \BibitemOpen
  \bibfield  {author} {\bibinfo {author} {\bibfnamefont {T.~G.}\ \bibnamefont
  {Thurtell}}\ and\ \bibinfo {author} {\bibfnamefont {A.}~\bibnamefont
  {Miyake}},\ }\bibfield  {title} {\bibinfo {title} {Optimizing one-axis twists
  for variational bayesian quantum metrology},\ }\href
  {https://doi.org/10.1103/PhysRevResearch.6.023179} {\bibfield  {journal}
  {\bibinfo  {journal} {Physical Review Research}\ }\textbf {\bibinfo {volume}
  {6}},\ \bibinfo {pages} {023179} (\bibinfo {year} {2024})}\BibitemShut
  {NoStop}%
\bibitem [{\citenamefont {Liu}\ \emph {et~al.}(2022)\citenamefont {Liu},
  \citenamefont {Wu}, \citenamefont {Yang}, \citenamefont {Li}, \citenamefont
  {Zhou}, \citenamefont {Chen}, \citenamefont {Yuan}, \citenamefont {Peng},\
  and\ \citenamefont {Du}}]{liu2022variational}%
  \BibitemOpen
  \bibfield  {author} {\bibinfo {author} {\bibfnamefont {R.}~\bibnamefont
  {Liu}}, \bibinfo {author} {\bibfnamefont {Z.}~\bibnamefont {Wu}}, \bibinfo
  {author} {\bibfnamefont {X.}~\bibnamefont {Yang}}, \bibinfo {author}
  {\bibfnamefont {Y.}~\bibnamefont {Li}}, \bibinfo {author} {\bibfnamefont
  {H.}~\bibnamefont {Zhou}}, \bibinfo {author} {\bibfnamefont {Y.}~\bibnamefont
  {Chen}}, \bibinfo {author} {\bibfnamefont {H.}~\bibnamefont {Yuan}}, \bibinfo
  {author} {\bibfnamefont {X.}~\bibnamefont {Peng}},\ and\ \bibinfo {author}
  {\bibfnamefont {J.}~\bibnamefont {Du}},\ }\bibfield  {title} {\bibinfo
  {title} {Variational quantum metrology with loschmidt echo},\ }\href
  {https://arxiv.org/abs/2211.12296} {\bibfield  {journal} {\bibinfo  {journal}
  {arXiv preprint arXiv:2211.12296}\ } (\bibinfo {year} {2022})}\BibitemShut
  {NoStop}%
\bibitem [{\citenamefont {Le}\ \emph {et~al.}(2023)\citenamefont {Le},
  \citenamefont {Nguyen},\ and\ \citenamefont {Ho}}]{Le2023variational}%
  \BibitemOpen
  \bibfield  {author} {\bibinfo {author} {\bibfnamefont {T.~K.}\ \bibnamefont
  {Le}}, \bibinfo {author} {\bibfnamefont {H.~Q.}\ \bibnamefont {Nguyen}},\
  and\ \bibinfo {author} {\bibfnamefont {L.~B.}\ \bibnamefont {Ho}},\
  }\bibfield  {title} {\bibinfo {title} {Variational quantum metrology for
  multiparameter estimation under dephasing noise},\ }\href
  {https://doi.org/10.1038/s41598-023-44786-0} {\bibfield  {journal} {\bibinfo
  {journal} {Scientific Reports}\ }\textbf {\bibinfo {volume} {13}},\ \bibinfo
  {pages} {17775} (\bibinfo {year} {2023})}\BibitemShut {NoStop}%
\bibitem [{\citenamefont {Meyer}\ \emph {et~al.}(2021)\citenamefont {Meyer},
  \citenamefont {Borregaard},\ and\ \citenamefont
  {Eisert}}]{meyer2020variational}%
  \BibitemOpen
  \bibfield  {author} {\bibinfo {author} {\bibfnamefont {J.~J.}\ \bibnamefont
  {Meyer}}, \bibinfo {author} {\bibfnamefont {J.}~\bibnamefont {Borregaard}},\
  and\ \bibinfo {author} {\bibfnamefont {J.}~\bibnamefont {Eisert}},\
  }\bibfield  {title} {\bibinfo {title} {A variational toolbox for quantum
  multi-parameter estimation},\ }\href
  {https://doi.org/10.1038/s41534-021-00425-y} {\bibfield  {journal} {\bibinfo
  {journal} {NPJ Quantum Information}\ }\textbf {\bibinfo {volume} {7}},\
  \bibinfo {pages} {1} (\bibinfo {year} {2021})}\BibitemShut {NoStop}%
\bibitem [{\citenamefont {Kaubruegger}\ \emph {et~al.}(2023)\citenamefont
  {Kaubruegger}, \citenamefont {Shankar}, \citenamefont {Vasilyev},\ and\
  \citenamefont {Zoller}}]{kaubruegger2023optimal}%
  \BibitemOpen
  \bibfield  {author} {\bibinfo {author} {\bibfnamefont {R.}~\bibnamefont
  {Kaubruegger}}, \bibinfo {author} {\bibfnamefont {A.}~\bibnamefont
  {Shankar}}, \bibinfo {author} {\bibfnamefont {D.~V.}\ \bibnamefont
  {Vasilyev}},\ and\ \bibinfo {author} {\bibfnamefont {P.}~\bibnamefont
  {Zoller}},\ }\bibfield  {title} {\bibinfo {title} {Optimal and variational
  multiparameter quantum metrology and vector-field sensing},\ }\href
  {https://doi.org/10.1103/PRXQuantum.4.020333} {\bibfield  {journal} {\bibinfo
   {journal} {PRX Quantum}\ }\textbf {\bibinfo {volume} {4}},\ \bibinfo {pages}
  {020333} (\bibinfo {year} {2023})}\BibitemShut {NoStop}%
\bibitem [{\citenamefont {Marciniak}\ \emph {et~al.}(2022)\citenamefont
  {Marciniak}, \citenamefont {Feldker}, \citenamefont {Pogorelov},
  \citenamefont {Kaubruegger}, \citenamefont {Vasilyev}, \citenamefont {van
  Bijnen}, \citenamefont {Schindler}, \citenamefont {Zoller}, \citenamefont
  {Blatt},\ and\ \citenamefont {Monz}}]{marciniak2022optimal}%
  \BibitemOpen
  \bibfield  {author} {\bibinfo {author} {\bibfnamefont {C.~D.}\ \bibnamefont
  {Marciniak}}, \bibinfo {author} {\bibfnamefont {T.}~\bibnamefont {Feldker}},
  \bibinfo {author} {\bibfnamefont {I.}~\bibnamefont {Pogorelov}}, \bibinfo
  {author} {\bibfnamefont {R.}~\bibnamefont {Kaubruegger}}, \bibinfo {author}
  {\bibfnamefont {D.~V.}\ \bibnamefont {Vasilyev}}, \bibinfo {author}
  {\bibfnamefont {R.}~\bibnamefont {van Bijnen}}, \bibinfo {author}
  {\bibfnamefont {P.}~\bibnamefont {Schindler}}, \bibinfo {author}
  {\bibfnamefont {P.}~\bibnamefont {Zoller}}, \bibinfo {author} {\bibfnamefont
  {R.}~\bibnamefont {Blatt}},\ and\ \bibinfo {author} {\bibfnamefont
  {T.}~\bibnamefont {Monz}},\ }\bibfield  {title} {\bibinfo {title} {Optimal
  metrology with programmable quantum sensors},\ }\href
  {https://doi.org/10.1038/s41586-022-04435-4} {\bibfield  {journal} {\bibinfo
  {journal} {Nature}\ }\textbf {\bibinfo {volume} {603}},\ \bibinfo {pages}
  {604} (\bibinfo {year} {2022})}\BibitemShut {NoStop}%
\bibitem [{\citenamefont {Direkci}\ \emph {et~al.}(2024)\citenamefont
  {Direkci}, \citenamefont {Finkelstein}, \citenamefont {Endres},\ and\
  \citenamefont {Gefen}}]{direkci2024heisenberglimited}%
  \BibitemOpen
  \bibfield  {author} {\bibinfo {author} {\bibfnamefont {S.}~\bibnamefont
  {Direkci}}, \bibinfo {author} {\bibfnamefont {R.}~\bibnamefont
  {Finkelstein}}, \bibinfo {author} {\bibfnamefont {M.}~\bibnamefont
  {Endres}},\ and\ \bibinfo {author} {\bibfnamefont {T.}~\bibnamefont
  {Gefen}},\ }\bibfield  {title} {\bibinfo {title} {Heisenberg-limited bayesian
  phase estimation with low-depth digital quantum circuits},\ }\href
  {https://arxiv.org/abs/2407.06006} {\bibfield  {journal} {\bibinfo  {journal}
  {arXiv preprint arXiv:2407.06006}\ } (\bibinfo {year} {2024})}\BibitemShut
  {NoStop}%
\bibitem [{\citenamefont {Castro}\ \emph {et~al.}(2024)\citenamefont {Castro},
  \citenamefont {Larson}, \citenamefont {Narayanan}, \citenamefont {Colussi},
  \citenamefont {Perlin},\ and\ \citenamefont
  {Lewis-Swan}}]{castro2024variational}%
  \BibitemOpen
  \bibfield  {author} {\bibinfo {author} {\bibfnamefont {J.~C.~Z.}\
  \bibnamefont {Castro}}, \bibinfo {author} {\bibfnamefont {J.}~\bibnamefont
  {Larson}}, \bibinfo {author} {\bibfnamefont {S.~H.~K.}\ \bibnamefont
  {Narayanan}}, \bibinfo {author} {\bibfnamefont {V.~E.}\ \bibnamefont
  {Colussi}}, \bibinfo {author} {\bibfnamefont {M.~A.}\ \bibnamefont
  {Perlin}},\ and\ \bibinfo {author} {\bibfnamefont {R.~J.}\ \bibnamefont
  {Lewis-Swan}},\ }\href {https://arxiv.org/abs/2406.01859} {\bibinfo {title}
  {Variational quantum state preparation for quantum-enhanced metrology in
  noisy systems}} (\bibinfo {year} {2024})\BibitemShut {NoStop}%
\bibitem [{\citenamefont {MacLellan}\ \emph {et~al.}(2024)\citenamefont
  {MacLellan}, \citenamefont {Roztocki}, \citenamefont {Czischek},\ and\
  \citenamefont {Melko}}]{maclellan2024end}%
  \BibitemOpen
  \bibfield  {author} {\bibinfo {author} {\bibfnamefont {B.}~\bibnamefont
  {MacLellan}}, \bibinfo {author} {\bibfnamefont {P.}~\bibnamefont {Roztocki}},
  \bibinfo {author} {\bibfnamefont {S.}~\bibnamefont {Czischek}},\ and\
  \bibinfo {author} {\bibfnamefont {R.~G.}\ \bibnamefont {Melko}},\ }\bibfield
  {title} {\bibinfo {title} {End-to-end variational quantum sensing},\ }\href
  {https://arxiv.org/abs/2403.02394} {\bibfield  {journal} {\bibinfo  {journal}
  {arXiv preprint arXiv:2403.02394}\ } (\bibinfo {year} {2024})}\BibitemShut
  {NoStop}%
\bibitem [{\citenamefont {{Huerta Alderete}}\ \emph {et~al.}(2022)\citenamefont
  {{Huerta Alderete}}, \citenamefont {{Gordon}}, \citenamefont {{Sauvage}},
  \citenamefont {{Sone}}, \citenamefont {{Sornborger}}, \citenamefont
  {{Coles}},\ and\ \citenamefont {{Cerezo}}}]{huerta2022inference}%
  \BibitemOpen
  \bibfield  {author} {\bibinfo {author} {\bibfnamefont {C.}~\bibnamefont
  {{Huerta Alderete}}}, \bibinfo {author} {\bibfnamefont {M.~H.}\ \bibnamefont
  {{Gordon}}}, \bibinfo {author} {\bibfnamefont {F.}~\bibnamefont {{Sauvage}}},
  \bibinfo {author} {\bibfnamefont {A.}~\bibnamefont {{Sone}}}, \bibinfo
  {author} {\bibfnamefont {A.~T.}\ \bibnamefont {{Sornborger}}}, \bibinfo
  {author} {\bibfnamefont {P.~J.}\ \bibnamefont {{Coles}}},\ and\ \bibinfo
  {author} {\bibfnamefont {M.}~\bibnamefont {{Cerezo}}},\ }\bibfield  {title}
  {\bibinfo {title} {Inference-based quantum sensing},\ }\href
  {https://doi.org/10.1103/PhysRevLett.129.190501} {\bibfield  {journal}
  {\bibinfo  {journal} {Phys. Rev. Lett.}\ }\textbf {\bibinfo {volume} {129}},\
  \bibinfo {pages} {190501} (\bibinfo {year} {2022})}\BibitemShut {NoStop}%
\bibitem [{\citenamefont {Yamamoto}\ \emph {et~al.}(2022)\citenamefont
  {Yamamoto}, \citenamefont {Endo}, \citenamefont {Hakoshima}, \citenamefont
  {Matsuzaki},\ and\ \citenamefont {Tokunaga}}]{yamamoto2022error}%
  \BibitemOpen
  \bibfield  {author} {\bibinfo {author} {\bibfnamefont {K.}~\bibnamefont
  {Yamamoto}}, \bibinfo {author} {\bibfnamefont {S.}~\bibnamefont {Endo}},
  \bibinfo {author} {\bibfnamefont {H.}~\bibnamefont {Hakoshima}}, \bibinfo
  {author} {\bibfnamefont {Y.}~\bibnamefont {Matsuzaki}},\ and\ \bibinfo
  {author} {\bibfnamefont {Y.}~\bibnamefont {Tokunaga}},\ }\bibfield  {title}
  {\bibinfo {title} {Error-mitigated quantum metrology via virtual
  purification},\ }\href {https://doi.org/10.1103/PhysRevLett.129.250503}
  {\bibfield  {journal} {\bibinfo  {journal} {Physical Review Letters}\
  }\textbf {\bibinfo {volume} {129}},\ \bibinfo {pages} {250503} (\bibinfo
  {year} {2022})}\BibitemShut {NoStop}%
\bibitem [{\citenamefont {Hama}\ and\ \citenamefont
  {Nishi}(2023)}]{hama2023quantum}%
  \BibitemOpen
  \bibfield  {author} {\bibinfo {author} {\bibfnamefont {Y.}~\bibnamefont
  {Hama}}\ and\ \bibinfo {author} {\bibfnamefont {H.}~\bibnamefont {Nishi}},\
  }\bibfield  {title} {\bibinfo {title} {Quantum-error-mitigation circuit
  groups for noisy quantum metrology},\ }\href
  {https://arxiv.org/abs/2303.01820} {\bibfield  {journal} {\bibinfo  {journal}
  {arXiv preprint arXiv:2303.01820}\ } (\bibinfo {year} {2023})}\BibitemShut
  {NoStop}%
\bibitem [{\citenamefont {Chen}\ and\ \citenamefont
  {Jing}(2024)}]{chen2024qubitassisted}%
  \BibitemOpen
  \bibfield  {author} {\bibinfo {author} {\bibfnamefont {P.}~\bibnamefont
  {Chen}}\ and\ \bibinfo {author} {\bibfnamefont {J.}~\bibnamefont {Jing}},\
  }\href@noop {} {\bibinfo {title} {Qubit-assisted quantum metrology}}
  (\bibinfo {year} {2024}),\ \Eprint {https://arxiv.org/abs/2404.12649}
  {arXiv:2404.12649 [quant-ph]} \BibitemShut {NoStop}%
\bibitem [{\citenamefont {Dyke}\ \emph {et~al.}(2024)\citenamefont {Dyke},
  \citenamefont {White},\ and\ \citenamefont {Quiroz}}]{dyke2024mitigating}%
  \BibitemOpen
  \bibfield  {author} {\bibinfo {author} {\bibfnamefont {J.~S.~V.}\
  \bibnamefont {Dyke}}, \bibinfo {author} {\bibfnamefont {Z.}~\bibnamefont
  {White}},\ and\ \bibinfo {author} {\bibfnamefont {G.}~\bibnamefont
  {Quiroz}},\ }\bibfield  {title} {\bibinfo {title} {Mitigating errors in dc
  magnetometry via zero-noise extrapolation},\ }\href
  {https://arxiv.org/abs/2402.16949} {\bibfield  {journal} {\bibinfo  {journal}
  {arXiv preprint arXiv:2402.16949}\ } (\bibinfo {year} {2024})}\BibitemShut
  {NoStop}%
\bibitem [{\citenamefont {Koczor}(2021)}]{koczor2020exponential}%
  \BibitemOpen
  \bibfield  {author} {\bibinfo {author} {\bibfnamefont {B.}~\bibnamefont
  {Koczor}},\ }\bibfield  {title} {\bibinfo {title} {Exponential error
  suppression for near-term quantum devices},\ }\href
  {https://doi.org/https://doi.org/10.1103/PhysRevX.11.031057} {\bibfield
  {journal} {\bibinfo  {journal} {Physical Review X}\ }\textbf {\bibinfo
  {volume} {11}},\ \bibinfo {pages} {031057} (\bibinfo {year}
  {2021})}\BibitemShut {NoStop}%
\bibitem [{\citenamefont {Huggins}\ \emph {et~al.}(2021)\citenamefont
  {Huggins}, \citenamefont {McArdle}, \citenamefont {O’Brien}, \citenamefont
  {Lee}, \citenamefont {Rubin}, \citenamefont {Boixo}, \citenamefont {Whaley},
  \citenamefont {Babbush},\ and\ \citenamefont {McClean}}]{huggins2020virtual}%
  \BibitemOpen
  \bibfield  {author} {\bibinfo {author} {\bibfnamefont {W.~J.}\ \bibnamefont
  {Huggins}}, \bibinfo {author} {\bibfnamefont {S.}~\bibnamefont {McArdle}},
  \bibinfo {author} {\bibfnamefont {T.~E.}\ \bibnamefont {O’Brien}}, \bibinfo
  {author} {\bibfnamefont {J.}~\bibnamefont {Lee}}, \bibinfo {author}
  {\bibfnamefont {N.~C.}\ \bibnamefont {Rubin}}, \bibinfo {author}
  {\bibfnamefont {S.}~\bibnamefont {Boixo}}, \bibinfo {author} {\bibfnamefont
  {K.~B.}\ \bibnamefont {Whaley}}, \bibinfo {author} {\bibfnamefont
  {R.}~\bibnamefont {Babbush}},\ and\ \bibinfo {author} {\bibfnamefont {J.~R.}\
  \bibnamefont {McClean}},\ }\bibfield  {title} {\bibinfo {title} {Virtual
  distillation for quantum error mitigation},\ }\href
  {https://doi.org/10.1103/PhysRevX.11.041036} {\bibfield  {journal} {\bibinfo
  {journal} {Physical Review X}\ }\textbf {\bibinfo {volume} {11}},\ \bibinfo
  {pages} {041036} (\bibinfo {year} {2021})}\BibitemShut {NoStop}%
\bibitem [{\citenamefont {Hayashi}(2004)}]{hayashi2004quantum}%
  \BibitemOpen
  \bibfield  {author} {\bibinfo {author} {\bibfnamefont {M.}~\bibnamefont
  {Hayashi}},\ }\href {https://doi.org/10.1007/978-3-662-49725-8} {\emph
  {\bibinfo {title} {Quantum Information Theory: Mathematical Foundation (2nd
  edition)}}}\ (\bibinfo  {publisher} {Springer},\ \bibinfo {year}
  {2004})\BibitemShut {NoStop}%
\bibitem [{\citenamefont {Liu}\ \emph {et~al.}(2020)\citenamefont {Liu},
  \citenamefont {Yuan}, \citenamefont {Lu},\ and\ \citenamefont
  {Wang}}]{liu2020quantum}%
  \BibitemOpen
  \bibfield  {author} {\bibinfo {author} {\bibfnamefont {J.}~\bibnamefont
  {Liu}}, \bibinfo {author} {\bibfnamefont {H.}~\bibnamefont {Yuan}}, \bibinfo
  {author} {\bibfnamefont {X.-M.}\ \bibnamefont {Lu}},\ and\ \bibinfo {author}
  {\bibfnamefont {X.}~\bibnamefont {Wang}},\ }\bibfield  {title} {\bibinfo
  {title} {Quantum fisher information matrix and multiparameter estimation},\
  }\href {https://doi.org/10.1088/1751-8121/ab5d4} {\bibfield  {journal}
  {\bibinfo  {journal} {Journal of Physics A: Mathematical and Theoretical}\
  }\textbf {\bibinfo {volume} {53}},\ \bibinfo {pages} {023001} (\bibinfo
  {year} {2020})}\BibitemShut {NoStop}%
\bibitem [{\citenamefont {Cai}\ \emph {et~al.}(2023)\citenamefont {Cai},
  \citenamefont {Babbush}, \citenamefont {Benjamin}, \citenamefont {Endo},
  \citenamefont {Huggins}, \citenamefont {Li}, \citenamefont {McClean},\ and\
  \citenamefont {O’Brien}}]{cai2022quantum}%
  \BibitemOpen
  \bibfield  {author} {\bibinfo {author} {\bibfnamefont {Z.}~\bibnamefont
  {Cai}}, \bibinfo {author} {\bibfnamefont {R.}~\bibnamefont {Babbush}},
  \bibinfo {author} {\bibfnamefont {S.~C.}\ \bibnamefont {Benjamin}}, \bibinfo
  {author} {\bibfnamefont {S.}~\bibnamefont {Endo}}, \bibinfo {author}
  {\bibfnamefont {W.~J.}\ \bibnamefont {Huggins}}, \bibinfo {author}
  {\bibfnamefont {Y.}~\bibnamefont {Li}}, \bibinfo {author} {\bibfnamefont
  {J.~R.}\ \bibnamefont {McClean}},\ and\ \bibinfo {author} {\bibfnamefont
  {T.~E.}\ \bibnamefont {O’Brien}},\ }\bibfield  {title} {\bibinfo {title}
  {Quantum error mitigation},\ }\href
  {https://doi.org/10.1103/RevModPhys.95.045005} {\bibfield  {journal}
  {\bibinfo  {journal} {Reviews of Modern Physics}\ }\textbf {\bibinfo {volume}
  {95}},\ \bibinfo {pages} {045005} (\bibinfo {year} {2023})}\BibitemShut
  {NoStop}%
\bibitem [{\citenamefont {Li}\ and\ \citenamefont
  {Benjamin}(2017)}]{li2017efficient}%
  \BibitemOpen
  \bibfield  {author} {\bibinfo {author} {\bibfnamefont {Y.}~\bibnamefont
  {Li}}\ and\ \bibinfo {author} {\bibfnamefont {S.~C.}\ \bibnamefont
  {Benjamin}},\ }\bibfield  {title} {\bibinfo {title} {Efficient variational
  quantum simulator incorporating active error minimization},\ }\href
  {https://doi.org/10.1103/PhysRevX.7.021050} {\bibfield  {journal} {\bibinfo
  {journal} {Phys. Rev. X}\ }\textbf {\bibinfo {volume} {7}},\ \bibinfo {pages}
  {021050} (\bibinfo {year} {2017})}\BibitemShut {NoStop}%
\bibitem [{\citenamefont {Temme}\ \emph {et~al.}(2017)\citenamefont {Temme},
  \citenamefont {Bravyi},\ and\ \citenamefont {Gambetta}}]{temme2017error}%
  \BibitemOpen
  \bibfield  {author} {\bibinfo {author} {\bibfnamefont {K.}~\bibnamefont
  {Temme}}, \bibinfo {author} {\bibfnamefont {S.}~\bibnamefont {Bravyi}},\ and\
  \bibinfo {author} {\bibfnamefont {J.~M.}\ \bibnamefont {Gambetta}},\
  }\bibfield  {title} {\bibinfo {title} {Error mitigation for short-depth
  quantum circuits},\ }\href {https://doi.org/10.1103/PhysRevLett.119.180509}
  {\bibfield  {journal} {\bibinfo  {journal} {Physical review letters}\
  }\textbf {\bibinfo {volume} {119}},\ \bibinfo {pages} {180509} (\bibinfo
  {year} {2017})}\BibitemShut {NoStop}%
\bibitem [{\citenamefont {Giurgica-Tiron}\ \emph {et~al.}(2020)\citenamefont
  {Giurgica-Tiron}, \citenamefont {Hindy}, \citenamefont {LaRose},
  \citenamefont {Mari},\ and\ \citenamefont {Zeng}}]{giurgica2020digital}%
  \BibitemOpen
  \bibfield  {author} {\bibinfo {author} {\bibfnamefont {T.}~\bibnamefont
  {Giurgica-Tiron}}, \bibinfo {author} {\bibfnamefont {Y.}~\bibnamefont
  {Hindy}}, \bibinfo {author} {\bibfnamefont {R.}~\bibnamefont {LaRose}},
  \bibinfo {author} {\bibfnamefont {A.}~\bibnamefont {Mari}},\ and\ \bibinfo
  {author} {\bibfnamefont {W.~J.}\ \bibnamefont {Zeng}},\ }\bibfield  {title}
  {\bibinfo {title} {Digital zero noise extrapolation for quantum error
  mitigation},\ }\href {https://doi.org/10.1109/QCE49297.2020.00045} {\bibfield
   {journal} {\bibinfo  {journal} {2020 IEEE International Conference on
  Quantum Computing and Engineering (QCE)}\ ,\ \bibinfo {pages} {306}}
  (\bibinfo {year} {2020})}\BibitemShut {NoStop}%
\bibitem [{\citenamefont {Endo}\ \emph {et~al.}(2018)\citenamefont {Endo},
  \citenamefont {Benjamin},\ and\ \citenamefont {Li}}]{endo2018practical}%
  \BibitemOpen
  \bibfield  {author} {\bibinfo {author} {\bibfnamefont {S.}~\bibnamefont
  {Endo}}, \bibinfo {author} {\bibfnamefont {S.~C.}\ \bibnamefont {Benjamin}},\
  and\ \bibinfo {author} {\bibfnamefont {Y.}~\bibnamefont {Li}},\ }\bibfield
  {title} {\bibinfo {title} {Practical quantum error mitigation for near-future
  applications},\ }\href
  {https://journals.aps.org/prx/abstract/10.1103/PhysRevX.8.031027} {\bibfield
  {journal} {\bibinfo  {journal} {Physical Review X}\ }\textbf {\bibinfo
  {volume} {8}},\ \bibinfo {pages} {031027} (\bibinfo {year}
  {2018})}\BibitemShut {NoStop}%
\bibitem [{\citenamefont {Czarnik}\ \emph {et~al.}(2021)\citenamefont
  {Czarnik}, \citenamefont {Arrasmith}, \citenamefont {Coles},\ and\
  \citenamefont {Cincio}}]{czarnik2020error}%
  \BibitemOpen
  \bibfield  {author} {\bibinfo {author} {\bibfnamefont {P.}~\bibnamefont
  {Czarnik}}, \bibinfo {author} {\bibfnamefont {A.}~\bibnamefont {Arrasmith}},
  \bibinfo {author} {\bibfnamefont {P.~J.}\ \bibnamefont {Coles}},\ and\
  \bibinfo {author} {\bibfnamefont {L.}~\bibnamefont {Cincio}},\ }\bibfield
  {title} {\bibinfo {title} {Error mitigation with {C}lifford quantum-circuit
  data},\ }\href {https://doi.org/10.22331/q-2021-11-26-592} {\bibfield
  {journal} {\bibinfo  {journal} {{Quantum}}\ }\textbf {\bibinfo {volume}
  {5}},\ \bibinfo {pages} {592} (\bibinfo {year} {2021})}\BibitemShut {NoStop}%
\bibitem [{\citenamefont {Seif}\ \emph {et~al.}(2023)\citenamefont {Seif},
  \citenamefont {Cian}, \citenamefont {Zhou}, \citenamefont {Chen},\ and\
  \citenamefont {Jiang}}]{seif2022shadow}%
  \BibitemOpen
  \bibfield  {author} {\bibinfo {author} {\bibfnamefont {A.}~\bibnamefont
  {Seif}}, \bibinfo {author} {\bibfnamefont {Z.-P.}\ \bibnamefont {Cian}},
  \bibinfo {author} {\bibfnamefont {S.}~\bibnamefont {Zhou}}, \bibinfo {author}
  {\bibfnamefont {S.}~\bibnamefont {Chen}},\ and\ \bibinfo {author}
  {\bibfnamefont {L.}~\bibnamefont {Jiang}},\ }\bibfield  {title} {\bibinfo
  {title} {Shadow distillation: Quantum error mitigation with classical shadows
  for near-term quantum processors},\ }\href
  {https://doi.org/10.1103/PRXQuantum.4.010303} {\bibfield  {journal} {\bibinfo
   {journal} {PRX Quantum}\ }\textbf {\bibinfo {volume} {4}},\ \bibinfo {pages}
  {010303} (\bibinfo {year} {2023})}\BibitemShut {NoStop}%
\bibitem [{\citenamefont {Otten}\ and\ \citenamefont
  {Gray}(2019)}]{otten2019recovering}%
  \BibitemOpen
  \bibfield  {author} {\bibinfo {author} {\bibfnamefont {M.}~\bibnamefont
  {Otten}}\ and\ \bibinfo {author} {\bibfnamefont {S.~K.}\ \bibnamefont
  {Gray}},\ }\bibfield  {title} {\bibinfo {title} {Recovering noise-free
  quantum observables},\ }\href {https://doi.org/10.1103/PhysRevA.99.012338}
  {\bibfield  {journal} {\bibinfo  {journal} {Physical Review A}\ }\textbf
  {\bibinfo {volume} {99}},\ \bibinfo {pages} {012338} (\bibinfo {year}
  {2019})}\BibitemShut {NoStop}%
\bibitem [{\citenamefont {Montanaro}\ and\ \citenamefont
  {Stanisic}(2021)}]{montanaro2021error}%
  \BibitemOpen
  \bibfield  {author} {\bibinfo {author} {\bibfnamefont {A.}~\bibnamefont
  {Montanaro}}\ and\ \bibinfo {author} {\bibfnamefont {S.}~\bibnamefont
  {Stanisic}},\ }\bibfield  {title} {\bibinfo {title} {Error mitigation by
  training with fermionic linear optics},\ }\href
  {https://arxiv.org/abs/2102.02120} {\bibfield  {journal} {\bibinfo  {journal}
  {arXiv preprint arXiv:2102.02120}\ } (\bibinfo {year} {2021})}\BibitemShut
  {NoStop}%
\bibitem [{\citenamefont {Acampora}\ \emph {et~al.}(2021)\citenamefont
  {Acampora}, \citenamefont {Grossi},\ and\ \citenamefont
  {Vitiello}}]{acampora2021genetic}%
  \BibitemOpen
  \bibfield  {author} {\bibinfo {author} {\bibfnamefont {G.}~\bibnamefont
  {Acampora}}, \bibinfo {author} {\bibfnamefont {M.}~\bibnamefont {Grossi}},\
  and\ \bibinfo {author} {\bibfnamefont {A.}~\bibnamefont {Vitiello}},\
  }\bibfield  {title} {\bibinfo {title} {Genetic algorithms for error
  mitigation in quantum measurement},\ }in\ \href
  {https://doi.org/10.1109/CEC45853.2021.9504796} {\emph {\bibinfo {booktitle}
  {2021 IEEE congress on evolutionary computation (CEC)}}}\ (\bibinfo
  {organization} {IEEE},\ \bibinfo {year} {2021})\ pp.\ \bibinfo {pages}
  {1826--1832}\BibitemShut {NoStop}%
\bibitem [{\citenamefont {Lolur}\ \emph {et~al.}(2023)\citenamefont {Lolur},
  \citenamefont {Skogh}, \citenamefont {Dobrautz}, \citenamefont {Warren},
  \citenamefont {Bizn{\'a}rov{\'a}}, \citenamefont {Osman}, \citenamefont
  {Tancredi}, \citenamefont {Wendin}, \citenamefont {Bylander},\ and\
  \citenamefont {Rahm}}]{lolur2023reference}%
  \BibitemOpen
  \bibfield  {author} {\bibinfo {author} {\bibfnamefont {P.}~\bibnamefont
  {Lolur}}, \bibinfo {author} {\bibfnamefont {M.}~\bibnamefont {Skogh}},
  \bibinfo {author} {\bibfnamefont {W.}~\bibnamefont {Dobrautz}}, \bibinfo
  {author} {\bibfnamefont {C.}~\bibnamefont {Warren}}, \bibinfo {author}
  {\bibfnamefont {J.}~\bibnamefont {Bizn{\'a}rov{\'a}}}, \bibinfo {author}
  {\bibfnamefont {A.}~\bibnamefont {Osman}}, \bibinfo {author} {\bibfnamefont
  {G.}~\bibnamefont {Tancredi}}, \bibinfo {author} {\bibfnamefont
  {G.}~\bibnamefont {Wendin}}, \bibinfo {author} {\bibfnamefont
  {J.}~\bibnamefont {Bylander}},\ and\ \bibinfo {author} {\bibfnamefont
  {M.}~\bibnamefont {Rahm}},\ }\bibfield  {title} {\bibinfo {title}
  {Reference-state error mitigation: A strategy for high accuracy quantum
  computation of chemistry},\ }\href {https://doi.org/10.1021/acs.jctc.2c00807}
  {\bibfield  {journal} {\bibinfo  {journal} {Journal of Chemical Theory and
  Computation}\ }\textbf {\bibinfo {volume} {19}},\ \bibinfo {pages} {783}
  (\bibinfo {year} {2023})}\BibitemShut {NoStop}%
\bibitem [{\citenamefont {Tsubouchi}\ \emph {et~al.}(2024)\citenamefont
  {Tsubouchi}, \citenamefont {Mitsuhashi}, \citenamefont {Sharma},\ and\
  \citenamefont {Yoshioka}}]{tsubouchi2024symmetric}%
  \BibitemOpen
  \bibfield  {author} {\bibinfo {author} {\bibfnamefont {K.}~\bibnamefont
  {Tsubouchi}}, \bibinfo {author} {\bibfnamefont {Y.}~\bibnamefont
  {Mitsuhashi}}, \bibinfo {author} {\bibfnamefont {K.}~\bibnamefont {Sharma}},\
  and\ \bibinfo {author} {\bibfnamefont {N.}~\bibnamefont {Yoshioka}},\
  }\href@noop {} {\bibinfo {title} {Symmetric clifford twirling for
  cost-optimal quantum error mitigation in early ftqc regime}} (\bibinfo {year}
  {2024}),\ \Eprint {https://arxiv.org/abs/2405.07720} {arXiv:2405.07720
  [quant-ph]} \BibitemShut {NoStop}%
\bibitem [{\citenamefont {Saxena}\ and\ \citenamefont
  {Kyaw}(2024)}]{saxena2024error}%
  \BibitemOpen
  \bibfield  {author} {\bibinfo {author} {\bibfnamefont {G.}~\bibnamefont
  {Saxena}}\ and\ \bibinfo {author} {\bibfnamefont {T.~H.}\ \bibnamefont
  {Kyaw}},\ }\href {https://arxiv.org/abs/2409.06636} {\bibinfo {title} {Error
  mitigation by restricted evolution}} (\bibinfo {year} {2024}),\ \Eprint
  {https://arxiv.org/abs/2409.06636} {arXiv:2409.06636 [quant-ph]} \BibitemShut
  {NoStop}%
\bibitem [{\citenamefont {Giovannetti}\ \emph {et~al.}(2006)\citenamefont
  {Giovannetti}, \citenamefont {Lloyd},\ and\ \citenamefont
  {Maccone}}]{giovannetti2006quantum}%
  \BibitemOpen
  \bibfield  {author} {\bibinfo {author} {\bibfnamefont {V.}~\bibnamefont
  {Giovannetti}}, \bibinfo {author} {\bibfnamefont {S.}~\bibnamefont {Lloyd}},\
  and\ \bibinfo {author} {\bibfnamefont {L.}~\bibnamefont {Maccone}},\
  }\bibfield  {title} {\bibinfo {title} {Quantum metrology},\ }\href
  {https://doi.org/10.1103/PhysRevLett.96.010401} {\bibfield  {journal}
  {\bibinfo  {journal} {Physical Review Letters}\ }\textbf {\bibinfo {volume}
  {96}},\ \bibinfo {pages} {010401} (\bibinfo {year} {2006})}\BibitemShut
  {NoStop}%
\bibitem [{\citenamefont {Stilck~Fran{\c{c}}a}\ and\ \citenamefont
  {Garcia-Patron}(2021)}]{franca2020limitations}%
  \BibitemOpen
  \bibfield  {author} {\bibinfo {author} {\bibfnamefont {D.}~\bibnamefont
  {Stilck~Fran{\c{c}}a}}\ and\ \bibinfo {author} {\bibfnamefont
  {R.}~\bibnamefont {Garcia-Patron}},\ }\bibfield  {title} {\bibinfo {title}
  {Limitations of optimization algorithms on noisy quantum devices},\ }\href
  {https://doi.org/10.1038/s41567-021-01356-3} {\bibfield  {journal} {\bibinfo
  {journal} {Nature Physics}\ }\textbf {\bibinfo {volume} {17}},\ \bibinfo
  {pages} {1221} (\bibinfo {year} {2021})}\BibitemShut {NoStop}%
\bibitem [{\citenamefont {M{\"u}ller-Hermes}\ \emph {et~al.}(2016)\citenamefont
  {M{\"u}ller-Hermes}, \citenamefont {Stilck~Fran{\c{c}}a},\ and\ \citenamefont
  {Wolf}}]{muller2016relative}%
  \BibitemOpen
  \bibfield  {author} {\bibinfo {author} {\bibfnamefont {A.}~\bibnamefont
  {M{\"u}ller-Hermes}}, \bibinfo {author} {\bibfnamefont {D.}~\bibnamefont
  {Stilck~Fran{\c{c}}a}},\ and\ \bibinfo {author} {\bibfnamefont {M.~M.}\
  \bibnamefont {Wolf}},\ }\bibfield  {title} {\bibinfo {title} {Relative
  entropy convergence for depolarizing channels},\ }\href
  {https://aip.scitation.org/doi/10.1063/1.4939560} {\bibfield  {journal}
  {\bibinfo  {journal} {Journal of Mathematical Physics}\ }\textbf {\bibinfo
  {volume} {57}},\ \bibinfo {pages} {022202} (\bibinfo {year}
  {2016})}\BibitemShut {NoStop}%
\bibitem [{\citenamefont {Krebsbach}\ \emph {et~al.}(2022)\citenamefont
  {Krebsbach}, \citenamefont {Trauzettel},\ and\ \citenamefont
  {Calzona}}]{krebsbach2022optimization}%
  \BibitemOpen
  \bibfield  {author} {\bibinfo {author} {\bibfnamefont {M.}~\bibnamefont
  {Krebsbach}}, \bibinfo {author} {\bibfnamefont {B.}~\bibnamefont
  {Trauzettel}},\ and\ \bibinfo {author} {\bibfnamefont {A.}~\bibnamefont
  {Calzona}},\ }\bibfield  {title} {\bibinfo {title} {Optimization of
  richardson extrapolation for quantum error mitigation},\ }\href
  {https://doi.org/10.1103/PhysRevA.106.062436} {\bibfield  {journal} {\bibinfo
   {journal} {Physical Review A}\ }\textbf {\bibinfo {volume} {106}},\ \bibinfo
  {pages} {062436} (\bibinfo {year} {2022})}\BibitemShut {NoStop}%
\bibitem [{\citenamefont {Hoel}\ and\ \citenamefont
  {Levine}(1964)}]{Hoel1994optimal}%
  \BibitemOpen
  \bibfield  {author} {\bibinfo {author} {\bibfnamefont {P.~G.}\ \bibnamefont
  {Hoel}}\ and\ \bibinfo {author} {\bibfnamefont {A.}~\bibnamefont {Levine}},\
  }\bibfield  {title} {\bibinfo {title} {Optimal spacing and weighting in
  polynomial prediction},\ }\href {http://www.jstor.org/stable/2238291}
  {\bibfield  {journal} {\bibinfo  {journal} {The Annals of Mathematical
  Statistics}\ }\textbf {\bibinfo {volume} {35}},\ \bibinfo {pages} {1553}
  (\bibinfo {year} {1964})}\BibitemShut {NoStop}%
\bibitem [{\citenamefont {Saywell}\ \emph {et~al.}(2023)\citenamefont
  {Saywell}, \citenamefont {Carey}, \citenamefont {Light}, \citenamefont
  {Szigeti}, \citenamefont {Milne}, \citenamefont {Gill}, \citenamefont {Goh},
  \citenamefont {Perunicic}, \citenamefont {Wilson}, \citenamefont {Macrae}
  \emph {et~al.}}]{saywell2023enhancing}%
  \BibitemOpen
  \bibfield  {author} {\bibinfo {author} {\bibfnamefont {J.~C.}\ \bibnamefont
  {Saywell}}, \bibinfo {author} {\bibfnamefont {M.~S.}\ \bibnamefont {Carey}},
  \bibinfo {author} {\bibfnamefont {P.~S.}\ \bibnamefont {Light}}, \bibinfo
  {author} {\bibfnamefont {S.~S.}\ \bibnamefont {Szigeti}}, \bibinfo {author}
  {\bibfnamefont {A.~R.}\ \bibnamefont {Milne}}, \bibinfo {author}
  {\bibfnamefont {K.~S.}\ \bibnamefont {Gill}}, \bibinfo {author}
  {\bibfnamefont {M.~L.}\ \bibnamefont {Goh}}, \bibinfo {author} {\bibfnamefont
  {V.~S.}\ \bibnamefont {Perunicic}}, \bibinfo {author} {\bibfnamefont {N.~M.}\
  \bibnamefont {Wilson}}, \bibinfo {author} {\bibfnamefont {C.~D.}\
  \bibnamefont {Macrae}}, \emph {et~al.},\ }\bibfield  {title} {\bibinfo
  {title} {Enhancing the sensitivity of atom-interferometric inertial sensors
  using robust control},\ }\href {https://doi.org/10.1038/s41467-023-43374-0}
  {\bibfield  {journal} {\bibinfo  {journal} {Nature Communications}\ }\textbf
  {\bibinfo {volume} {14}},\ \bibinfo {pages} {7626} (\bibinfo {year}
  {2023})}\BibitemShut {NoStop}%
\bibitem [{\citenamefont {Borràs}\ \emph {et~al.}(2024)\citenamefont
  {Borràs}, \citenamefont {Carpenter}, \citenamefont {Žaper}, \citenamefont
  {Rao}, \citenamefont {Couet}, \citenamefont {Munsch}, \citenamefont
  {Maletinsky},\ and\ \citenamefont {Rickhaus}}]{borras2024aquantumsensing}%
  \BibitemOpen
  \bibfield  {author} {\bibinfo {author} {\bibfnamefont {V.~J.}\ \bibnamefont
  {Borràs}}, \bibinfo {author} {\bibfnamefont {R.}~\bibnamefont {Carpenter}},
  \bibinfo {author} {\bibfnamefont {L.}~\bibnamefont {Žaper}}, \bibinfo
  {author} {\bibfnamefont {S.}~\bibnamefont {Rao}}, \bibinfo {author}
  {\bibfnamefont {S.}~\bibnamefont {Couet}}, \bibinfo {author} {\bibfnamefont
  {M.}~\bibnamefont {Munsch}}, \bibinfo {author} {\bibfnamefont
  {P.}~\bibnamefont {Maletinsky}},\ and\ \bibinfo {author} {\bibfnamefont
  {P.}~\bibnamefont {Rickhaus}},\ }\bibfield  {title} {\bibinfo {title} {A
  quantum sensing metrology for magnetic memories},\ }\bibfield  {journal}
  {\bibinfo  {journal} {npj Spintronics}\ }\textbf {\bibinfo {volume} {2}},\
  \href {https://doi.org/10.1038/s44306-024-00016-5}
  {10.1038/s44306-024-00016-5} (\bibinfo {year} {2024})\BibitemShut {NoStop}%
\bibitem [{\citenamefont {Koczor}\ \emph {et~al.}(2024)\citenamefont {Koczor},
  \citenamefont {Morton},\ and\ \citenamefont
  {Benjamin}}]{koczor2024probabilistic}%
  \BibitemOpen
  \bibfield  {author} {\bibinfo {author} {\bibfnamefont {B.}~\bibnamefont
  {Koczor}}, \bibinfo {author} {\bibfnamefont {J.~J.}\ \bibnamefont {Morton}},\
  and\ \bibinfo {author} {\bibfnamefont {S.~C.}\ \bibnamefont {Benjamin}},\
  }\bibfield  {title} {\bibinfo {title} {Probabilistic interpolation of quantum
  rotation angles},\ }\href {https://arxiv.org/abs/2305.19881} {\bibfield
  {journal} {\bibinfo  {journal} {Physical Review Letters}\ }\textbf {\bibinfo
  {volume} {132}},\ \bibinfo {pages} {130602} (\bibinfo {year}
  {2024})}\BibitemShut {NoStop}%
\bibitem [{\citenamefont {Rubio}\ and\ \citenamefont
  {Dunningham}(2019)}]{rubio2019quantummetrology}%
  \BibitemOpen
  \bibfield  {author} {\bibinfo {author} {\bibfnamefont {J.}~\bibnamefont
  {Rubio}}\ and\ \bibinfo {author} {\bibfnamefont {J.}~\bibnamefont
  {Dunningham}},\ }\bibfield  {title} {\bibinfo {title} {Quantum metrology in
  the presence of limited data},\ }\href
  {https://doi.org/10.1088/1367-2630/ab098b} {\bibfield  {journal} {\bibinfo
  {journal} {New Journal of Physics}\ }\textbf {\bibinfo {volume} {21}},\
  \bibinfo {pages} {043037} (\bibinfo {year} {2019})}\BibitemShut {NoStop}%
\bibitem [{\citenamefont {Champ}\ and\ \citenamefont
  {Sills}(2022)}]{champ2022generalized}%
  \BibitemOpen
  \bibfield  {author} {\bibinfo {author} {\bibfnamefont {C.~W.}\ \bibnamefont
  {Champ}}\ and\ \bibinfo {author} {\bibfnamefont {A.~V.}\ \bibnamefont
  {Sills}},\ }\href@noop {} {\bibinfo {title} {The generalized law of total
  covariance}} (\bibinfo {year} {2022}),\ \Eprint
  {https://arxiv.org/abs/2205.14525} {arXiv:2205.14525 [math.PR]} \BibitemShut
  {NoStop}%
\bibitem [{\citenamefont {Goh}\ and\ \citenamefont
  {Koczor}(2024)}]{goh2024direct}%
  \BibitemOpen
  \bibfield  {author} {\bibinfo {author} {\bibfnamefont {M.~L.}\ \bibnamefont
  {Goh}}\ and\ \bibinfo {author} {\bibfnamefont {B.}~\bibnamefont {Koczor}},\
  }\bibfield  {title} {\bibinfo {title} {Direct estimation of the density of
  states for fermionic systems},\ }\href {https://arxiv.org/abs/2407.03414}
  {\bibfield  {journal} {\bibinfo  {journal} {arXiv preprint arXiv:2407.03414}\
  } (\bibinfo {year} {2024})}\BibitemShut {NoStop}%
\bibitem [{\citenamefont {Kim}\ \emph {et~al.}(2023)\citenamefont {Kim},
  \citenamefont {Eddins}, \citenamefont {Anand}, \citenamefont {Wei},
  \citenamefont {Van Den~Berg}, \citenamefont {Rosenblatt}, \citenamefont
  {Nayfeh}, \citenamefont {Wu}, \citenamefont {Zaletel}, \citenamefont {Temme}
  \emph {et~al.}}]{kim2023evidence}%
  \BibitemOpen
  \bibfield  {author} {\bibinfo {author} {\bibfnamefont {Y.}~\bibnamefont
  {Kim}}, \bibinfo {author} {\bibfnamefont {A.}~\bibnamefont {Eddins}},
  \bibinfo {author} {\bibfnamefont {S.}~\bibnamefont {Anand}}, \bibinfo
  {author} {\bibfnamefont {K.~X.}\ \bibnamefont {Wei}}, \bibinfo {author}
  {\bibfnamefont {E.}~\bibnamefont {Van Den~Berg}}, \bibinfo {author}
  {\bibfnamefont {S.}~\bibnamefont {Rosenblatt}}, \bibinfo {author}
  {\bibfnamefont {H.}~\bibnamefont {Nayfeh}}, \bibinfo {author} {\bibfnamefont
  {Y.}~\bibnamefont {Wu}}, \bibinfo {author} {\bibfnamefont {M.}~\bibnamefont
  {Zaletel}}, \bibinfo {author} {\bibfnamefont {K.}~\bibnamefont {Temme}},
  \emph {et~al.},\ }\bibfield  {title} {\bibinfo {title} {Evidence for the
  utility of quantum computing before fault tolerance},\ }\href
  {https://doi.org/10.1038/s41586-023-06096-3} {\bibfield  {journal} {\bibinfo
  {journal} {Nature}\ }\textbf {\bibinfo {volume} {618}},\ \bibinfo {pages}
  {500} (\bibinfo {year} {2023})}\BibitemShut {NoStop}%
\bibitem [{\citenamefont {Jones}\ \emph {et~al.}(2019)\citenamefont {Jones},
  \citenamefont {Brown}, \citenamefont {Bush},\ and\ \citenamefont
  {Benjamin}}]{jones2019quest}%
  \BibitemOpen
  \bibfield  {author} {\bibinfo {author} {\bibfnamefont {T.}~\bibnamefont
  {Jones}}, \bibinfo {author} {\bibfnamefont {A.}~\bibnamefont {Brown}},
  \bibinfo {author} {\bibfnamefont {I.}~\bibnamefont {Bush}},\ and\ \bibinfo
  {author} {\bibfnamefont {S.~C.}\ \bibnamefont {Benjamin}},\ }\bibfield
  {title} {\bibinfo {title} {Quest and high performance simulation of quantum
  computers},\ }\href {https://doi.org/10.1038/s41598-019-47174-9} {\bibfield
  {journal} {\bibinfo  {journal} {Scientific reports}\ }\textbf {\bibinfo
  {volume} {9}},\ \bibinfo {pages} {10736} (\bibinfo {year}
  {2019})}\BibitemShut {NoStop}%
\bibitem [{\citenamefont {Jones}\ and\ \citenamefont
  {Benjamin}(2020)}]{jones2020questlink}%
  \BibitemOpen
  \bibfield  {author} {\bibinfo {author} {\bibfnamefont {T.}~\bibnamefont
  {Jones}}\ and\ \bibinfo {author} {\bibfnamefont {S.}~\bibnamefont
  {Benjamin}},\ }\bibfield  {title} {\bibinfo {title} {Questlink—mathematica
  embiggened by a hardware-optimised quantum emulator},\ }\href
  {https://doi.org/10.1088/2058-9565/ab8506} {\bibfield  {journal} {\bibinfo
  {journal} {Quantum Science and Technology}\ }\textbf {\bibinfo {volume}
  {5}},\ \bibinfo {pages} {034012} (\bibinfo {year} {2020})}\BibitemShut
  {NoStop}%
\bibitem [{\citenamefont {Richards}(2015)}]{oxfordARC}%
  \BibitemOpen
  \bibfield  {author} {\bibinfo {author} {\bibfnamefont {A.}~\bibnamefont
  {Richards}},\ }\href {https://doi.org/10.5281/zenodo.22558} {\bibinfo {title}
  {University of oxford advanced research computing}},\ \bibinfo {howpublished}
  {DOI: 10.5281/zenodo.22558} (\bibinfo {year} {2015})\BibitemShut {NoStop}%
\bibitem [{\citenamefont {Atkinson}(1991)}]{atkinson1991introduction}%
  \BibitemOpen
  \bibfield  {author} {\bibinfo {author} {\bibfnamefont {K.}~\bibnamefont
  {Atkinson}},\ }\href {https://books.google.co.uk/books?id=wbCjEAAAQBAJ}
  {\emph {\bibinfo {title} {An Introduction to Numerical Analysis}}}\ (\bibinfo
   {publisher} {Wiley},\ \bibinfo {year} {1991})\BibitemShut {NoStop}%
\bibitem [{\citenamefont {Jackson}(1913)}]{jackson1913accuracy}%
  \BibitemOpen
  \bibfield  {author} {\bibinfo {author} {\bibfnamefont {D.}~\bibnamefont
  {Jackson}},\ }\bibfield  {title} {\bibinfo {title} {On the accuracy of
  trigonometric interpolation},\ }\href {https://doi.org/10.2307/1988698}
  {\bibfield  {journal} {\bibinfo  {journal} {Transactions of the American
  Mathematical Society}\ }\textbf {\bibinfo {volume} {14}},\ \bibinfo {pages}
  {453} (\bibinfo {year} {1913})}\BibitemShut {NoStop}%
\bibitem [{\citenamefont {Mitarai}\ \emph {et~al.}(2018)\citenamefont
  {Mitarai}, \citenamefont {Negoro}, \citenamefont {Kitagawa},\ and\
  \citenamefont {Fujii}}]{mitarai2018quantum}%
  \BibitemOpen
  \bibfield  {author} {\bibinfo {author} {\bibfnamefont {K.}~\bibnamefont
  {Mitarai}}, \bibinfo {author} {\bibfnamefont {M.}~\bibnamefont {Negoro}},
  \bibinfo {author} {\bibfnamefont {M.}~\bibnamefont {Kitagawa}},\ and\
  \bibinfo {author} {\bibfnamefont {K.}~\bibnamefont {Fujii}},\ }\bibfield
  {title} {\bibinfo {title} {Quantum circuit learning},\ }\href
  {https://doi.org/10.1103/PhysRevA.98.032309} {\bibfield  {journal} {\bibinfo
  {journal} {Physical Review A}\ }\textbf {\bibinfo {volume} {98}},\ \bibinfo
  {pages} {032309} (\bibinfo {year} {2018})}\BibitemShut {NoStop}%
\bibitem [{\citenamefont {Schuld}\ \emph {et~al.}(2019)\citenamefont {Schuld},
  \citenamefont {Bergholm}, \citenamefont {Gogolin}, \citenamefont {Izaac},\
  and\ \citenamefont {Killoran}}]{schuld2019evaluating}%
  \BibitemOpen
  \bibfield  {author} {\bibinfo {author} {\bibfnamefont {M.}~\bibnamefont
  {Schuld}}, \bibinfo {author} {\bibfnamefont {V.}~\bibnamefont {Bergholm}},
  \bibinfo {author} {\bibfnamefont {C.}~\bibnamefont {Gogolin}}, \bibinfo
  {author} {\bibfnamefont {J.}~\bibnamefont {Izaac}},\ and\ \bibinfo {author}
  {\bibfnamefont {N.}~\bibnamefont {Killoran}},\ }\bibfield  {title} {\bibinfo
  {title} {Evaluating analytic gradients on quantum hardware},\ }\href
  {https://doi.org/10.1103/PhysRevA.99.032331} {\bibfield  {journal} {\bibinfo
  {journal} {Physical Review A}\ }\textbf {\bibinfo {volume} {99}},\ \bibinfo
  {pages} {032331} (\bibinfo {year} {2019})}\BibitemShut {NoStop}%
\bibitem [{\citenamefont {Banchi}\ and\ \citenamefont
  {Crooks}(2021)}]{Banchi2021measuring}%
  \BibitemOpen
  \bibfield  {author} {\bibinfo {author} {\bibfnamefont {L.}~\bibnamefont
  {Banchi}}\ and\ \bibinfo {author} {\bibfnamefont {G.~E.}\ \bibnamefont
  {Crooks}},\ }\bibfield  {title} {\bibinfo {title} {Measuring {A}nalytic
  {G}radients of {G}eneral {Q}uantum {E}volution with the {S}tochastic
  {P}arameter {S}hift {R}ule},\ }\href
  {https://doi.org/10.22331/q-2021-01-25-386} {\bibfield  {journal} {\bibinfo
  {journal} {{Quantum}}\ }\textbf {\bibinfo {volume} {5}},\ \bibinfo {pages}
  {386} (\bibinfo {year} {2021})}\BibitemShut {NoStop}%
\bibitem [{\citenamefont {Rahimi}\ and\ \citenamefont
  {Recht}(2007)}]{rahimi2007random}%
  \BibitemOpen
  \bibfield  {author} {\bibinfo {author} {\bibfnamefont {A.}~\bibnamefont
  {Rahimi}}\ and\ \bibinfo {author} {\bibfnamefont {B.}~\bibnamefont {Recht}},\
  }\bibfield  {title} {\bibinfo {title} {Random features for large-scale kernel
  machines},\ }\href
  {https://papers.nips.cc/paper_files/paper/2007/hash/013a006f03dbc5392effeb8f18fda755-Abstract.html}
  {\bibfield  {journal} {\bibinfo  {journal} {Advances in neural information
  processing systems}\ }\textbf {\bibinfo {volume} {20}} (\bibinfo {year}
  {2007})}\BibitemShut {NoStop}%
\bibitem [{\citenamefont {Cucker}\ and\ \citenamefont
  {Smale}(2002)}]{cucker2002mathematical}%
  \BibitemOpen
  \bibfield  {author} {\bibinfo {author} {\bibfnamefont {F.}~\bibnamefont
  {Cucker}}\ and\ \bibinfo {author} {\bibfnamefont {S.}~\bibnamefont {Smale}},\
  }\bibfield  {title} {\bibinfo {title} {On the mathematical foundations of
  learning},\ }\href
  {https://www.ams.org/journals/bull/2002-39-01/S0273-0979-01-00923-5/S0273-0979-01-00923-5.pdf}
  {\bibfield  {journal} {\bibinfo  {journal} {Bulletin of the American
  mathematical society}\ }\textbf {\bibinfo {volume} {39}},\ \bibinfo {pages}
  {1} (\bibinfo {year} {2002})}\BibitemShut {NoStop}%
\bibitem [{\citenamefont {Barber}(2024)}]{barber2024hoeffding}%
  \BibitemOpen
  \bibfield  {author} {\bibinfo {author} {\bibfnamefont {R.~F.}\ \bibnamefont
  {Barber}},\ }\href {https://arxiv.org/abs/2404.06457} {\bibinfo {title}
  {Hoeffding and bernstein inequalities for weighted sums of exchangeable
  random variables}} (\bibinfo {year} {2024}),\ \Eprint
  {https://arxiv.org/abs/2404.06457} {arXiv:2404.06457 [math.ST]} \BibitemShut
  {NoStop}%
\bibitem [{\citenamefont {Van Den~Berg}\ \emph {et~al.}(2023)\citenamefont {Van
  Den~Berg}, \citenamefont {Minev}, \citenamefont {Kandala},\ and\
  \citenamefont {Temme}}]{berg2022probabilistic}%
  \BibitemOpen
  \bibfield  {author} {\bibinfo {author} {\bibfnamefont {E.}~\bibnamefont {Van
  Den~Berg}}, \bibinfo {author} {\bibfnamefont {Z.~K.}\ \bibnamefont {Minev}},
  \bibinfo {author} {\bibfnamefont {A.}~\bibnamefont {Kandala}},\ and\ \bibinfo
  {author} {\bibfnamefont {K.}~\bibnamefont {Temme}},\ }\bibfield  {title}
  {\bibinfo {title} {Probabilistic error cancellation with sparse
  pauli--lindblad models on noisy quantum processors},\ }\href
  {https://doi.org/10.1038/s41567-023-02042-2} {\bibfield  {journal} {\bibinfo
  {journal} {Nature Physics}\ ,\ \bibinfo {pages} {1}} (\bibinfo {year}
  {2023})}\BibitemShut {NoStop}%
\end{thebibliography}
%

\newpage\clearpage

\section*{Appendices for ``More buck-per-shot: Why learning trumps mitigation in noisy quantum sensing''}
Here we present additional details and proofs for the results in the main text. The appendices are organized as follows. First, in Appendix~\ref{appendix:notation_summary_table}, we provide a full table summarizing notation used in this manuscript. In Appendix~\ref{appendix:zne}, we outline the use of Richardson extrapolation for ZNE in the context of our framework, and the selection of performant hyper-parameters for this protocol. Appendix~\ref{appendix:Inference-based-sensing} details the learning-based inference protocol of Ref.~\cite{huerta2022inference}, reproducing key theorems in our notation and framework. Expanding upon this, in Appendix~\ref{appendix:theorems_and_proofs} we present various theorems and proofs relevant to the main results, including improved bounds upon inference-based sensing and a generalization of these results to error-mitigated response functions. In Appendix~\ref{appendix:error_bound_exact_expressions}, we detail the assumptions and expressions used to directly compare analytic error bounds in Sec.~\ref{sec:analytic_error_comparison}. Finally, in Appendix \ref{appendix:noise_models}, we provide details of all hardware noise models used throughout this work.

\section{Summary of notation}
\label{appendix:notation_summary_table}

\begin{table*}[h]
  \centering
  \resizebox{\textwidth}{!}{\begin{tabular}{@{}|c|c|| c|c|@{}}
    \hline
    \textbf{Notation}  & \textbf{Meaning}  & \textbf{Notation}  & \textbf{Meaning}  \\ \hline 
    $n$ & Number of qubits/system size & $N$ & Total shot budget available \\ \hline
    $R(\theta)$  &  Noiseless response at $\theta$ & $R_{\lambda}(\theta)$  &  Noisy response at $\theta$ \\ \hline
    $L$ & Lipschitz constant of $R(\theta)$ & $L_{\lambda}$ & Lipschitz constant of $R_{\lambda}(\theta)$ \\ \hline
    $m$ & Number of noise nodes used in ZNE & $\gamma(x)$ & Lagrange basis polynomials used in ZNE \\ \hline
    $\Tilde{R}_{\lambda}(\theta)$  &  Inferred noisy response at $\theta$ & $\Tilde{R}_{\lambda}(\theta,N_I)$  &  Inferred noisy response at $\theta$ obtained using $N_I$ shots \\ \hline
    $R_{M}(\theta)$  &  Error-mitigated response at $\theta$ & $\Tilde{R}_{M}(\theta)$  &  Error-mitigated inferred response at $\theta$ \\ \hline
    $\overline{R}_{\lambda}(\theta,N)$  &  N-shot estimated noisy response at $\theta$ & $\overline{R}_{M}(\theta,N)$  &  Error-mitigated response at $\theta$ where ZNE uses N shots \\ \hline
    $\Delta^2R_{\lambda}$ & Variance in noisy response distribution &     $\Delta^2\overline{R}_{\lambda}$ & Variance in shot-limited noisy response distribution\\ \hline
    $\Delta^2\Tilde{R}_{\lambda}$ & Variance in inferred noisy response distribution & $\Delta^2\Tilde{R}_{M}$ & Variance in mitigated inferred response distribution\\ \hline
    $\Theta^*$ & Random variable representing the unknown phase & $\theta^*$ & Represents members of the sample space of $\Theta^*$ \\ \hline
    $\thetaest$  &  Estimated value of the unknown phase & $\MSE[\thetaest]$ & Mean-squared-error in phase estimate $\thetaest$ \\ \hline
    $\mathcal{P}_{\Theta^*}$ & Probability density function of $\Theta^*$ & $\mathbb{E}_{\Theta^*}$ & Average over multiple realizations of $\Theta^*$ \\ \hline
    $\BMSE[\thetaest]$  & Bayesian MSE in $\thetaest$ & $\CMSE[\thetaest|\theta^*]$ &  Conditional MSE in $\thetaest$ given $\Theta^* = \theta^*$ \\ \hline
    $\Bias[\thetaest | \theta^*]$ & Conditional bias in $\thetaest$ given $\Theta^* = \theta^*$ & $\Var[\thetaest | \theta^*]$ & Conditional variance in $\thetaest$ given $\Theta^* = \theta^*$ \\ \hline
    $\mathbb{E}_{N_I}$ & Average over multiple realizations of inference  & $\mathbb{E}_{N_E}$ & Average over multiple realizations of estimation \\ \hline
    $\epsilon$ & Largest sampling error in a set of noisy responses & $\chi$ & Largest sampling error in a set of mitigated responses \\
    \hline
  \end{tabular}}
  \caption{\textbf{Full summary of notation.}}
\end{table*}

\section{Zero-noise extrapolation via Richardson extrapolation}
\label{appendix:zne}

In Sec.~\ref{section:ZNE_introduction}, we introduced ZNE via the Richardson extrapolation method. Here, we present an in-depth review of the optimized version of this protocol.

\subsection{Richardson extrapolation protocol}
\label{appendix:richardson_extrapolation}

Error mitigation via zero-noise extrapolation requires observing a noisy quantum system at various noise levels in order to estimate the zero-noise response value. As shown in Fig.~\ref{fig:ZNE_demo}, when using ZNE for sensing, we assume that $\Theta^* = \theta^*$ is kept fixed while the noise level is gradually increased from its minimum (base) noise level of $\lambda$. We assume that we can control the noise level; $\lambda \rightarrow x \lambda$, where $x>1$ is a (real-number) multiplier that parametrizes boosted noise levels. Suppose we have a noisy state $\rho_{x\lambda}$ at noise level $x \lambda$ produced by some known preparation channel, as outlined in Sec.~\ref{section:noisy_sensing_introduction}. Next, we measure the observable $O$ at the fixed phase value of $\theta^*$, leading to the noise boosted response $R_{x\lambda}(\theta^*)$. From these, we aim to determine the zero-noise value $R(\theta^*)$ as accurately as possible. The boosted noise response can be rewritten as a power series in noise as
\begin{equation}
R_{x\lambda}(\theta^*) = R(\theta^*) + \sum_{k=1}^\infty a_k (x \lambda)^k\,,
\label{eq:ResponsePowerSeriesExpansion}
\end{equation}
where constants $a_k$ can depend on system size and evolution time \cite{temme2017error}. 

For a fixed $\theta^*$ value, the system response is observed at $m+1$ distinct noise levels. These levels can be set to 
\begin{align*}
\lambda_j = x_j \lambda\,,
\end{align*}
where $x_0< x_1 < ,\dots, < x_m$ and $x_j \in [1, \infty)\; \forall j$. The corresponding Lagrange polynomial of degree $m$ that interpolates the set $\{ \lambda_j, R_{\lambda_j}(\theta^*) \}_{j=0}^{m}$ is given as
\begin{align} \label{eq:REpolynomial}
 \sum_{j=0}^{m} \gamma_j(x) R_{\lambda_j}(\theta^*) \,,
\end{align}
where $\gamma_j(x)$ are the Lagrange basis polynomials
\begin{align}
\label{eqn:lagrange_basis_polynomial_def}
     \gamma_j(x) = \prod_{0 \leq l \leq m ; l \neq j} \frac{x - x_l}{x_j - x_l}\,.
\end{align}
These polynomials satisfy the following conditions 
\begin{align}
    \sum_j \gamma_j(x) &= \prod_{0 \leq l \leq m} (x-x_l)\sum_j \prod_{l \neq j} \frac{1}{(x_j - x_l)}\frac{1}{(x-x_j)} =1\,,  \quad\qquad  \text{and}  \quad\qquad
    \sum_{j=0}^{m} \gamma_j x_{j}^{k}=0, \quad  k = 1, \dots, m \,.
\end{align}
Now, to obtain the zero-noise response estimate at $\theta^*$, which we call $R_M(\theta^*)$, we evaluate the Lagrange interpolating polynomial above at $x=0$
\begin{align}
\label{eq:DefinitionMitigatedPolynomial}
    R_{M}(\theta^*) &= \sum_{j=0}^{m} \gamma_j \; R_{\lambda_j}(\theta^*)\,,
\end{align}
where we have defined $\gamma_j \equiv \gamma_j(x=0)$. When $\lambda$ is sufficiently small (as it will be in the relevant settings), we can write each $R_{\lambda_j}(\theta^*)$ in Eq.~\eqref{eq:DefinitionMitigatedPolynomial} as a power series of the form of Eq.~\eqref{eq:ResponsePowerSeriesExpansion} to obtain
\begin{align}
\label{eq:ResponseMitigatedTruncationStart}
    R_{M}(\theta^*) 
    &= \sum_{j=0}^{m} \gamma_j \; R_{x_j\lambda}(\theta^*) = \sum_{j=0}^{m} \gamma_j \left[ R(\theta^*) + \sum_{k=1}^{\infty}a_k x_j^k \lambda^k \right] = \underbrace{\sum_{j=0}^{m} \gamma_j}_{\sum_j \gamma_j =1} R(\theta^*) + \sum_{j=0}^{m} \gamma_j \left[\sum_{k=1}^{\infty}a_k x_j^k \lambda^k \right]  \\
    &= R(\theta^*) + \sum_{j=0}^{m} \sum_{k=1}^{\infty}  \gamma_j a_k x_j^k \lambda^k \nonumber\,,
\end{align}
where we have used the previously stated properties of Lagrange polynomials to simplify the expressions. We can choose some order $m$ in the expansion and discard higher-order terms to obtain
\begin{align}
    R_{M}(\theta^*) &=
    R(\theta^*) + \underbrace{\sum_{j=0}^{m}\sum_{k=1}^{m} \gamma_j a_k x_{j}^{k}\lambda^k}_{0} + \sum_{j=0}^{m}\sum_{k=m+1}^{\infty} \gamma_j a_k x_{j}^{k}\lambda^k= R(\theta^*) + \sum_{j=0}^{m} \sum_{k=m+1}^{\infty} \gamma_j a_k x_{j}^{k}  \lambda^k = R(\theta^*) + \Theta(\lambda^{m+1}) \,.\nonumber
\end{align}
Hence, for a small $\lambda$, Richardson extrapolation works well and reduces the correction term in the zero-noise estimate from $\Theta(\lambda)$ in the unmitigated case to $\Theta(\lambda^{m+1})$ in the mitigated case.

\subsection{Optimizing hyper-parameters}
\label{appendix:ZNE_hyperparameters}

Let us begin by recalling that Richardson extrapolation has the following free parameters: the order $m$, the node positions $x_j$, and in the case of quantum observables, the number of shots $N_j$ allocated to estimating each $\overline{R}_{\lambda_j}$, constrained by some total shot budget $N= \sum_j N_j$. However, the optimal values for some of these hyper-parameters is known, and we will henceforth fix their values when such  optimal solution is available. 

For instance,  the optimal distribution of shots $N_j$ is known from Lemma 1 in Ref.~\cite{Hoel1994optimal} to be
\begin{equation}
\label{eqn:richardson_shot_distribution}
	N_j = N \frac{|\gamma_j|}{\sum_{j=0}^m |\gamma_j|} \,,
\end{equation}
where $\sum_j \gamma_j =1$ holds. In  Ref.~\cite{Hoel1994optimal}, the authors prove that this shot budget allocation method will minimize the variance of the predicted value of a polynomial regression curve at any given point beyond the interval on which observations are made. In the present setting, while $R_M(\theta^*)$ is obtained by interpolating responses obtained on $[x_0\lambda, x_m \lambda]$, this shot allocation minimizes $\Delta^2\overline{R}_M(\theta^*)$ of the Lagrange polynomial at $x=0$. This holds for any chosen set of nodes $\{x_j\}$, regardless of how they are spaced. 

Next, we follow Ref.~\cite{krebsbach2022optimization}, where an in-depth study on optimizing the spacing between noise nodes for Richardson extrapolation was conducted, such that it keeps both the bias in mitigation and the variance in the zero-noise estimate minimal. Following these results, in our setting, the optimal performance is achieved when using tilted Chebyshev nodes
\begin{equation}
\label{eqn:tilted_chebychev_nodes}
	x_j = 1+\frac{\sin^2\left(\frac{j}{n+1}\frac{\pi}{2}\right)}{\sin^2\left(\frac{1}{n+1}\frac{\pi}{2}\right)}(x_1-1)\,.
\end{equation}
This leaves two hyper-parameters: the number of nodes (extrapolation order) $m$, and the position $x_1$ of the first boosted node in Eq.~\eqref{eqn:tilted_chebychev_nodes}. To make a fair comparison between ZNE and other methods, we partially optimize these hyper-parameters such that the method reaches a good performance. However, here it is important to highlight the fact that the optimal choice of hyper-parameters depends heavily on the noise model and total shot budget. Moreover, re-optimizing the hyper-parameters for every new regime is expensive, and crucially, requires knowledge of the system. As explored in depth in this work, the quantum resources expended on parameter tunning could be more effectively utilized, e.g., for learning response functions via inference. Thus, for a fair comparison, we find a single set of hyper-parameters that are performant across many relevant regimes.

\begin{figure}
	\centering
	\includegraphics[width=0.55\textwidth]{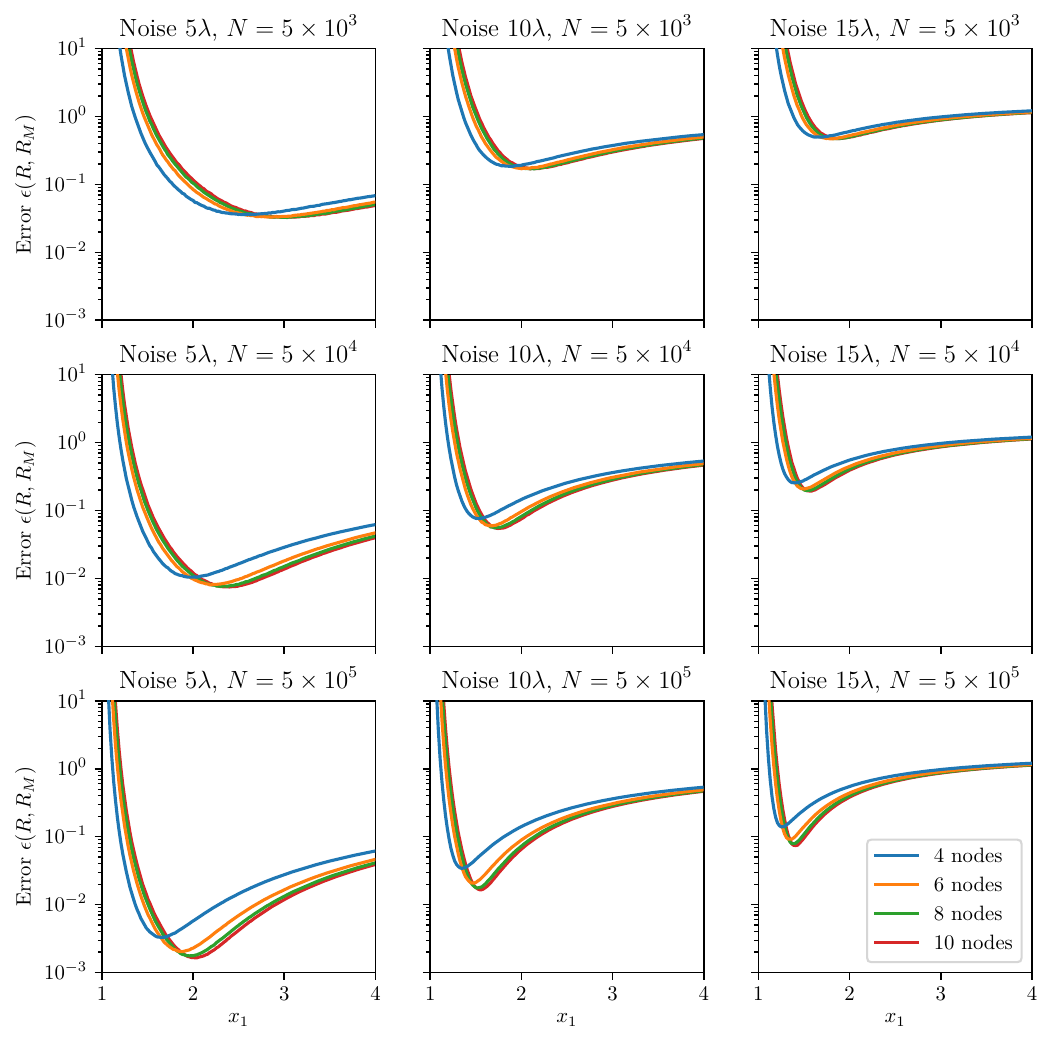}
	\caption{\textbf{Choosing hyper-parameters for ZNE.} We plot the overall mitigated response function $\epsilon_{x_1,m}(R,R_M)$ versus the first boosted node position $x_1$ at varying node counts $m$, for a range of different base noise levels $\lambda$ and overall shot budgets $N$. Confidence intervals are too small to visualize. }
	\label{fig:hyperparameters}
\end{figure}

In many cases, the experimenter may not know the field for which the noisy response reaches its highest sensitivity. Therefore, it is crucial to optimize ZNE across the whole response function domain. We therefore quantify the performance of ZNE by the objective function
\begin{equation}
	\epsilon_{x_1,m}(R,R_M)\equiv\mathbb{E}\left[\int_{0}^{2\pi} d\theta \left|R(\theta)-R_M(\theta)\right|^2\right]\,.
\end{equation}
In Fig.~\ref{fig:hyperparameters}, we plot the objective function $\epsilon_{x_1,m}(R,R_M)$ across a range of overall shot budgets, base noise levels, and hyper-parameter choices.  Here, one can qualitatively observe that higher shot budgets and higher noise levels lead to optima at lower $x_1$. Although the optimal hyper-parameters depend on noise profile and shot budget, we find that setting $x_1=1.75$ and $m=4$ yields good performance for a variety of scenarios typical for our investigation, and we utilize these throughout. Given $x_1$, the set of noise nodes $\{ x_j\}$ can be fixed using Eq.~\eqref{eqn:tilted_chebychev_nodes}, and the corresponding Lagrange basis polynomials $\gamma_j(x)$ can be determined using Eq.~\eqref{eqn:lagrange_basis_polynomial_def}.

\section{Inference-based sensing}
\label{appendix:Inference-based-sensing}

In Sec.~\ref{Subsection:InferenceIntroduction}, we introduced an inference-based sensing following the framework of Ref.~\cite{huerta2022inference}. Here we recall said protocol, as well as review the derived error-bounds.

\subsection{Inference protocol}

Consider the general response function $R_{\lambda}(\theta) = \Tr[\mathcal{D}_{\lambda} \circ \mathcal{S}_{\theta} \circ \mathcal{E}_{\lambda} (\rho_{in}) O]$ as in Eq.~\eqref{eq:response} for a single-parameter sensing setting. If the Hamiltonians which encodes the phase satisfies $H= \sum_j h_j$ with $h^{2}_j = \id$ and $[h_j, h_{j^{'}}]=0$ $\forall j, j^{'}$, and the observable $O$ is such that $\|O\|_{\infty}\leq 1$, then following Theorem 1 in Ref.~\cite{huerta2022inference}, we know that the noisy response function is a $n$-degree trigonometric polynomial; $R_{\lambda}(\theta)= \sum_{s=1}^{n} [a_s \cos(s\theta)+ b_s \sin(s\theta)] + c$. 

Hence, given $2n+1$ distinct response observations, the unknown variables $a_s, b_s, c$ can be exactly obtained by solving a linear system of equations. In practice, however, one has a finite shot budget $N_I$ and can only approximately infer the trigonometric response. We represent this approximate function as $\tilde{R}_{\lambda}(\theta, N_I)= \sum_{s=1}^{n} [\tilde{a}_s \cos(s\theta)+ \tilde{b}_s \sin(s\theta)] + \tilde{c}$. Using the shot-limited response observation dataset $\{\theta_k,  \overline{R}_{\lambda}(\theta_k, N_k) \}_{k=1}^{2n+1}$, where each inference node $\theta_k$ is allocated $N_k$ shots for measurement such that $\sum_k N_k = N_I$ is satisfied. 

Since the dataset consists of shot-limited estimates of the $R_\lambda(\theta_k)$, the approximate equality $\Tilde{R}_{\lambda}(\theta, N_I) \approx R_{\lambda}(\theta)$ becomes exact as $N_I\to \infty$. We note that the above matrix with trigonometric terms is closely related to the Vandermonde matrix. Hence, this trigonometric interpolation problem can be transformed into a polynomial interpolation problem \cite{atkinson1991introduction}, implying that the interpolant is unique. The error in approximation is minimized by a uniform spacing over the $2\pi$ period \cite{huerta2022inference}, such that
\begin{equation}
\label{eqn:inference_phase_distribution}
    \theta_{i+1}- \theta_{i}= \frac{2\pi}{2n+1}.
\end{equation}

\subsection{Bounds on the error in inference}
\label{appendix:inference_error_bound_theorems}

In this section we review one of the main results of Ref.~\cite{huerta2022inference}, which bounds the inference-error .

\begin{theorem}
\label{thm:inference_sampling_bound}
    Let $R_\lambda(\theta)$ be the exact noisy response function, and $\overline{R}_{\lambda}(\theta_k, N_k)$ be the shot-limited response observations at each inference node. Let $\tilde{R}_\lambda(\theta, N_I)$ be the trigonometric polynomial approximation to $R_\lambda(\theta)$ obtained from $\{ \theta_k, \overline{R}_\lambda(\theta_k, N_k) \}_{k=1}^{2n+1}$ via the inference scheme of Sec.~\ref{Subsection:InferenceIntroduction}, where $N_k$ shots are used per inference node such that $\sum_{k=1}^{2n+1} N_k = N_I$ is satisfied. Defining the maximum sampling error as $\epsilon \equiv \underset{\theta_k\in \{ \theta_k \}}{\max} |R_\lambda(\theta_k)-\overline{R}_\lambda(\theta_k, N_k)|$, then we have that for all $\theta$, the inference error is upper bounded as
	\begin{equation}
 \label{eqn:interpolation_error_bound_with_sampling_error}
		|R_\lambda(\theta)- \tilde{R}_\lambda(\theta, N_I) | \leq \mathcal{O}(\epsilon \log(n)).
	\end{equation}
\end{theorem}

\begin{specialproofsketch}
    Note that $R_\lambda(\theta)$ and $\tilde{R}_\lambda(\theta, N_I)$ are both $n$-degree trigonometric polynomials, and therefore so is their difference $R_\lambda(\theta)-\tilde{R}_\lambda(\theta,N_I)$. Write the difference in this form, and upper-bound using the known bound on trigonometric interpolation errors \cite{jackson1913accuracy}. The full proof is given in Appendix IV of Ref.~\cite{huerta2022inference}.
\end{specialproofsketch}

\subsection{Hoeffding bound on shots required per node in inference}
\label{appendix:N_k_lower_bound_hoeffding_proof}

\begin{corollary}
To ensure with a (constant) high probability that the inference error does not exceed $\delta$, for all $\theta$, the number of shots $N_k$ used at any $\theta_k$ should satisfy $N_k \in \Omega(\frac{\log^3(n)}{\delta^2})$.
\label{corollary:inference_error_shots}
\end{corollary}
\begin{specialproofsketch}
	Consider the sampling error $|R_\lambda(\theta_k)-\overline{R}_\lambda(\theta_k, N_k)|$ at a single inference point $\theta_k$. First bound the probability that this error exceeds a threshold $\epsilon$ using Hoeffding's inequality. Then, use Boole's inequality to bound the probability that \emph{any} of the $2n+1$ sampling errors (one per $\theta_k$) exceed $\epsilon$. Finally, use Theorem \ref{thm:inference_sampling_bound} to relate the sampling error to inference error, which results in
	\begin{equation}
		N_k\geq \frac{50\log^2(n)\log[(4n+2)/a]}{\delta^2}
	\end{equation}
	for some small constant failure probability $a$. The full proof is given in Appendix V of Ref.~\cite{huerta2022inference}.
\end{specialproofsketch}

\section{Theorems and proofs}
\label{appendix:theorems_and_proofs}

\subsection{Bias-Variance decomposition of conditional mean squared error}
\label{appendix:conditional_MSE}

\begin{lemma}
    Given the random variable that represents the target phase $\Theta^*$ takes the value $\theta^*$ from its sample space, the conditional mean squared error in the corresponding phase estimate $\thetaest$, $\CMSE[\thetaest | \theta^*]$, can be broken down into bias and variance components as $\CMSE[\thetaest | \theta^*] = \Var_{N}[\thetaest | \theta^*] + (\Bias_N[\thetaest | \theta^*])^2$. 
\end{lemma}

\begin{specialproof}
    We define the Bayesian mean squared error in the estimated parameter $\thetaest$ as 
    \begin{align}
        \BMSE[\thetaest] & \equiv \mathbb{E}_{\Theta^*}\left[\mathbb{E}_{N}\left[(\thetaest-\theta^*)^2 \big| \Theta^* = \theta^* \right] \right] \,,
    \end{align}
    where $\Theta^*$ is a random variable representing the unknown phase, $\mathbb{E}_{N}$ represents taking expectation over multiple N-shot protocol realizations and $\mathbb{E}_{\Theta^*}$ represents averaging over the distribution of $\Theta^*$. Next, we define the conditional mean squared error as
    \begin{align}
        \CMSE[\thetaest \mid \theta^*] & \equiv \mathbb{E}_{N}\left[(\thetaest-\theta^*)^2 \big| \Theta^* = \theta^* \right] \,.
    \end{align}
    Hence, for  $\Theta^* = \theta^*$, we can find the bias-variance decomposition of the $\CMSE$ as follows
    \begin{align}
        \CMSE[\thetaest \mid \theta^*] &= \mathbb{E}_{N}\left[(\thetaest-\theta^*)^2 \mid \theta^* \right]\\
        &= \mathbb{E}_{N}\left[\biggl(\thetaest - \mathbb{E}_{N}[\thetaest \mid \theta^*] + \mathbb{E}_{N}[\thetaest \mid \theta^*] - \theta^*\biggr)^2 \bigg|~  \theta^* \right] \nonumber \\ 
        &= \mathbb{E}_{N} \left[ \biggl(\thetaest - \mathbb{E}_{N}[\thetaest \mid \theta^*] \biggr)^2 + 2\biggl(\thetaest - \mathbb{E}_{N}[\thetaest \mid \theta^*] \biggr) \biggl(\mathbb{E}_{N}[\thetaest \mid \theta^*] - \theta^*\biggr) + \biggl(\mathbb{E}_{N}[\thetaest \mid \theta^*] - \theta^* \biggr)^2 
     \bigg|~ \theta^* \right] \nonumber \\
          &= \mathbb{E}_{N} \left[ (\thetaest - \mathbb{E}_{N}[\thetaest \mid \theta^*])^2 \bigg|~ \theta^* \right] + (\mathbb{E}_{N}[\thetaest \mid \theta^*] - \theta^*)^2 \nonumber \\
          &= \Var_{N}[\thetaest \mid \theta^*] + (\Bias_N[\thetaest \mid \theta^*])^2 \nonumber\,,
    \end{align}
    where the manipulations between the third and fourth line are the same as those used in the standard proof of bias-variance decomposition for the mean-squared error of an estimator. The two terms in the last step represent the conditional variance and conditional bias in the parameter estimate, respectively. 
\end{specialproof}

\subsection{Lipschitz constants of response functions}
\label{appendix:noiseless_lipschitz_bound}

\begin{theorem} Given the general response function $R_{\lambda}(\theta) = \Tr[\mathcal{D}_{\lambda} \circ \mathcal{S}_{\theta} \circ \mathcal{E}_{\lambda} (\rho_{in}) O]$ as in Eq.~\eqref{eq:response}, if the Hamiltonian which encodes the phase satisfied  $H= \sum_j h_j$ with $h^{2}_j = \id$ and $[h_j, h_{j^{'}}]=0$ $\forall j, j^{'}$, and observable $O$ is such that $\|O\|_{\infty}\leq 1$, then the Lipschitz constant of the response function satisfies $L_{\lambda} \in \mathcal{O}(n)$.
\end{theorem}

\begin{specialproof}
        We use the parameter-shift-rule for this proof ~\cite{mitarai2018quantum, schuld2019evaluating, Banchi2021measuring}. For generality, let us consider the case where there are $d$ sensing parameters instead of one. Assuming this multi-parameter setting, using  $\vec{\theta}$ to represent the $d$-dim vector, the response function $R_{\lambda}(\vec{\theta})$ is given by
        \begin{align}
            R_{\lambda}(\vec{\theta}) &= \Tr[\mathcal{D}_{\lambda} \circ \mathcal{S}_{\vec{\theta}} \circ \mathcal{E}_{\lambda} (\rho_{in}) O] =\Tr[\mathcal{D}_{\lambda} \circ \mathcal{S}_{\vec{\theta}}(\rho) O] = \Tr[ \mathcal{D}_{\lambda}\bigg( U(\vec{\theta})\rho U^{\dagger}(\vec{\theta}) \bigg) O] \nonumber\,,
        \end{align}
        where $\rho \equiv \mathcal{E}_{\lambda} (\rho_{in})$. 
        
        We now use the Kraus decomposition of  $\mathcal{D}_{\lambda}$ channel, which satisfy $\sum_{k}E^{\dagger}_{k}(\lambda)E_{k}(\lambda) = \id$. Note that this notation clearly shows how the Kraus operators $E_{k}(\lambda)$ depend on ${\lambda}$ but have no dependence on $\theta$. Hence, the response function $R_{\lambda}(\vec{\theta})$ becomes
        \begin{align}
            R_{\lambda}(\vec{\theta}) &= \sum_{k} \Tr[E_{k}(\lambda) U(\vec{\theta})\rho U^{\dagger}(\vec{\theta})E^{\dagger}_{k}(\lambda)O]\,,
        \end{align}
        where we have exploited the linearity of trace. Now, let us consider the $i$-th component in the gradient vector of the response function $\nabla R_{\lambda}(\vec{\theta})$
        \begin{align}
            \frac{\partial R_{\lambda}(\vec{\theta})}{\partial \theta_i} &= \sum_{k} \Tr[E_{k}(\lambda) \frac{\partial U({\theta}_i)}{\partial \theta_i}\rho U^{\dagger}(\theta_{i})E^{\dagger}_{k}(\lambda)O] + \sum_{k} \Tr[E_{k}(\lambda) U({\theta}_{i})\rho \frac{\partial U^{\dagger}(\theta_i)}{\partial \theta_i}E^{\dagger}_{k}(\lambda)O]\,,
        \end{align}
        where we define $U(\theta_i) \equiv e^{-i \theta_i H} = e^{-i \theta_i \sum_{j=1}^{n} h_j}$. Given this unitary, its  partial derivative with respect to parameter $\theta_i$ is given as
        \begin{align}
            \frac{\partial U(\theta_i)}{\partial \theta_i} &= \sum_{j=1}^{n}{\frac{\partial}{\partial \theta_i}}e^{-i\theta_i h_j}\prod_{q \neq j} e^{-i\theta_i h_q}= \sum_j{-i h_je^{-i\theta_i h_j}}\prod_{q \neq j} e^{-i\theta_i h_q} = \sum_j{-i h_j U(\theta_i)} \nonumber\,.
        \end{align}

        Inserting the previous result back into the gradient of the response function, we get
        \begin{align}
            \frac{\partial R_{\lambda}(\vec{\theta})}{\partial \theta_i} &= \sum_j \biggl[ \sum_{k} \Tr[E_{k}(\lambda)(-ih_j) U({\theta}_i)\rho U^{\dagger}(\theta_{i})E^{\dagger}_{k}(\lambda)O] + \sum_{k} \Tr[E_{k}(\lambda) U({\theta}_{i})\rho U^{\dagger}(\theta_i) (i h_j) E^{\dagger}_{k}(\lambda)O] \biggr]\,.
        \end{align}

        Using the cyclic property of trace, we can move the observable $O$ as follows
        \begin{align}
            \frac{\partial R_{\lambda}(\vec{\theta})}{\partial \theta_i} &= \sum_j \biggl[ \sum_{k} \Tr[  E^{\dagger}_{k}(\lambda) O E_{k}(\lambda)(-ih_j) U({\theta}_i)\rho U^{\dagger}(\theta_{i})] + \sum_{k} \Tr[ E^{\dagger}_{k}(\lambda) O E_{k}(\lambda) U({\theta}_{i})\rho U^{\dagger}(\theta_i) (i h_j)] \biggr]\\
            &= \sum_j \biggl[  \Tr\biggl[ \underbrace{\sum_{k} E^{\dagger}_{k}(\lambda) O E_{k}(\lambda)}_{\Tilde{O}(\lambda)}(-ih_j) U({\theta}_i)\rho U^{\dagger}(\theta_{i})\biggr] + \Tr\biggl[ \underbrace{\sum_{k} E^{\dagger}_{k}(\lambda) O E_{k}(\lambda)}_{\Tilde{O}(\lambda)} U({\theta}_{i})\rho U^{\dagger}(\theta_i) (i h_j)\biggr] \biggr] \nonumber \\
            &= \sum_j \biggl[  \Tr[ \Tilde{O}(\lambda)(-ih_j) U({\theta}_i)\rho U^{\dagger}(\theta_{i})] + \Tr[ \Tilde{O}(\lambda) U({\theta}_{i})\rho U^{\dagger}(\theta_i) (i h_j)] \biggr] \nonumber\,,
        \end{align}
        where $\Tilde{O}(\lambda) \equiv \sum_{k} E^{\dagger}_{k}(\lambda) O E_{k}(\lambda)$ represents the observable $O$ evolving under the adjoint of channel $\mathcal{D}_{\lambda}$ in the Heisenberg picture. Hence, the gradient can be simplified  as
        \begin{align}
            \frac{\partial R_{\lambda}(\vec{\theta})}{\partial \theta_i} &= \sum_j \biggl[ i \Tr[U(\theta_i) \rho U^{\dagger}(\theta_i) [h_j, \Tilde{O}(\lambda)] ]\biggr]\,,
        \end{align}
        where $[h_j, \Tilde{O}(\lambda)]$ represents the commutator between the two operators. By using the following mathematical identity for commutators containing involutory operators \cite{mitarai2018quantum}
        \begin{align*}
            [h_j, \Tilde{O}] = -i \biggl( U^{\dagger}\left(\frac{\pi}{2}\right) \Tilde{O} U\left(\frac{\pi}{2}\right) - U^{\dagger}\left(\frac{-\pi}{2}\right) \Tilde{O} U\left(\frac{-\pi}{2}\right) \biggr) \,, 
        \end{align*}
        we can update the gradients as
        \begin{align}
            \frac{\partial R_{\lambda}(\vec{\theta})}{\partial \theta_i} &= \sum_j \biggl[ \Tr[ U\left(\theta_i + \frac{\pi}{2}\right) \rho U^{\dagger}\left(\theta_i + \frac{\pi}{2}\right) \Tilde{O}(\lambda)] - \Tr[ U\left(\theta_i - \frac{\pi}{2}\right) \rho U^{\dagger}\left(\theta_i - \frac{\pi}{2}\right) \Tilde{O}(\lambda)] \biggr] \\
            &= \sum_j \biggl[ R_{\lambda}\left(\theta_i + \frac{\pi}{2}\right) - R_{\lambda}\left(\theta_i - \frac{\pi}{2}\right) \biggr] \nonumber\,,
        \end{align}
        where in the last step, we recover the original response function but with shifted parameter values. This is because the expectation values do not change whether we evolve the state $\rho$ under channel $\mathcal{D}_{\lambda}$ and then measure $O$ (Schrodinger picture) or if we evolve the observable $O$ under the adjoint of channel $\mathcal{D}_{\lambda}$ and then measure it with $\rho$ (Heisenberg picture). Looking at the magnitude of this partial derivative, we can see that
        \begin{align}
            \biggl|\frac{\partial R_{\lambda}(\vec{\theta})}{\partial \theta_i} \biggr| &= \sum_{j=1}^{n} \biggl| R_{\lambda}(\theta_i + \frac{\pi}{2}) - R_{\lambda}(\theta_i - \frac{\pi}{2}) \biggr| \leq \sum_{j=1}^{n} \biggl( \biggl| R_{\lambda}(\theta_i + \frac{\pi}{2})\biggr| + \biggl|R_{\lambda}(\theta_i - \frac{\pi}{2})\biggr| \biggr)  \leq 2n \nonumber\,,
        \end{align}
        where the second step uses the triangle inequality. In the last step, we have used the fact that $|R_{\lambda}(\theta_i)| \leq 1$ as we have restricted ourselves to observables that satisfy $ \lVert O \rVert_{\infty} \leq 1$. Note that $ \lVert \Tilde{O}(\lambda) \rVert_{\infty} \leq 1$ is also satisfied since the CPTP map is norm preserving.
        
        Hence, the magnitude of the gradient vector is bounded as
        \begin{align}
            \lVert \nabla R_{\lambda}(\vec{\theta}) \rVert_2 &= \sqrt{ \bigg( \frac{\partial R_{\lambda}(\vec{\theta})}{\partial \theta_1} \bigg)^2 ~+~ \dots ~+~ \bigg( \frac{\partial R_{\lambda}(\vec{\theta})}{\partial \theta_d} \bigg)^2}  \leq 2 \sqrt{d} n \nonumber\,.
        \end{align}
        
        This means that the magnitude of the gradient is $\mathcal{O}(n)$. As the response functions in the present framework are trigonometric polynomial functions, from the definition of the Lipschitz constant $L_{\lambda}$, $\lVert R_{\lambda}(\vec{\theta_1}) - R_{\lambda}( \vec{\theta_2}) \rVert_2 = L_{\lambda} \lVert \vec{\theta_1} - \vec{\theta_2} \rVert_2$, we know that $L_{\lambda} = \lVert \nabla R_{\lambda}(\vec{\theta^*}) \rVert_2 $, where $\vec{\theta^*} = \arg\max_{\vec{\theta}} \lVert \nabla R_{\lambda}(\vec{\theta})\rVert_2$. Hence, the Lipschitz constant $L_{\lambda}$ will be a function of both noise and number of qubits and will be upper bounded by $\mathcal{O}(n)$. Note that for the single-parameter setting ($d=1$) for noiseless ($\lambda=0$) magnetometry with GHZ states, $L=n$ is exact. 
\end{specialproof}

\subsection{Uncertainty of Observable at different ZNE noise nodes}
\label{appendix:Variance_R_M_noise_nodes}

In this section we show the following theorem which states that  for small noise level $\lambda$, one can consider the variances at different Richardson extrapolation nodes as being constant.

\begin{lemma}
    The variance in system responses (variance of the response’s distribution) measured at different noise levels during Richardson extrapolation method can be approximated to be roughly similar; $\Delta^2 R_{\lambda_j}(\theta^*) \approx \Delta^2 R_{\lambda}(\theta^*)$, where $\lambda$ represents the minimum possible noise level.
\end{lemma}

\begin{specialproof}
    In \cite{krebsbach2022optimization}, the authors claim that the variances at different noise levels will be of the same order of magnitude and hence can be treated to be approximately equal. Here, we attempt to provide a formal proof for this claim. Let us assume that $\lambda$ is small and the observable satisfies $O^2= \id$. Then for the base noise level ($x=1)$ we can write 
    \begin{align*}
        \Delta^2 R_{\lambda} &= \Tr[\rho O^2] - \Tr[\rho O]^2 = 1- [\sum_{k=0}^{\infty} a_k  (x \lambda)^k]^2= 1- [a_0 + a_1\lambda + a_2\lambda^2+ \dots]^2 \approx 1 - a_0^2 - 2a_0a_1\lambda\,,
    \end{align*}
    where we have expanded the response in a power series in noise (Eq.~\eqref{eq:ResponsePowerSeriesExpansion}) and have ignored $\lambda^2$ terms and higher. Similarly for higher noise levels where $x>1$, we get 
    \begin{align*}
        \Delta^2 R_{x \lambda} &= \Tr[\rho O^2] - \Tr[\rho O]^2 = 1- [\sum_{k=0}^{\infty}a_k x^k \lambda^k]^2= 1- [a_0 + a_1 x \lambda + a_2 x^2 \lambda^2+ \dots]^2  \approx 1 - a_0^2 - 2a_0a_1 x \lambda\,.
    \end{align*}
    The difference between the previous variances is
    \begin{align*}
        \Delta^2 R_{x \lambda} - \Delta^2 R_{\lambda} &= 2a_0a_1 \lambda (1-x)\approx 2a_0a_1 \lambda \exp(-x)\,,
    \end{align*}
    where this difference decreases exponentially and quickly approaches close-to-zero values as we go farther from $x=1$.  
\end{specialproof}

\subsection{Shots required per node in inference-based sensing}
\label{appendix:N_k_lower_bound_general_proof}

\begin{theorem} To ensure with a (constant) high probability, and for all $\theta$,
that the maximum sampling error $| R_{\lambda}(\theta_k) - \overline{R}_{\lambda}(\theta_k, N_k)| \leq \epsilon$, number of shots $N_k$ used at any $\theta_k$ during inference should satisfy $N_k \in \Omega \biggr( \frac{1}{\epsilon^2} \log (\frac{n}{\epsilon}) \biggl)$. 
\end{theorem}

\begin{specialproof}
     This proof is formulated along the lines of Claim 1 in \cite{rahimi2007random}. That is, we will  bound the probability of obtaining a sampling error larger than $\epsilon$ for any $\theta \in \Xi$, where $\Xi$ is the parameter domain. Hence, this proof will hold even when $\{ \theta_k \}$ are randomly chosen from $\Xi$ and for any target phase if $N_E \geq N_k$ holds for the estimation step. For generality, we can consider a multi-parameter setting with $\theta \in \mathbb{R}^d$. Let $D(\Xi)$ be the diameter of the desired parameter set $\Xi$. It is defined as $D(\Xi) \equiv \sup \{\lVert \theta_1 - \theta_2 \rVert_{2} : \theta_1, \theta_2 \in \Xi \}$. Let us define a real-valued function $f(\theta): \mathbb{R}^d \rightarrow \mathbb{R}$ as $f(\theta) \equiv R_{\lambda}(\theta) - \overline{R}_{\lambda}(\theta)$. The first two things we can note are that $\mathbb{E}[f(\theta)]=0$ and $|f(\theta)| \leq 2$. \\ 
     
     We know that given a set of radius $Y$ in a finite-dimensional Banach space of dimension $d$, the covering number of balls of radius $r$ for this set is $X \leq (\frac{4Y}{r})^d$ \cite{cucker2002mathematical}. For the set $\Xi$, we consider the radius of the corresponding minimum enclosing ball (the smallest ball that covers all points in the set). Assuming the worst-case scenario, we can set the radius of the minimum enclosing ball to be $D(\Xi)$. Note that this radius can be made slightly tighter using Jung's theorem. Hence, we obtain the covering number as $X \leq (\frac{4D(\Xi)}{r})^d$.\\ 
     
    Let $\Xi_c = \{ \theta_i \}_{i=1}^X$ be the set of the centers of these balls; the net cover of $\Xi$. This means that $\forall \theta \in \Xi, ~\exists i: \lVert \theta - \theta_i \rVert_2 \leq r$. Let $L_f$ be the Lipschitz constant of the function $f(\theta)$, i.e., $\lVert f(\theta_1) - f(\theta_2) \rVert_2 = L_f \lVert \theta_1 - \theta_2 \rVert_2$. Here we note that since the response functions are differentiable, $f(\theta)$ is also differentiable. As the response functions are trigonometric polynomial functions, we know that $L_f = \lVert \nabla f(\theta^*) \rVert_2 $, where $\theta^* = \underset{\theta \in \Xi}{\mathrm{argmax}} \lVert \nabla f(\theta)\rVert_2$. We note that
     \begin{align}
         \mathbb{E}[L_f ^2] &= \mathbb{E}[\lVert \nabla f(\theta^*) \rVert_2^2]\\
         &= \mathbb{E}[\lVert \nabla R_{\lambda}(\theta^*) - \nabla \overline{R}_{\lambda}(\theta^*)\rVert_2^2] \nonumber \\
         &= \mathbb{E}[\lVert \nabla R_{\lambda}(\theta^*)\rVert_2^2 + \lVert \nabla \overline{R}_{\lambda}(\theta^*)\rVert_2^2 -2\nabla^T R_{\lambda}(\theta^*)\nabla \overline{R}_{\lambda}(\theta^*) ] \nonumber \\
         &= \mathbb{E}[\lVert \nabla R_{\lambda}(\theta^*)\rVert_2^2] - \mathbb{E}[\lVert \nabla \overline{R}_{\lambda}(\theta^*)\rVert_2^2] \nonumber \\  
         & \leq \mathbb{E}[\lVert \nabla R_{\lambda}(\theta^*)\rVert_2^2]  \nonumber \\
         & \leq L_{\lambda}^2 \nonumber\,,
     \end{align}
     where $L_{\lambda}$ is the Lipschitz constant of the noisy response function $R_{\lambda}(\theta)$. In the third step, we have used the fact that $\mathbb{E}[f(\theta)]=0$. In the last step, we have again used the fact that for a Lipschitz-continuous function, the Lipschitz constant is the magnitude of its maximum gradient. Hence, $\forall \theta \in \Xi$, $\lVert \nabla R_{\lambda}(\theta)\rVert_2 \leq L_{\lambda} $. 
     
    Now we will bound the error made within each ball. If we set $L_f < \frac{\epsilon}{2r}$, then for any $\theta_i , \theta_j \in \Xi_c$, we observe that
    \begin{align*}
        \lVert f(\theta_i) - f(\theta_j) \rVert_2 &= L_f \lVert \theta_i - \theta_j \rVert_2< \frac{\epsilon}{2r} \underbrace{\lVert \theta_i - \theta_j \rVert_2}_{=2r}< \epsilon\,.
    \end{align*}

    Hence, given the Lipschitz constant satisfies $L_f < \frac{\epsilon}{2r}$, the function $f(\theta)$ changes at most by $\frac{\epsilon}{2}$ over the radius of a ball. We summarize this statement in the notational form as $| f(\theta_i) | < \frac{\epsilon}{2}$, where $\theta_i \in \Xi_c$. From the definition of the net cover $\Xi_c$, this means that to ensure that $\forall \theta \in \Xi$, $|R_{\lambda}(\theta) - \overline{R}_{\lambda}(\theta)| < \epsilon$ is satisfied with high probability, we need the Lipschitz constant to take this value with high probability. From Markov's inequality, we know that
    \begin{align}
    \label{eqn:Lifschitz_bound_markovian}
        \mathbb{P} \left[ L_f > \frac{\epsilon}{2r} \right] &\leq \mathbb{P} \left[ L_f^2 > (\frac{\epsilon}{2r})^2 \right] \leq \frac{\mathbb{E}[L_f^2]}{(\frac{\epsilon}{2r})^2}   \leq \biggl( \frac{2rL_{\lambda}}{\epsilon} \biggr)^2 \,.
    \end{align}

    Another way to find the probability of a certain error is by limiting the probability of that error in each ball using Hoeffding's inequality. Given that we use $N_k$ shots to obtain $\overline{R}_{\lambda}$, using the union bound in Hoeffding's inequality over the net cover $\Xi_c$ gives
    \begin{align}
    \label{eqn:Hoeffding_bound_cover}
        \mathbb{P} \left[ \cup_{i=1}^
        X |f(\theta_i)| > \frac{\epsilon}{2} \right] &\leq 2 X \exp \left( \frac{-N_k \epsilon^2}{8} \right)  \leq 2 \biggl( \frac{4D(\Xi)}{r} \biggr)^d \exp \left( \frac{-N_k \epsilon^2}{8} \right) \,,
    \end{align}
    where we have used the fact that $|f(\theta_i)| \leq 2$. By adding Eq.~\eqref{eqn:Lifschitz_bound_markovian} and Eq.~\eqref{eqn:Hoeffding_bound_cover}, we can account for various ways of limiting the probability of a certain sampling error ( in terms of the free variable $r$)
    \begin{align}
        \mathbb{P} \left[ \sup_{\theta \in \Xi}|f(\theta)| > \epsilon \right] &\leq 2 \biggl( \frac{4D(\Xi)}{r} \biggr)^d \exp \left( \frac{-N_k \epsilon_{\lambda_0}^2}{8} \right) ~ + ~ \biggl( \frac{2rL_{\lambda}}{\epsilon} \biggr)^2\,.
    \end{align}

Rewriting the above by expressing $r$ in terms of other constants gives us \cite{rahimi2007random} 
    \begin{align}
    \label{eq:HoeffdingGeneral}
    \mathbb{P} \left[ \sup_{\theta \in \Xi}|R_{\lambda}(\theta) - \overline{R}_{\lambda}(\theta, N_k)| > \epsilon \right] &\leq 2^8 \left( \frac{L_{\lambda} D(\Xi)}{\epsilon}\right)^2 \exp \left( \frac{- N_k \epsilon^2}{4(d+2)}\right)\,.
    \end{align}

Thus, for a given $\epsilon$, to ensure that $\mathbb{P}\left[ \underset{\theta \in \Xi}{\mathrm{sup}} ~ |R_{\lambda}(\theta) - \overline{R}_{\lambda}(\theta, N_k)| > \epsilon \right]  \leq \beta$, where $\beta$ is some small positive real number in $ [0, 1]$, we require that number of shots $N_k$ to satisfy
\begin{align}
\label{eq:N_kHoeffdoingGeneral}
    N_k \geq \frac{4(d+2)}{\epsilon^2}\log \biggl( \frac{2^8 L_{\lambda}^2 D(\Xi)^2}{\beta \epsilon^2}  \biggr)\,.
\end{align}

Hence, with a large constant probability of $1-\beta$, for any parameter $\theta \in \Xi$, we are guaranteed that the sampling error at that parameter does not exceed $\epsilon$. This gives us a lower bound on $N_k$. Moreover, since in our framework, $L_{\lambda} \in \Theta(n)$, we can simplify the above to get $N_k \in \Omega \biggr( \frac{1}{\epsilon^2} \log (\frac{n}{\epsilon}) \biggl)$. This bound is asympotically tighter in $n$ compared to the Hoeffding-inspired lower bound of Corollary 1 in Ref. \cite{huerta2022inference}. Hoeffding bounds can generally be extremely loose since they only consider the extent of the range of the bounded random variables rather than their variances. Hence, other concentration inequalities like Bernstein's bounds are generally tighter.  
\end{specialproof}

\subsection{Shots required per node in error-mitigated inference}
\label{appendix:N_k_hoeffding_error_mitigation_inference_proof}

\begin{lemma}
To ensure that $| R_{M}(\theta_k) - \overline{R}_{M}(\theta_k, N_k)| \leq \chi$ is satisfied with a (constant) high probability, the number of shots $N_k$ used at any $\theta_k$ should satisfy $N_k \in \Omega \biggr( \frac{\Lambda^3}{\chi^2 max_j|\gamma_j|} \log(n) \biggl)$.
\end{lemma}

\begin{specialproof}
    We recall that the maximum sampling error in a set of mitigated responses is defined as
    \begin{align*}
        \chi \equiv \max_{\theta_k\in \{ \theta_k \}} |R_{M}(\theta_k) - \overline{R}_{M}(\theta_k, N_k)|\,.
    \end{align*}

    We will first try to drive a Hoeffding-based bound on the probability of exceeding $\chi$ in one protocol run but we will see that it is very loose. We then give a tighter bound with Bernstein's inequality. Let us first recall the known Hoeffding inequality for a weighted sum of random variables (Theorem 1 in \cite{barber2024hoeffding})
    \begin{align}
    \label{eq:hoeffding_weighted_sum_def}
        \mathbb{P} \left[ \sum_{j=1}^{m} w_j \biggl( Z_j - \mathbb{E}[Z_j] \biggr) \geq \chi \right] \leq \exp(\frac{- \chi^2}{2 \lVert w \rVert_2^2})\,,
    \end{align}
    where the distribution of each random variable $Z_j$ is supported on $[-1,1]$. Let us also recall the definition of mitigated response (Eq.~\eqref{eq:EMpoly})
    \begin{align*}
    R_M(\theta_k) = \sum_{j=0}^{m} \gamma_j R_{\lambda_j}(\theta_k)\,,
    \end{align*}
     where $\gamma_j \equiv \gamma_j(x=0)$ are the Lagrange basis polynomials evaluated at zero noise (Eq.~\eqref{eq:DefinitionMitigatedPolynomial}). Next, shot-limited mitigated response (Eq.~\eqref{eq:EMpolyEst}) is defined as
     \begin{align*}
     \overline{R}_M(\theta_k, N_k) = \sum_{j=0}^{m} \gamma_j \overline{R}_{\lambda_j}(\theta_k, N_j)
     \end{align*}
    such that $\sum_{j=0}^{m} N_j = N_k$ is satisfied at each node. Now, we can apply Eq.~\eqref{eq:hoeffding_weighted_sum_def} to error-mitigated inference. For each inference node $\theta_k$
    \begin{align}
        \mathbb{P} \left[ \biggl| \overline{R}_M(\theta_k, N_k) - R_M(\theta_k) \biggr| \geq \chi \right] &= \mathbb{P} \left[ \biggl| \sum_{j=0}^{m} \gamma_j \biggl( \overline{R}_{\lambda_j}(\theta_k, N_j) - R_{\lambda_j}(\theta_k) \biggr) \biggr| \geq \chi \right] \leq 2 \exp(\frac{- \chi^2}{2 \lVert \gamma \rVert_2^2}) \nonumber  \leq 2 \exp(\frac{- \chi^2}{2 \Lambda^2}) \nonumber\,,
    \end{align}   
    where we again use $\lVert O \rVert_{\infty} \leq 1$ and the definition of sampling overhead $\Lambda \equiv \sum_{j=0}^{m}|\gamma_j|$.
    
    Now, using Boole's inequality, we can further imply that
    \begin{align}    
    \mathbb{P} \biggl[ \cup_{k=1}^{2n+1} \biggl| \overline{R}_M(\theta_k, N_k) - R_M(\theta_k) \biggr| \geq \chi \biggr] & \leq \sum_{k=1}^{2n+1} 2\exp \left( \frac{- \chi^2}{2 \Lambda^2}\right)\leq (4n+2) \exp \left( \frac{- \chi^2}{2\Lambda^2}\right) \nonumber\,.
    \end{align}   

    The previous result can be used to derive a lower bound on the sampling overhead; $\Lambda^2 \in \Omega(\frac{\chi^2}{\log(n)})$. However, since we know the variance of the random variables, we can use Bernstein's inequality to exploit this extra information to get a significantly tighter bound. We recall Theorem 3 in \cite{barber2024hoeffding}
    \begin{align}
    \label{eq:bernstein_weighted_sum_def}
        \mathbb{P} \left[ \sum_{j=1}^{m} w_j \biggl( Z_j - \mathbb{E}[Z_j] \biggr) \geq \chi \right] \leq \exp(\frac{- \chi^2}{\frac{8\lVert w \rVert_{\infty}^2}{9} + \frac{4\chi\lVert w \rVert_{\infty}}{3} + 2 \lVert w \rVert_2^2 ~ \Delta^2 Z})\,,
    \end{align}
    where $\Delta^2 Z$ is the variance of random variable $Z_j$. Keeping only the dominant term in the denominator in the exponential, we apply this to our setting to get a slightly looser Bernstein bound
    \begin{align}
        \mathbb{P} \left[ \biggl| \overline{R}_M(\theta_k, N_k) - R_M(\theta_k) \biggr| \geq \chi \right] &= \mathbb{P} \left[ \biggl| \sum_{j=0}^{m} \gamma_j \biggl( \overline{R}_{\lambda_j}(\theta_k, N_j) - R_{\lambda_j}(\theta_k) \biggr) \biggr| \geq \chi \right] \\
        &\leq 2 \exp(\frac{- \chi^2}{2 \lVert \gamma \rVert_2^2 ~ \Delta^2 \overline{R}_{\lambda}(\theta_k, N_j)}) \nonumber \\
        & \leq 2 \exp(\frac{- \chi^2 N_k \lVert \gamma \rVert_{\infty} }{2 \Lambda^3 \Delta^2 R_{\lambda}(\theta_k)}) \nonumber\,,
    \end{align} 
    where we assume optimal shot allocation during Richardson extrapolation, $\lVert \gamma \rVert_{\infty} \equiv \max_{j}{|\gamma_j|}$ and we have used the approximation $\Delta^2 R_{\lambda_j}(\theta_k) \approx \Delta^2 R_{\lambda}(\theta_k)$ (see Appendix~\ref{appendix:Variance_R_M_noise_nodes} for proof). Using Boole's inequality again, we get:
    \begin{align}    
    \mathbb{P} \biggl[ \cup_{k=1}^{2n+1} \biggl| \overline{R}_M(\theta_k, N_k) - R_M(\theta_k) \biggr| \geq \chi \biggr] & \leq \sum_{k=1}^{2n+1} 2 \exp(\frac{- \chi^2 N_k \lVert \gamma \rVert_{\infty} }{2 \Lambda^3 \Delta^2 R_{\lambda}(\theta_k)}) \,.
    \end{align}

    Thus, for a given $\chi$, to ensure that $\mathbb{P}\left[ \cup_{k=1}^{2n+1}|\overline{R}_M(\theta_k, N_k) - R_M(\theta_k)| \geq \chi \right]  \leq \alpha$, where $\alpha$ is some small positive real number in $[ 0, 1 ]$, we require that number of shots $N_k$ satisfy
    \begin{align} 
        N_k \geq \frac{2 \Lambda^3 ~ \Delta^2 R_{\lambda}(\theta_k) ~ \log(\frac{4n+2}{\alpha})}{\chi^2 \lVert \gamma \rVert_{\infty}}\,.
    \end{align}
    
    Hence, with a large constant probability of $1-\alpha$, for any inference node $\theta_k$, we are guaranteed that the sampling error in mitigated responses at that node does not exceed $\chi$ given $N_k \in \Omega \biggr( \frac{\Lambda^3}{\max_j|\gamma_j|\chi^2} \log(n) \biggl)$.    

\end{specialproof}

\begin{corollary}
To ensure with a (constant) high probability that the inference error does not exceed $\delta$, for all $\theta$, the number of shots $N_k$ used at any $\theta_k$ should satisfy $N_k \in \bigl( \frac{\Lambda^3 \log^3(n)}{\max_j |\gamma_j|\delta^2} \bigr)$.
\label{corollary:error_mitigated_inference_error_shots}
\end{corollary}

\begin{specialproof}
Similar to the proof of Corollary \ref{corollary:inference_error_shots}, we can set a desired accuracy in inference, $|R_M(\theta) - \Tilde{R}_M(\theta, N_I)| \leq \delta$, by controlling the sampling error in mitigated responses. That is, if the probability that maximum sampling error in a given set of mitigated responses exceeds $\chi$ is bounded by $\alpha$, $\mathbb{P} \left [ |R_M(\theta_k) - \overline{R}_M(\theta_k, N_k)| \geq \chi \right] \leq \alpha$, then the probability that inference error exceeds $5 \chi \log(n)$ is also bounded by $\alpha$
\begin{align*}
    \mathbb{P} \left[ |R_M(\theta) - \Tilde{R}_M(\theta, N_I)| \geq 5 \chi \log(n) \right] \leq \alpha\,.
\end{align*}

We can then rewrite the required number of shots in terms of $\delta$ as
\begin{align} 
    N_k \geq \frac{50 \Lambda^3 \Delta^2 R_{\lambda}(\theta_k) \log^2(n)\log(\frac{4n+2}{\alpha})}{\max_j |\gamma_j|\delta^2}\,.
\end{align}

Hence, given the number of shots per inference node satisfy $N_k \in \Omega \bigl( \frac{\Lambda^3 \log^3(n)}{\max_j |\gamma_j|\delta^2} \bigr)$, we can ensure that $|R_M(\theta) - \Tilde{R}_M(\theta, N_I)| \leq \delta$ with constant probability of $1-\alpha$. 

\end{specialproof}

\section{Error bound comparison}
\label{appendix:error_bound_exact_expressions}
In Sec.~\ref{section:theoretical}, the error bounds for the protocols considered in this work are given to varying levels of approximation. We compare these error bounds in Sec.~\ref{sec:analytic_error_comparison}. Here, we detail exactly which expressions are numerically evaluated in the comparison, and how their terms are computed. We use a global depolarizing noise model with fault rate $\lambda$, such that $R_{\lambda}(\theta)=(1-p_\lambda) \cos(n\theta)$, $\Delta^2R_\lambda(\theta)=1- (1-p_\lambda)^2 \cos^2(n\theta)$, and $|\partial_\theta R_{\lambda}(\theta)|=(1-p_\lambda)n|\sin(n\theta)|$ (Sec.~\ref{sec:global_depol_toy_model}).

\subsubsection{Noise-aware noisy quantum sensing}
The first line of Eq.~\eqref{eqn:noisy_meas_noisy_inv_error} gives
\begin{equation}
    \CMSE[\thetaest | \theta^*] = \frac{\Delta^2 R_{\lambda}(\theta^*)}{N(\partial_\theta R_{\lambda}(\theta)|_{\theta=\theta^*})^2}\,,
\end{equation}
where all terms are exactly known. Since for the purposes of this computation we know $\partial_\theta R_\lambda(\theta)$, we elect not to approximate the gradient on the denominator by its Lipschitz constant, since this can underestimate the error in case we are far from the point of maximal gradient.

\subsubsection{Naive noisy quantum sensing}
We again elect not to approximate the gradient by its Lipschitz constant (c.f. Eq.~\eqref{eq:mse_noisy_lower_bound}), yielding
\begin{align}
	\CMSE[\thetaest |\theta^*]  = \frac{\Delta^2 R_{\lambda}(\theta^*)}{N (\partial_\theta R(\theta)|_{\theta=\theta^*})^2}+ ~ \bigg( \frac{ | R_{\lambda}(\theta^*) - R(\theta^*) |}{\partial_\theta R(\theta)|_{\theta=\theta^*}} \bigg)^2 \,,
\end{align}
where all terms are exactly known.

\subsubsection{Noisy sensing mitigated by zero-noise extrapolation}
Here we compute
\begin{equation}
    \CMSE[\thetaest |\theta^*] = \Var_N[\thetaest | \theta^*] + (\Bias_N[\thetaest | \theta^*])^2\,.
\end{equation}
The variance term can be computed exactly by substituting Eq.~\eqref{eq:varRM} into Eq.~\eqref{eqn:error_mitigated_variance_defn}, yielding
\begin{equation}
    \Var_N[\thetaest | \theta^*] = \sum_{j=0}^{m} \gamma_j^2 \; \frac{\Delta^2 R_{\lambda_j}(\theta^*)}{N_j(\partial_\theta R(\theta)|_{\theta=\theta^*})^2}\,.
\end{equation}
Compared to the looser bound on the variance in Eq.~\eqref{eq:mse_mitigated_lower_bound}, we have omitted the approximation of the gradient by its Lipschitz constant, as well as the assumption of equal variances $\Delta^2 R_{\lambda_j}(\theta^*) \approx \Delta^2 R_{\lambda}(\theta^*)$ (Appendix~\ref{appendix:Variance_R_M_noise_nodes}). Nonetheless, we observed numerically that the latter approximation holds quite well in this regime. The values of $\gamma_j$ are chosen from those selected in the hyperparameter optimization of Appendix~\ref{appendix:ZNE_hyperparameters}, and thus are the same as those used in numerical experiments. The bias term is approximated as
\begin{equation}
    \Bias_N[\thetaest | \theta^*] \approx \frac{\lambda^{m+1}}{\partial_\theta R(\theta)|_{\theta=\theta^*}}\,,
\end{equation}
where, compared to Eq.~\eqref{eqn:bias_mitigated_sensing}, we have approximated the mitigation bias as $|R_M(\theta^*)-R(\theta^*)| \in \Theta(\lambda^{m+1})\approx \lambda^{m+1}$, and not approximated the denominator.

\subsubsection{Inference-based noisy sensing}
\label{appendix:numerical_evaluation_inference}
As outlined in Sec.~\ref{sec:inference_error}, the CMSE is given as
\begin{equation}
\label{eqn:appendix_cmse_inference}
    \CMSE[\thetaest |\theta^*] = (\Bias[\thetaest | \theta^*])^2 + \underbrace{\mathbb{E}_{N_I}\left[ \Var_{N_E} \left[\thetaest \big| \theta^*, f \right]  \big| ~ \theta^* \right]  
    + \Var_{N_I}\left[ \mathbb{E}_{N_E}\left[\thetaest \big| \theta^*, f \right] \big| ~ \theta^* \right] }_{  \Var[\thetaest | \theta^*]}\,.
\end{equation}
Following the fourth line of Eq.~\eqref{eqn:inference_bias_derivation_steps}, the bias term is given as
\begin{align}
    \Bias[\thetaest | \theta^*] &\approx \frac{\mathbb{E}_{N_I}[|R_\lambda(\theta^*)-\tilde{R}_\lambda(\theta^*)|]}{|\partial_{\theta}R_{\lambda}(\theta)|_{\theta=\theta^*}|} \leq \frac{\delta}{|\partial_{\theta}R_{\lambda}(\theta)|_{\theta=\theta^*}|}
    \label{eqn:inference_bias_bound}\,,
\end{align}
where the last inequality is satisfied with high probability due to Corollary \ref{corollary:inference_error_shots}. We note that this is a very pessimistic estimate: Eq.~\eqref{eqn:inference_bias_bound} assumes that the worst-case bound $|R_\lambda(\theta^*)-\tilde{R}_\lambda(\theta^*)|<\delta$ is saturated everywhere, but in general the average-case inference error $|R_\lambda(\theta^*)-\tilde{R}_\lambda(\theta^*)|$ will be much smaller. Thus, we expect our analytic approximations (i.e., Fig.~\ref{fig:analytic_error_bounds}) to overestimate the error of inference-based methods, an intuition that we numerically verify in Sec.~\ref{sec:numerics}. Corollary \ref{corollary:inference_error_shots} demonstrates that we require $N_k\in \Omega(\log^3(n)/\delta^2)$ shots to achieve an inference error bound of $\delta$ with high probability. In approximating this term numerically, we therefore assume that $\delta\approx \sqrt{\log^3(n)/N_k}$. The estimation variance (second term in Eq.~\eqref{eqn:appendix_cmse_inference}) follows from the second line of Eq.~\eqref{eqn:estimation_variance_inference} as
\begin{equation}
    \mathbb{E}_{N_I}\left[ \Var_{N_E}\left[\thetaest \big| \theta^*, f\right] \big| ~ \theta^* \right] \approx \frac{\Delta^2 R_{\lambda}(\theta^*)}{N_E (\partial_\theta R_{\lambda}(\theta)|_{\theta=\theta^*})^2}\,.
\end{equation}
The inferred response function variance (third term in Eq.~\eqref{eqn:appendix_cmse_inference}) follows from the first line of Eq.~\eqref{eqn:err_fluc_variance_inference} as
\begin{equation}
    \Var_{N_I}\left[ \mathbb{E}_{N_E}\left[\thetaest \big| \theta^*, f \right] \big| ~ \theta^* \right] \leq \frac{\delta^2}{(\partial_\theta R_{\lambda}(\theta)|_{\theta=\theta^*})^2} \nonumber \\
\end{equation}
which is computed in the same manner as the bias term.
\subsubsection{Inference-based noisy sensing mitigated by zero-noise extrapolation}
As shown in Sec.~\ref{sec:error_mitigated_inference_bound}, adding error mitigation to inference leads to error bounds of a similar functional form (see Eqs.~\eqref{eq:mse_inferred_lower_bound} and~\eqref{eq:mse_inferred_mitigated_lower_bound}). This bound is evaluated in the same manner as Sec.~\ref{appendix:numerical_evaluation_inference}, with the following changes:
\begin{itemize}
    \item The estimation variance term is scaled by a factor of $\Lambda^2$ (see Eq.~\eqref{eqn:estimation_variance_error_mitigated_inference}, c.f. Eq.~\eqref{eqn:estimation_variance_inference}).
    \item The bound on inference error $\delta$ is estimated from the result of Corollary \ref{corollary:error_mitigated_inference_error_shots}, and therefore we assume that it may be approximated as $\delta\approx\sqrt{\frac{\Lambda^3 \log^3(n)}{\max_j |\gamma_j|N_k}}$. This is a factor of $\sqrt{\Lambda^3/\max_j |\gamma_j|}$ worse than the standard inference error reported in Corollary \ref{corollary:inference_error_shots}.
\end{itemize}

\section{Noise models}
\label{appendix:noise_models}
\begin{figure}
\centering
\includegraphics[width=0.4\textwidth]{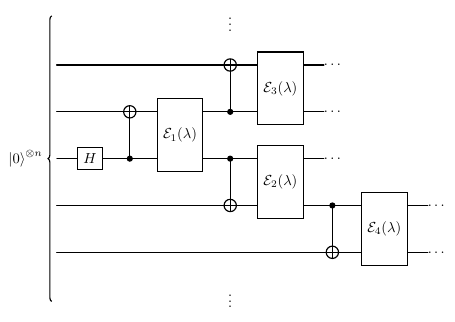}
\caption{\textbf{GHZ state preparation channel for numerical experiments.} To prepare a GHZ state using linear-chain CNOT connectivity, we apply a Hadamard at the center of the chain, followed by CNOTs outwards in a ladder fashion. Each CNOT is accompanied by a noise channel $\mathcal{E}_m(\lambda)$, which could be site-dependent.}
\label{fig:NumericalNoiseChannel}
\end{figure}

In all numerical experiments, we consider noisy preparation of an $n$-qubit GHZ probe state. We assume linear-chain CNOT connectivity and consider the standard preparation circuit involving one Hadamard gate (applied at the center of the chain) and $n-1$ CNOT gates. Each two-qubit gate is accompanied by a CPTP channel $\mathcal{E}_m(\lambda)$ implementing decoherence, and we allow the channels associated with each site to differ in general to model device calibration effects. Where possible, we use the more realistic model outlined in Appendix~\ref{appendix:ibm_eagle_model}, resorting to the local depolarizing model of \ref{appendix:local_depolarizing_noise} when dealing with large systems. In both cases, the state preparation channel depicted in Fig.~\ref{fig:NumericalNoiseChannel} is utilized, with different error channels $\mathcal{E}_m(\lambda)$ depending on context. We detail the specifics of these channels below.

\subsection{IBM Eagle-based model}
\label{appendix:ibm_eagle_model}
For simulations of modest system size (i.e., simulable with density matrix methods on modern hardware), we use a realistic noise model based on the IBM Eagle processor. We include differing noise for each CNOT in the preparation channel, with channels sampled from a sparse Pauli-Lindblad model previously used for studies involving probabilistic error cancellation \cite{berg2022probabilistic,kim2023evidence}. This error model was previously used in Ref.~\cite{goh2024direct}; we reproduce the outline here for convenience.

Each pair of qubits has a unique error model, in order to capture the effects of differing calibration properties of each qubit (which crucially can induce asymmetries of the GHZ magnetometry response function studied in this work). We consider 2-qubit noise channels associated with each CNOT (i.e., no crosstalk effects), constructed in the sparse Pauli-Lindblad form outlined in Ref.~\cite{berg2022probabilistic}. Each channel is generated by a site-dependent Lindbladian $\mathcal{L}_m$, where
\begin{equation}
\mathcal{E}_m(\rho)=\exp\left[\mathcal{L}_m\right](\rho),\quad \mathcal{L}_m(\cdot)=\lambda\sum_{k\in\mathcal{K}}\gamma_{km}(P_k\cdot P_k^\dagger-I\cdot I)
\end{equation}
for $\gamma_k\geq 0$, and where $\lambda$ is an overall parametrization of the base noise level and $\mathcal{K}$ is the set of 2-qubit Pauli operators for the qubits acted upon by the CNOT.

This corresponds to a diagonal Pauli transfer matrix with diagonal terms $f_j=\prod_{k\in\mathcal{K}}w_k+(1-w_k)(-1)^{\langle j,k\rangle}$, where $w_k\equiv(1+e^{-2\lambda\gamma_k})/2$ and $\langle a,b \rangle$ is the binary symplectic product
\begin{equation}
\langle a, b \rangle =  
\begin{cases}
0 \textrm{  when  } [P_a, P_b]=0\\
1 \textrm{  when  } \{P_a, P_b\}=0
\end{cases}\,.
\label{eq:sympl_product}
\end{equation}
One can easily write this in Kraus operator form
\begin{equation}
    \Lambda(\cdot)=\sum_j c_j P_j \cdot P_j^\dagger
    \label{eqn:kraus_form}
\end{equation}
by the Walsh-Hadamard type transform
\begin{equation}
    c_b =\frac{1}{4^n}\sum_a (-1)^{\langle a,b \rangle}f_a\,.
\end{equation}
In our numerical simulations, we generate the Kraus forms of the noise channels Eq.~\eqref{eqn:kraus_form} for a given noise level $\lambda$ and apply these directly to the density matrix. Different $\gamma_{km}$ are chosen for each qubit pairing $m$, sampled randomly from a set of amplitudes learned directly from CNOT processes at different sites in a 127-qubit IBM Eagle device \cite{kim2023evidence}.

\subsection{Depolarizing noise}
\label{appendix:local_depolarizing_noise}
For larger system sizes, we consider a simplified model where two-qubit gates in state preparation are accompanied by a local two-qubit depolarizing noise of fault probability $p$ - that is, for all sites we have $\mathcal{E}_m(\rho)=(1-p)\rho+p\openone$. We write the $m$-qubit GHZ state as $\ket{\operatorname{GHZ}_m}$. After applying a Hadamard, the state is $\ket{\operatorname{GHZ}_1}\bra{\operatorname{GHZ}_1}\otimes(\ket{0}\bra{0})^{\otimes{n-1}}$. Applying a CNOT followed by two-qubit depolarizing channel, the state becomes
\begin{equation}
    \left(\left(1-p\right)\ket{\operatorname{GHZ}_2}\bra{\operatorname{GHZ}_2}+p\frac{\mathbbm{1}}{2^2}\right)\otimes\left(\ket{0}\bra{0}\right)^{\otimes n-2}\,.
\end{equation}
Applying the second CNOT and depolarizing channel yields
\begin{equation}
    \left(\left(1-p\right)^2\ket{\operatorname{GHZ}_3}\bra{\operatorname{GHZ}_3}+\left(1-p\right)\ket{\operatorname{GHZ}_1}\bra{\operatorname{GHZ}_1}\otimes\frac{\mathbbm{1}}{2^2}+p\frac{\mathbbm{1}}{2}\otimes\left(\ket{00}\bra{00}+\ket{11}\bra{11}\right)\right)\otimes\left(\ket{0}\bra{0}\right)^{\otimes n-3}\,,
\end{equation}
and so forth. Iterating this yields a state
\begin{equation}
    \left(1-p\right)^{n-1}\ket{\operatorname{GHZ}_n}\bra{\operatorname{GHZ}_n}+\dots\,,
\end{equation}
where the remaining terms are maximally mixed on the subspace of at least one qubit, and thus have zero contribution to the parity measurement $X^{\otimes n}$. The noisy response function is thus
\begin{equation}
    R(\theta)=\Tr[\left(1-p\right)^{n-1}(e^{-i\theta Z/2})^{\otimes n}\ket{\operatorname{GHZ}_n}\bra{\operatorname{GHZ}_n}(e^{i\theta Z/2})^{\otimes n}X^{\otimes n}]=\left(1-p\right)^{n-1}\cos(n\theta)\,.
\end{equation}
Unlike the more realistic model of Appendix~\ref{appendix:ibm_eagle_model}, this yields a simple exponential reduction of fringe contrast, and all fringes are identical (that is, due to symmetry in the error model the response function has a period of $2\pi/n$, not $2\pi$).

\end{document}